\documentclass[aps,prb,superscriptaddress,showpacs]{revtex4-1}
\usepackage{graphics}
\usepackage[linktocpage=true,
  colorlinks=true,
  pdfborder={0 0 0},
  linkcolor=blue,
  citecolor=red,
  filecolor=yellow,
  bookmarks,
  pdftitle={Electron pairing in the presence of incipient bands in iron-based superconductors},
  pdfauthor={},
]{hyperref}
\usepackage{subfigure}
\usepackage{amssymb}
\usepackage{longtable}
\usepackage{graphicx}
\usepackage{pstricks}
\usepackage{amsmath}
\usepackage{dcolumn}
\usepackage{bm}

\def\k{{\bf k}}
\newcommand{\tbk}{{\tilde{\bf k}}\/}
\newcommand{\bk}{{\bf k}}
\newcommand{\bQ}{{\bf Q}}

\newcommand{\gtwid}{\mathrel{\raise.3ex\hbox{$>$\kern-.75em\lower1ex\hbox{$\sim$}}}}
\newcommand{\ltwid}{\mathrel{\raise.3ex\hbox{$<$\kern-.75em\lower1ex\hbox{$\sim$}}}}
\newcommand{\q}{\bf q}

\newcommand{\vk}{\mathbf{k}}
\newcommand{\vq}{\mathbf{q}}

\newcommand{\comment}[1]{}

\newcommand{\spp}{$s_{++}$}
\newcommand{\spm}{$s_\pm$}

\newcommand{\LOFFA}{LaO$_{1-x}$F$_x$FeAs}
\newcommand{\BFA}{BaFe$_2$As$_2$}
\newcommand{\NFA}{NaFeAs}
\newcommand{\BKFA}{Ba$_{1-x}$K$_x$Fe$_2$As$_2$}
\newcommand{\BFCA}{Ba(Fe$_{1-x}$Co$_x$)$_2$As$_2$}

\newcommand{\LFA}{LiFeAs}

\newcommand{\beq}{\begin{equation}}
\newcommand{\eeq}{\end{equation}}
\newcommand{\bea}{\begin{eqnarray}}
\newcommand{\eea}{\end{eqnarray}}
\newcommand{\D}{\Delta}
\begin{document}

\title{Using Gap Symmetry and Structure to Reveal the Pairing Mechanism in Fe-based Superconductors}

\author{P. J. Hirschfeld}
\affiliation{Department of Physics, University of Florida, Gainesville, FL 32611, U.S.A.}

\begin{abstract}
  I review theoretical ideas and implications of experiments for the gap structure
and symmetry of the Fe-based superconductors.  Unlike any other class of unconventional
superconductors, one has in these systems the possibility to tune the interactions
by small changes in pressure, doping or disorder.  Thus, measurements of order parameter evolution
  with these parameters should enable a deeper understanding of the underlying interactions.   I briefly
  review the "standard paradigm" for $s$-wave pairing in these systems, and then focus on developments in the past several years which have challenged this picture.   I discuss the reasons for the apparent close competition between pairing in s- and d-wave channels, particularly
in those systems where one type of Fermi surface pocket -- hole or electron -- is missing.
Observation of a transition between $s$- and $d$-wave symmetry, possibly via a time reversal symmetry breaking ``$s+id$" state,  would provide an important
confirmation of these ideas.  Several proposals for detecting these novel phases are discussed, including the appearance of order parameter collective modes in
  Raman and optical conductivities.  Transitions between two different types of $s$-wave states, involving various combinations of signs on Fermi surface pockets, can also proceed through a ${\cal T}$-breaking ``$s+is$" state.
I further discuss recent work that suggests pairing may take place away from the Fermi level over a surprisingly large energy range, as well as
the effect of  glide plane symmetry of the Fe-based systems
on  the superconductivity, including various exotic, time and translational invariance breaking
 pair states that have been proposed.
   Finally, I address  disorder issues, and the various ways systematic introduction of disorder can (and cannot) be used to
extract information on gap symmetry and structure.
\vskip .2cm
\today
\end{abstract}


\maketitle

\tableofcontents
\section{Introduction}
\label{sec:intro}
The primary interest in any newly discovered superconducting material is understanding the mechanism by which superconductivity arises.   Although not always sufficient in this regard, determining the symmetry and structure of the energy gap function in momentum space is often an important clue.   Over the past thirty years,  a number of experimental probes, together with simple theoretical modeling, have been devised to deduce information on the gap function in novel superconductors.  In this short review, I consider the Fe-based superconductors (FeSC), discovered in 2008 in the laboratory of Hideo Hosono\cite{Hosono_discovery}, with the intention of collecting what is known about the order parameter in these systems, and discussing the implications for the pairing mechanism\cite{HosonoKuroki2015,CH_PhysToday2015,ThomaleReview2014,HKMReview2011,ChubReview2012,WenLiReview2011}.

Both the question and the answer in Fe-based systems are inevitably more complicated than in, say cuprate superconductors, where it was possible within a few years of their discovery to make a fairly strong case that the pairing symmetry was $d$-wave, and furthermore rather close to the idealized form $\Delta_\k \propto \cos k_x - \cos k_y$ for $\k$ on the Fermi surface for a wide range of materials and dopings\cite{DJS_physrep,Tsuei}. The FeSC exhibit electronic structures which involve several Fe $d$-bands near the Fermi level, and small electron- and hole-like pockets which are quite sensitive to external perturbations such as pressure, doping, etc.  Thus it is perhaps not surprising that across the range of materials considered to fall in this category, the superconducting state  appears also to be quite sensitive, exhibiting e.g. in some materials a full gap, and in others clear evidence for low energy nodal quasiparticle excitations.

 The argument was made early on\cite{Kemper_sensitivity} that these observations of ``sensitivity" or "nonuniversality" of low-energy properties were  prima facie evidence for $s_\pm$ pairing, a superconducting state with the full crystal symmetry but with  sign change between Fermi surface sheets.  This form of order was proposed as a ground state for the Fe-pnictide superconductors  by several authors as a natural way of obtaining unconventional pairing from repulsive Coulomb interactions, based on the structure of the Fermi surface\cite{Mazin,Kuroki08}.     Nodes in an $s_\pm$ state are perforce {\it accidental}, i.e. determined not by symmetry but by details of the pairing interaction, and can thus be created or destroyed without any phase transition at finite temperatures.  Thus such a state  is itself a natural candidate to explain the variability of low-energy excitations across the space of materials.  However, it was also found within spin fluctuation pairing theories that $s_\pm$ pairing competes closely with $d$-wave pairing\cite{Kuroki08,Graser09,Maiti}, and thus possibilities for new states which mix $s$ and $d$ channels  in various ways, including time reversal symmetry breaking phases  are also allowed.  In recent years, theories which pair electrons in an orbital basis have emerged, and led to predictions of even more exotic states  which break parity or translational invariance.  I will argue in this review that there is little microscopic reason to believe that such states can be stabilized.

The general form of superconducting gap function $\Delta_\nu(\k)$, where $\nu$ is a band index and $\k$ is momentum on the Fermi surface, will be difficult to pin down using a single experimental probe.  As has been the case in all previous classes of unconventional superconductors, to the extent a consensus on the gap structure has been  established, it has been because of information gleaned from a variety of experimental probes of  thermodynamics, transport, and, only in some cases, phase-sensitive information from pair tunneling.   Choosing among the various candidates in the Fe-based system is certain to be a nontrivial task, and will require extensive, careful experiments on a wide range of materials.

  The reader new to the field may by now have the impression that the situation in Fe-based systems is so complicated that nothing useful can be learned by studying gap symmetry and structure.   However, the very {\it nonuniversality} of the superconducting state in the Fe-based systems, which makes them seem complicated, can be used to distinguish different gap structures and provide tests of theories of pairing.    Consider the variability of superconducting properties tuned by doping or pressure.  While any number of theories may predict an $s_\pm$ state generically (e.g., at optimal doping),  as is indeed currently the case (see Sec. \ref{sec:standardmodel} below), it is likely that only the correct, materials-specific pairing theory will be able to predict the correct doping and pressure dependence.  Thus, with respect e.g. to the cuprates, there are new types of data available that will constrain theory and thus provide information on how electrons pair.  In the process, one will certainly take steps towards a quantitative theory of some classes of unconventional superconductors that will aid in the search for new, higher-$T_c$ systems.

This is not intended as a comprehensive review of theory or experiment on the FeSC, nor even on the superconducting state of the FeSC.   Instead, I would like it to be provocative,  to collect some of the interesting concepts in unconventional superconductivity that have been inspired by these fascinating materials, with a special emphasis on those newer developments since the appearance of Ref. \onlinecite{HKMReview2011}.  It also represents a rather personal perspective on which ideas represent productive lines of inquiry.  In this spirit, I have certainly neglected a number of novel proposals, particularly those outside the framework of unconventional superconductivity due to  spin fluctuations of itinerant electrons.   With the same motivation, I comment only on those  experiments which seem best to illustrate the relevant concepts.  I apologize to the authors of the many important contributions neglected here.

\section{Concepts of Gap Symmetry and Structure in Fe-based systems}
\label{sec:symmetry}

\begin{figure}[tbp]
\includegraphics[angle=0,width=0.8\linewidth]{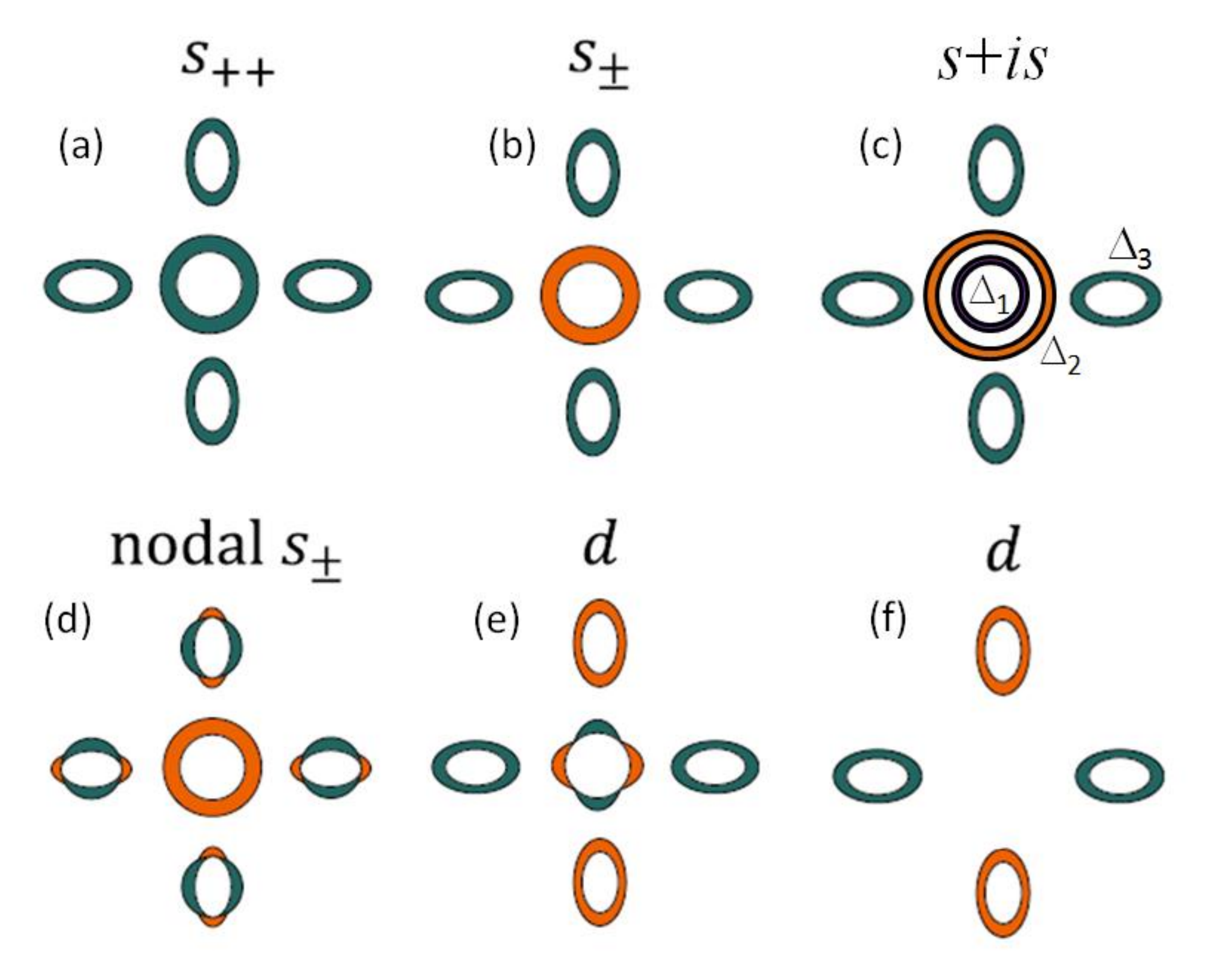}
\caption{(Color online.) Schematic gaps $\Delta(k)$ in FeSC.  Color represents phase of $\Delta(k)$.    (a)-(d) Model Fermi surface with   one hole and two electron pockets.  (a) Conventional $s$-wave ($s_{++}$) state; (b) $s_\pm$ state with gap on hole pocket minus that on electron pockets; (c) antiphase $s$-wave state possible when two or more hole pockets are present, showing gaps with three different phases $\Delta_i$;  (d) similar to (b), but with accidental nodes on electron pockets; (e) $d$-wave state; (f) $d$ wave state in situation with no central hole pocket.
    }
\label{fig:symmetry}
\end{figure}
The  classification of gap symmetries by group theory has  been  reviewed  elsewhere in the general case of unconventional superconductors\cite{SigristUeda}, as well as in the specific case of Fe-based systems, where the unusual glide plane symmetry of the Fe-pnictide or chalcogenide layer and the multiple orbital degrees of freedom introduce some new possibilities\cite{FeSCsymmetry}.  Roughly speaking, the possible states correspond to the irreducible representations of the space group of the crystal.   In the simplest cases, the representations of the point group play the essential role.    Let us consider first a very simple model of the
Fermi surface of a tetragonal crystal with 1 Fe/unit cell, as in Fig. \ref{fig:symmetry}a), with a single, $\Gamma$-centered hole pocket, and $X$- and $Y$-centered electron pockets, and postpone consideration of the subtleties of the glide plane symmetry to Section \ref{sec:eta}.
For the moment I assume spin singlet  pairing (see however Sec. \ref{subsec:singlet_triplet}).  I focus first on simple tetragonal point group symmetry. In a 2D tetragonal system, group theory allows only for four one-dimensional irreducible representations: $A_{1g}$ (``$s$-wave''), $B_{1g}$ (``$d$-wave'' [$x^2-y^2$]), $B_{2g}$ (``$d$-wave'' [$xy$]),  and $A_{2g}$ (``$g$-wave'' [$xy(x^2-y^2)$]),  depending on how the order parameter transforms under crystal rotations by 90$^\circ$ and other operations of the tetragonal group. In Figure~\ref{fig:symmetry} I have illustrated the two symmetry classes, namely $s$-wave and $d_{x^2-y^2}$-wave,
which are most relevant to the discussion of Fe-based systems.    Within the $s$-wave class there are several
 possibilities shown in Fig. \ref{fig:symmetry}a)-d): $s_{++}$, where the gap has the same sign on both electron and hole pockets,    $s_{\pm}$  where the sign changes between the pockets, a more complicated phase-changing $s$-wave
 state allowed if one has more than one hole pocket, and finally a nodal $s_{\pm}$ where the overall sign changes, but the gap varies around at least one set of pockets such that it has (accidental) zeros.

 Note all such $s$ wave cases have the same {\it symmetry}, i.e. none changes sign if the crystal axes are rotated by 90$^\circ$; the differences among the four examples are in gap {\it structure}.  By contrast, the $d$-wave states shown in Fig. \ref{fig:symmetry}e),f)  change sign under a 90$^\circ$ rotation.   If central hole pockets are present, the gap must have nodes on these
 sheets, as in e).  On the other hand,   if such pockets are altogether absent at the Fermi level, as seems to occur for several interesting materials, it is possible to obey $d$-wave symmetry without creating nodes, as shown in f)\cite{footnote_Mazindwave}.

In materials with  inversion symmetry the superconducting order parameter can be characterized by its parity, since the spatial part of
its wave function (the order parameter) can either change sign or remain the
same under the inversion operation. Since electrons obey the Pauli principle, the full wave
function must be antisymmetric under exchange, which also involves spin exchange,  requiring $\Delta_{\alpha\beta}(\k)=-\Delta_{\beta\alpha}(-\k)$, where $\alpha$ and $\beta$ are the spin indices of the two
    electrons\cite{footnote_oddfreq}.  Thus in the absence of spin-orbit coupling the pair wave function  $\Delta_{\alpha\beta}(\k)\propto f(\k)\chi_{\alpha\beta}$ can have  total spin $S=1$ (triplet), with the spatial part of the wave
function $f(\k)$ odd,  $\Delta_{\alpha\beta}(\k)=-\Delta_{\alpha\beta}(-\k)$ or   $S=0$ (singlet) with spatial part  even ($\Delta_{\alpha\beta}(\k)=\Delta_{\alpha\beta}(-\k)$).

Some interesting aspects of this argument in multiorbital systems have described by Fischer\cite{FeSCsymmetry}.  In essence, if one pairs in orbital space one can fulfill the Pauli principle by having the single-particle orbital degrees of freedom help enforce the antisymmetry under particle exchange.  For example, if $\Delta_{\alpha a;\beta b}(\k)$ describes a pairing between single-particle states characterized by (momentum, spin, orbital) $[\k,\alpha,a]$ and $[-\k,\beta,b]$, $\Delta_{\alpha a;\beta b}(\k)$ may  be taken $\propto f(\k)\chi_{\alpha,\beta}\phi(a,b)$ in the absence of spin-orbit coupling.  This allows one to construct states consistent with the Pauli principle which are, e.g. even parity, even under spin exchange, and odd under orbital exchange, or odd parity, odd under spin exchange, odd under orbital exchange.   Such exotic states are generally expected to be energetically disfavored, since they generically involve substantial amplitudes to pair electrons at different single-particle energies.  This is because after performing the unitary transformation to bring the orbital
 pair $ab$ into band space, one will inevitably generate linear combinations involving contributions from electrons in different bands.  e.g. one electron at the Fermi level at $\k\nu$, and one at $-\k\mu$, where $\mu\ne\nu$ are band indices.  Note that it is possible to imagine pair potentials which take advantage of such pairings to lower the energy of the system, but for standard BCS-like
 potentials such energy imbalances in pairing typically leads to strong suppression of the condensation energy.   It is for this reason that most theoretical researchers have made the ansatz of paired electrons in like bands at the Fermi level; however as I argue below in Sec. \ref{subsec:incipient}, there are circumstances in multiband systems where even standard pairing interactions can lead to pairing in incipient bands, and there may be similar arguments favoring the stability of interband pairs (see, e.g. Ref.
 \onlinecite{Sorella2013}).  These problems certainly deserve closer investigation.

%

\section{Experimental overview: SC state of iron pnictides}
\label{sec:expt:pnictides}
Here I briefly review the basic conclusions of Ref. \onlinecite{HKMReview2011} and many others regarding the basic
symmetry and structure aspects of the ``original" pnictide-based FeSC, i.e. the 1111, 122 and 111 arsenides and phosphides.

\subsection{Singlet vs. triplet}
\label{subsec:singlet_triplet}
\vskip .2cm


An equal-spin triplet superconductor 
($S_z=\pm 1$) should polarize in an external magnetic field, just as free spins in a normal metal.  Thus one expects in
such a system that the spin susceptibility (NMR Knight shift) should be featureless at $T_c$. Spin-orbit coupling to
the lattice can suppress  this for some directions, but not for  others.  In a singlet superconductor, on the other hand, or
for the $S_z=0$ component of the triplet, the bound pair cannot be polarized by the applied field, so that
magnetic susceptibility vanishes as $T\rightarrow 0$. Thus for singlet superconductivity one can expect the
uniform spin susceptibility to diminish below $T_{c}$, although the same occurs qualitatively for a state including an $S_z=0$
 triplet component, and vanishing susceptibility is often difficult to determine due to background van Vleck contributions.

 The conventional way to determine  the spin susceptibility is by measuring the Knight shift.  Early on, this
experiment was performed on several FeSC including \BFCA~\cite{Ning}, \LOFFA~\cite{Grafe}, PrFeAsO$_{0.89}$F$_{0.11}$~\cite{Matano}, \BKFA~\cite{MatanoBKFA, Yashima}, \LFA~\cite{Jeglic111, Li111}, and BaFe$_2$(As$_{0.67}$P$_{0.33}$)$_2$~\cite{NakaiPdoped}.
These works reported that the Knight shift decreases for all angles of the field with respect to the crystallographic axes. This effectively excluded triplet symmetries such as $p$-wave or $f$-wave, and there is a general consensus that all FeSC have spin singlet gap symmetry.  Some early reports regarding the intriguing material LiFeAs (Sec. \ref{subsec:LiFeAs}) suggested signs of proximate ferromagnetism, which, when combined with measurements of weak or absent $T$-dependence of the Knight shift below $T_c$ led to the proposal of possible spin triplet pairing in this material\cite{Buechner_LiFeAsNMR,vdBrink_triplet}.  However these measurements have not been confirmed on other samples to my knowledge and remain a mystery.

\subsection{\spm vs. \spp}
\label{subsec:spm_vs_spp}

\vskip .2cm
Thus for many, and probably all of the FeSC, the spin structure of the gap is of the singlet type.
In addition, there are strong theoretical reasons to believe that the momentum dependent part of the
gap function is even parity, and probably $s$-wave (see Sec. \ref{sec:standardmodel}).  Possible exceptions to this
last conclusion include
strongly overdoped systems, where there is some evidence for $d$-wave, discussed in Sec. \ref{subsec:sd_prox}).
Remember from Sec. \ref{sec:symmetry}, the term ``$s$-wave" is used loosely in a crystalline system to  refer
 to a gap with the full symmetry of the lattice.   This does not tell us, for example, whether it changes sign between
 Fermi surfaces, nor whether it has nodes on individual Fermi surfaces.      This question acquires a particular urgency when one recognizes that the theoretical community, while largely endorsing the spin-fluctuation driven $s_\pm$ pairing notion, includes a significant component arguing that orbital fluctuations play an essential role, and that they lead to $s_{++}$ pairing, at least in
 some FeSC (Sec. \ref{subsubsec:orbital_fluctuations}).    Thus this important question of gap structure is also directly related to questions about the origins of superconductivity.

Let us first ask what evidence exists that the gap changes sign at all.    In this regard, there were a few isolated  experiments  of relevance when the forerunner of this review was written in 2011: neutron spin resonance\cite{ChristiansonBKFA,Dai_resonance,Inosov, Lumsden, ChristiansonBFCA, Park, Argyriou, QiuFeSeTe, Babkevich}, Josephson $\pi$-junctions\cite{Chen},
and quasiparticle interference (QPI) experiments\cite{Hanaguri2010,Chi2014}.   In the meantime, further examples of these experiments have turned up on other materials, but no really noteworthy experimental results nor serious new ideas for ``smoking gun" experiments have arisen,  with the exception of some involving disorder.  These include a) new STM measurements of impurity bound states (Sec. \ref{subsec:boundstates}); b) a class of systematic disorder experiments employing electron irradiation (Sec. \ref{sec:gapevolution}); and c) some possible
new ways to analyze QPI data (Sec. \ref{subsec:qpi}).    It has also been noted that the observed coexistence of superconductivity and spin density wave order in some FeSC is strong evidence for $s_{\pm}$ character of the pair state\cite{Fernandes_coexistence_2010}.

 Replicating the pioneering cuprate phase sensitive  experiments\cite{Tsuei2000} that provided hard evidence for $d$-wave
pairing has proven extremely difficult,  for technical reasons related to fabricating clean junctions, but more importantly for
fundamental reasons.    Because the $s_{++}$ and $s_\pm$ states in question have the same angular symmetry, there is
no way to distinguish them by, e.g. tunneling into different faces of a crystal in the analog of the famous corner-junction experiment of Wollman et al.\cite{vanHarlingen}.   Although other schemes have been proposed, they mostly depend on quantitative model calculations of the  Josephson critical currents.  An exception is an test involving a thin film of a conventional $s$-wave superconductor grown on the surface of a putative $s_{\pm}$ proposed by Koshelev and Stanev\cite{Koshelev_Stanev_prox}, who showed that the signs of subdominant tunneling features could identify gap sign changes.  Coupling between such systems can also lead to controlled πjunctions, as shown by Linder et al.\cite{Linder09} Nevertheless, such proposals are not trivial to realize; thus, controlled impurity experiments, which are also in principle phase sensitive, have acquired a new importance.

\vskip .2cm
{\it Spin resonance.} The observation of a neutron spin resonance in the magnetic susceptibility $\chi"(\q,\omega)$  as a test of an $s_\pm$ state was proposed early on
theoretically, \cite{Korshunov2008,Maier,Maier2} and reported experimentally very quickly\cite{Inosov, ChristiansonBKFA, Lumsden, ChristiansonBFCA, Park, Argyriou, QiuFeSeTe, Babkevich} in all the FeSC under study at that time (except
LiFeAs, where it is  weak and occurs at an incommensurate wave vector\cite{BoothroydLiFeAs}).      The  familiar 1111, 122 and 111  materials  have strong magnetic fluctuations in the normal state at $\q=(\pi,0)$, corresponding to the ordered magnetic state of the parent compounds.    I do not review this experimental situation here, nor the various theoretical approaches, but refer the reader to several excellent recent reviews, with focus on both experimental \cite{DaiReview2015,InosovReview2015} and theoretical\cite{BasconesReview2015} issues.

The neutron spin resonance phenomenon is generally understood to represent a paramagnon mode of the system which sharpens only below $T_c$ because the corresponding pole
in the magnetic susceptibility $\chi(\q,\omega)$ becomes undamped due to the suppression of low-energy particle-hole excitations when the Fermi surface is gapped.  The mode is of interest in the current context because the term in $\chi''(\q,\omega)$ arising from the anomalous Green functions in the SC state  is proportional to the coherence factor
\begin{equation}
\sum_\k \left[1 - \frac{\Delta_\k \Delta_{\k+\q}}{E_{\k} E_{\k+\q}}\right]...
\label{eq:cohfac}
\end{equation}
where $...$ is the kernel of the BCS susceptibility.   A cursory examination shows that this factor vanishes if
$\Delta_\k$ and $\Delta_{\k+\q}$ have the same sign, and is maximized if the sign is changed.  Thus the observation
of a neutron resonance has generally been taken as a ``natural" indication of a sign change in the order parameter.  It should be noted, however, that in situations where the quasiparticle relaxation rate in the SC state has a strong energy dependence, a conventional $s_{++}$ state may also have a peak that sharpens somewhat below $T_c$\cite{Kontani_neutron}.   In a simple symmetric two-band model,  the peak is typically located above 2$\Delta$ in energy if generated by the scattering rate effect in an $s_{++}$ state, whereas in the  $s_\pm$ case, as for other unconventional pair states,  it is always located below 2$\Delta$.

\vskip .2cm
Further information on recent developments is available in the review article by D. Inosov in this volume\cite{InosovReview2015}.
\vskip .2cm

{\it Josephson tests.}  An early  experiment on a granular 1111 sample presented  indirect evidence that Josephson $\pi$ loops were present in
 the sample\cite{Chen}.  In this experiment Chen et al.  measured  a very large number of
randomly  pairs of contacting grains with a Nb ``fork",  in hope that some of them would accidentally
fulfil the condition that the two contacts have the required phase shift.   This can occur in a
state where the order parameter changes sign if the numbers that specify the Josephson current (orbital composition
of the wave function, relative gap sizes, etc)  work together  in such a way
that the current along one direction will be dominated by hole pockets and the other
by electron pockets.    A thick tunneling barrier will for example preferentially favor tunneling
from states at the zone center, i.e. hole states.   So a small fraction of the loops containing the fork
and two FeSC grains may contain a $\pi$ phase shift; and this was in fact observed in the experiment\cite{Chen}.
To my knowledge, no further experiments of this type have been performed.

A more promising technique that avoids the difficulties of in-plane tunneling was proposed in Ref. \onlinecite{ParkerPRL}, a ``sandwich" design  whereby an epitaxial film of a hole-doped FeSC is grown on top of an electron-doped one (or vice versa).
If the epitaxial growth is very good, the parallel momentum is conserved at the interface and both types of carriers
experience coherent c-axis transport. An electrical contact to the hole-doped
layer  will then be dominated by the hole current, and the  contact  to the electron-doped layer by the electron
current. If these contacts are now connected in a loop, there will be a  phase difference between the two layers
which will  directly reflect on the sign change of the SC order parameter between the hole and electron sheets.
Progress in fabricating such artificial heterostructures was slow for some time, but success with  interfaces with
SrTiO$_3$ (STO) (See e.g. Ref. \onlinecite{Eom} and citations in Sec. \ref{subsubsec:monolayer}) suggest that these problems are not insurmountable.

 \vskip .2cm
 {\it Disorder effects.}
 In Sec. \ref{subsec:boundstates} below, I  list some  STM  measurements performed in the past few years that have observed impurity bound states  in situations where it appears unlikely that the impurity has a predominantly magnetic character.  This then
 suggests that at least this set of materials has an $s_{\pm}$ character.  Other reports of weak or nonexistent bound states in other
 situations can easily be understood as consistent with $s_\pm$  in terms of simple models of impurity scattering in these systems, but
 there is no way to generate such a resonance in a $s_{++}$ state.  In Secs.  \ref{subsec:Tcsuppression} and \ref{sec:gapevolution}, I discuss the qualitatively new insights provided by a new series of electron irradiation experiments on bulk properties of FeSC.    These are important because they are the closest approximation to introducing purely pairbreaking random potentials to the system in a systematic way. Only in such a  case can results be compared directly to theory.   These experiments show a $T_c$ suppression rate relative to residual resistivity increase which is comparable to the Abrikosov-Gor'kov rate for magnetic impurities in a single-band $s$-wave system, but for nonomagnetic impurities, implying $s_{\pm}$.   Finally, in Sec.  \ref{subsec:qpi}, I discuss  some issues that have arisen regarding the interpretation of experiments purporting to show from Fourier transform scanning tunneling microscopy (STM), commonly referred to as quasiparticle interference experiments, that the order parameter in some systems is $s_\pm$.   The conclusion is probably still correct, but there are subtleties in the analysis, and other ways of analyzing the data than used heretofore are probably preferable.

\subsection{Fully gapped vs. nodal}
 As opposed to the $d$-wave case, where
 nodes are  mandated on the hole pockets by symmetry, in an extended $s$-wave scheme they
may appear on either type of pocket if amplitudes of higher harmonics in the angular expansion of the order parameter
are sufficiently large. As discussed in Section~\ref{sec:standardmodel},
there are microscopic reasons  why this may be the case.
Moreover, since nodeless $s_{\pm}$, nodeless $s_{++}$, and an extended $s$ with accidental nodes (either with average gap switching sign between pockets (``nodal $s_\pm$) or not (``nodal $s_{++}$")) all belong to the same symmetry class, the difference between them is only quantitative (but important).

Information on the existence of nodes, or large gap anisotropy on individual Fermi surface sheets without nodes, is generally easier to obtain than clear indications of gap sign changes.  This is because nodes or deep gap minima result in low-energy quasiparticle excitations, which can be detected in a number of thermodynamic and transport experiments.  Thermal conductivity measurements have played an important role in this debate, primarily because they can be extended to quite low temperatures and can thus often distinguish between true nodes and small gap minima.   Penetration depth experiments have also been quite useful\cite{ProzorovROPP}.    A number of the low-$T_c$ systems, particularly the phosphides, as well as KFe$_2$As$_2$, are found to be nodal.   A simple explanation within the spin fluctuation theory proposed by Kuroki and co-workers suggests that the larger Pn-Fe-Pn bond angle leads to a suppression of the $d_{xy}$ band at the Fermi surface, and thus the elimination of an important intraorbital stabilization of the isotropic  $s_\pm$ interaction (see Refs. \onlinecite{Maieranisotropy},\onlinecite{Kemper_sensitivity} and Sec. \ref{subsec:sd_prox}).   FeSe crystals\cite{ShibauchiFeSe} and thick films\cite{SongFeSe} appear to be nodal as well, and consistent
with predictions of spin fluctuation theory\cite{ChoubeyFeSe,Shantanu15}.

A higher $T_c$  system (optimal 30K) which has been established as nodal is BaFe$_2$(As,P)$_2$, thought to be doped by chemical pressure since As,P are isovalent\cite{footnote_chem_pressure}.   In this material, the $d_{xy}$ FS sheet is present, so alternative explanations for the strong anisotropy were required.   Early 3D spin fluctuation calculations\cite{Graser3D,Suzuki2011} suggested that nodes might occur on hole pockets due to rapid changes in orbital content near the top of the Brillouin zone, and  also exhibited deep gap minima on the electron sheets.  It appears more likely from phenomenological fits to angle-dependent magnetothermal conductivity measurements, however, that these minima descend in real systems  and become the primary nodes in this system in the
 form of electron pocket loops\cite{VekhterShibauchi}.

 Finally, gap anisotropy has been investigated intensively in the context of  BaFe$_2$As$_2$ overdoped with both K and Co.
 In both cases, anisotropy increases signifcantly as one overdopes away from optimal doping, as predicted by spin fluctuation theory\cite{Maiti,HKMReview2011}.   The K-doped side is particularly interesting as the system must make a transition from a fully gapped to a nodal state that may be $d$-wave, a discussion that I postpone until Sec. \ref{sec:s_vs_d}.  The electron-doped side is more anisotropic even at optimal doping, according to $c$-axis thermal conductivity measurements that indicate a low-$T$ linear term, which reflects the presence of nodes, growing immediately away from optimal doping\cite{Reid2010}.   These nodes have also been claimed from analysis of Raman data to be of the loop type\cite{MazinDevereaux}.

It is obviously extremely interesting to study the crossover between nodal and fully gapped systems.  While this is usually discussed in the context of doping or addition of disorder (see Secs. \ref{sec:gapevolution}), it can be driven by pressure or strain as well.   Recently Guguchia et al.\cite{Fernandes_pressure_nodelifting} studied such a transition in Rb-doped Ba122 by measuring changes in low-temperature penetration depth.  Deciding whether the nodal state in such a transition arises because of the onset of accidental nodes in an $s_\pm$ state or a transition to $d$-wave is always a delicate issue, and probes with angular resolution are probably required to make clear statements.

\section{Multiband superconductivity}
\label{sec:multiband}
To provide a background for a few theoretical topics on the frontier of the Fe-based superconductivity field, I first
 review basic ideas about multiband superconductivity within a simple model where interactions are constant on and between bands.  The general BCS gap equation for a multband system with  intra- and interband interactions $V_{\mu\nu}(\k,\k')$ is
\begin{eqnarray}
\Delta_\mu(\k)&=&-{\sum_{\k '\nu}}'V_{\mu\nu}(\k,\k'){\Delta_\nu(\k')\over 2 E_{\nu\k'}} \tanh {E_{\nu\k'}\over 2 T} \\
&\simeq &- \sum_{\nu}\Lambda_{\mu\nu}\Delta_{\nu}L(\Delta_{\nu},T),
\label{Deltak}
\end{eqnarray}
where $\mu,\nu$ are band indices, the prime on the sum indicates a restriction to states within a BCS cutoff energy $\omega_D$ of the Fermi level, and the quasiparticle energy is $E_{\nu\k}=\sqrt{\xi_{\nu\k}^2+\Delta_{\nu\k}^2}$, with $\xi_\k$ the single-particle dispersion measured with respect to the chemical potential.   In BCS theory  $\Delta_\k$ is restricted to states on the Fermi surface, so I can perform
  the energy integration perpendicular to the Fermi surface.    In the second line of  (\ref{Deltak}), following  Matthias, Suhl, and
Walker \cite{Suhl} and Moskalenko \cite{Moskal},    constant densities of states $N_\nu$ for each   Fermi surface sheet $\nu$ and isotropic interactions $V_{\mu\nu}$ are assumed.  In this case the theory may be expressed in simple algebraic form as in (\ref{Deltak}),  in terms of the dimensionless interaction matrix  $\Lambda$ with elements $\Lambda_{\mu\nu}=V_{\mu\nu}N_{\nu}$  and $L(\Delta,T)=\int_{0}^{\omega_{D}}d\xi\tanh(\frac{\sqrt{\xi^{2}+\Delta^{2}%
}}{2T})/\sqrt{\xi^{2}+\Delta^{2}}$.

 In the BCS weak coupling limit, one may derive an expression for $T_c$ in such a system analogous to the single band expression,  $T_{c}%
=\omega_{D}\exp(-1/\lambda_{eff}),$  where $\lambda_{eff}$ is the largest
eigenvalue.   The elements of the eigenvector are then the ratios of the individual order
parameters.
Unless all $V_{ij}$ are the same, the temperature dependence of individual gaps $\Delta_\mu (T)$  does not follow the usual BCS behavior. As an example, consider a two-band superconductor where $\Lambda_{\mu\mu} \gg \Lambda_{\mu\ne\nu}$: in the limit where the interband coupling is zero, there
 are two independent $T_c^0$'s  for $\Delta_1$ and $\Delta_2$.  However, even for very small interband coupling there can be only one thermodynamic transition, such that  the smaller gap opens initially very slowly, and only at a $T$  corresponding to the ``bare" superconducting transition ($\Lambda_{\mu\ne\nu}=0$) does it start to increase rapidly. As the interband coupling increases, the two gaps begin to aquire a similar $T$ dependence.

  Equations~(\ref{Deltak})  may have solutions even when all elements of the interaction matrices $\Lambda$ are
negative, (repulsive according to the above convention)\cite{AS}. The simplest case is a 2-band  repulsive interband interaction:
$V_{11}=V_{12}=0,$ $V_{12}=V_{21}=-V<0.$ In this case the solution reads:
$\lambda_{eff}=\sqrt{\Lambda_{12}\Lambda_{21}}=|V_{12}|\sqrt{N_{1}N_{2}},$ $\Delta_{1}(T_{c})/\Delta_{2}(T_{c})=-\sqrt{N_{2}/N_{1}}$.

The reader who wishes a more detailed account of multiband SC is urged to consult the reviews by Tanaka\cite{TanakaReview}
and Stanev\cite{StanevThesis}.

\section{Microscopics: the ``standard model" and its variants}
\label{sec:standardmodel}
In the first year of study of the Fe-based superconductors, it became clear that the multiband, small pocket aspect of their electronic structure was crucial, and that repulsive Coulomb interactions among the various small pockets found both  in DFT calculations and in ARPES was driving the superconductivity.   From a modeling perspective, it was desirable to begin with the simplest possible band structure, and then add local interactions.
What were the essential features to include in the band modeling?

First-principles calculations\cite{Lebegue2007,Singh2008,Mazin2008,Kuroki08,Cao2008,Ma2008,Haule2008}  showed the electronic
properties and the topology of Fermi surface consisting of hole/electron pockets at the Brillouin zone
center/corners and indicated the importance to superconductivity of Fe 3d orbitals, which give the  dominant density of states (DOS) at the Fermi level. The five Fe-3d orbitals split into the  $e_g$ doublet ($d_{z^2}$ , $d_{x^2−y^2}$ ) and the  $t_{2g}$ triplet ($d_{xz}$, $d_{yz}$, $d_{xy}$) under the tetrahedral crystal
field\cite{Li2008}, and it is the latter that typically dominate the states at the Fermi energy.
Several theoretical analyses\cite{Raghu2008,Dai2008,Han2008,Li2008,Daghofer2008,Graser2008,Qi2008}  of the superconductivity, as well as speculations regarding the symmetry of
the superconducting order parameters of LaO$_{1−x}$F$_x$FeAs based on two-orbital model of the electronic
structure appeared very soon after its discovery. Three-orbital\cite{Lee2008,Daghofer2010}  and more realistic five-orbital Fe-only
tight-binding models  were used together with a  spin-fluctuation interaction\cite{Kuroki08,Graser09},  predicting an $s_\pm$ gap\cite{Mazin} with
various momentum dependences (anisotropy and gap nodes) which are qualitatively consistent with experiments\cite{HKMReview2011}.  In some special cases discussed below, 10- and even 16-orbital calculations have been performed.

While faithful to the crystal symmetries of the DFT calculations from which they were derived, these calculations were performed using
a 1-Fe unit cell, appealing to  an exact unfolding of the bandstructure in the $k_z=0$ plane as discussed by Eschrig\cite{Eschrig2009}.
However, this treatment misses important 3D effects in some systems, particularly 122s, as well as the effect of the nonperturbative band folding potential  due to the out-of-plane As atoms\cite{Lin2011}
on the quasiparticle weights in the unfolded zone.     This has significant consequences for angle-resolved photoemission  in the normal state,
leading to the remarkable variation of spectral intensity around, e.g. the electron pockets\cite{Brouet2012}. In principle, pairing calculations involving all 10 Fe $d$-orbitals in the 2-Fe unit cell should capture these 1-particle effects properly; for a fuller account of some subtleties, see Sec. \ref{sec:eta} below.

\subsection{Spin fluctuation pairing theory}
\label{subsec:sf_pairing}

Virtually all calculations of this type have been performed using a  model (``multiband Hubbard" or ``Hubbard-Hund")
 Hamiltonian consisting of a multiband tight-binding kinetic energy $H_0$ as discussed above, plus an interaction $H_{int}$ containing
 all possible two-body {\it on-site}
interactions between electrons in  orbitals $\ell$,
\begin{eqnarray}\label{eq:H}
H & = & H_{0}+\bar{U}\sum_{i,\ell}n_{i\ell\uparrow}n_{i\ell\downarrow}+\bar{U}'\sum_{i,\ell'<\ell}n_{i\ell}n_{i\ell'}\\
 &  & +\bar{J}\sum_{i,\ell'<\ell}\sum_{\sigma,\sigma'}c_{i\ell\sigma}^{\dagger}c_{i\ell'\sigma'}^{\dagger}c_{i\ell\sigma'}c_{i\ell'\sigma} \nonumber \\
 &  & +\bar{J}'\sum_{i,\ell'\neq\ell}c_{i\ell\uparrow}^{\dagger}c_{i\ell\downarrow}^{\dagger}c_{i\ell'\downarrow}c_{i\ell'\uparrow} \nonumber
\end{eqnarray}
where $n_{i\ell} =  n_{i,\ell\uparrow} + n_{i\ell\downarrow}$. The Coulomb parameters ${\bar U}$, ${\bar U}'$, ${\bar J}$, and ${\bar J}'$ represent a Hubbard interaction, interorbital Hubbard interaction, Hund's rule exchange and
``pair hopping interaction", respectively.    The tight-binding model $H_0$ used in $H$ is typically downfolded from a full DFT  band structure
onto an Fe-only tight-binding model\cite{Eschrig2009}.

In Equation~(\ref{eq:H}), the intra- and inter-orbital
Coulomb repulsion $\bar U$ and $\bar U'$ are written as distinct, along with  the Hund's rule exchange $\bar J$ and
``pair hopping'' interaction $\bar J'$ for generality, but if they
are generated from a single two-body term with spin rotational
invariance they are related by $\bar U'=\bar U - 2 \bar J$ and
$\bar J'=\bar J$.  In a real crystal, such a local symmetry will not always hold.  Work until now has focussed on
generating tight-binding bandstructures for $H_0$ from DFT.  If Coulomb interactions are indeed local, the Hamiltonian  (\ref{eq:H}) may be expected to provide a good description of the physics of a given material, provided one knows or has a method of calculating the interaction parameters $\bar U,...$ (see below).  Attempts have been made to calculate these from first principles\cite{Anisimov,Imada}, but there are some subtle issues associated with the different Wannier bases which have been implemented.  Alternatively, these can be calculated within a GW framework\cite{Kutepov10}.

\begin{figure}[tbp]
\includegraphics[angle=0,width=0.8\linewidth]{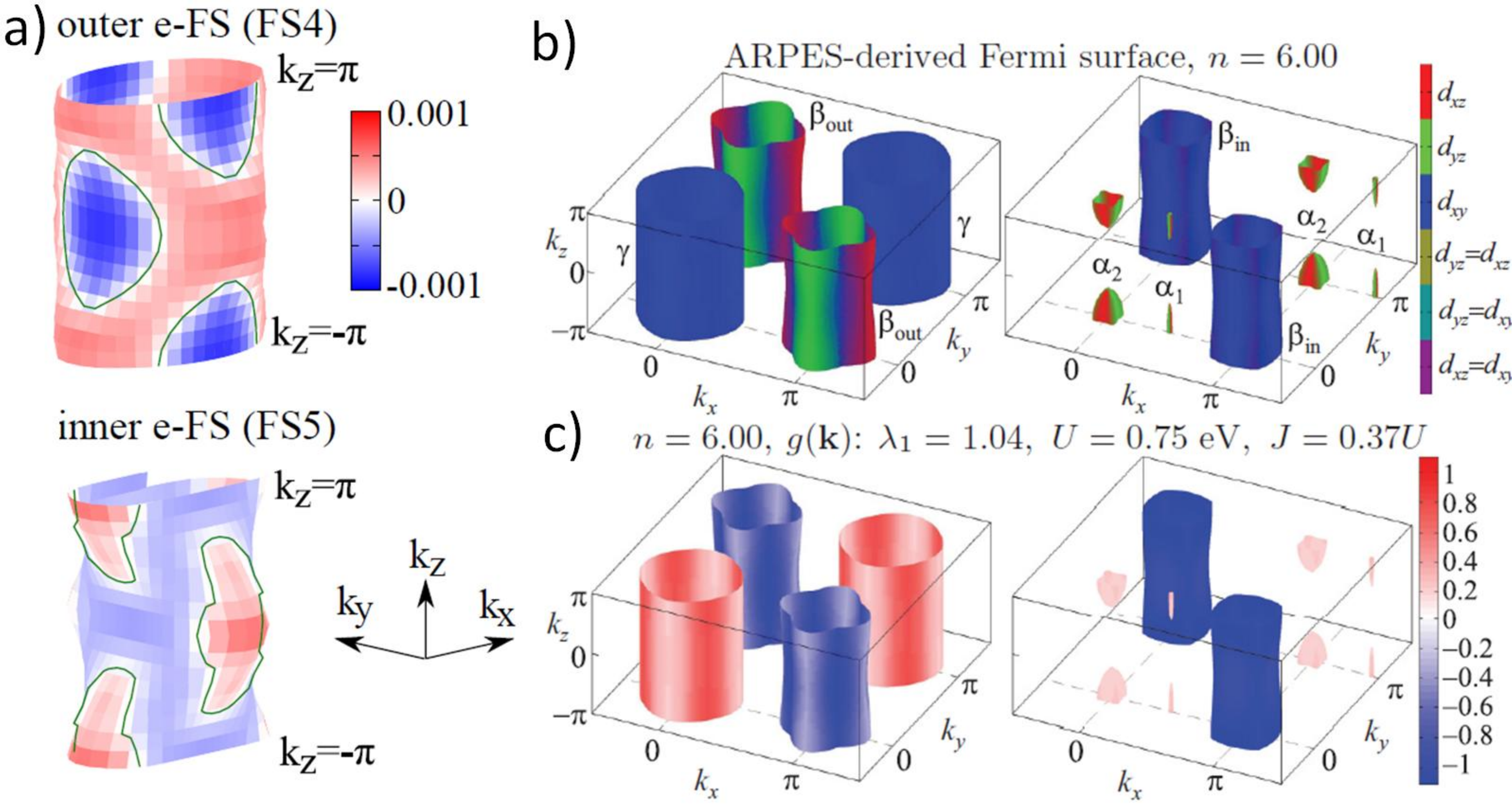}
\caption{ (Color online.) Gap functions on 3D Fermi surfaces of 122 FeSC from 10-orbital fluctuation exchange calculations.   a) on electron pockets of BaFe$_2$(As,P)$_2$   from Ref. \onlinecite{Saito13};  b),c) for LiFeAs from Ref. \onlinecite{Wang13}: b) Fermi surface with orbital content derived from ARPES measurements; c) gap function on these sheets.
    }
\label{fig:3Dsf}
\end{figure}

The generalization of  simple 1-band Berk-Schrieffer spin fluctuation theory\cite{BerkSchrieffer} to the multiorbital case was performed by many authors \cite{ref:Takimoto1,Kubo}.
The effective pair
interaction  vertex $\Gamma(\k,\k')$ between bands $i$ and $j$ in the singlet channel is
\begin{eqnarray}\label{eq:Gam_ij}
{\Gamma}_{ij} (\k,\k') & = & \mathrm{Re}\left[\sum_{\ell_1\ell_2\ell_3\ell_4} a_{i}^{\ell_1,*}(\k) a_{i}^{\ell_4,*}(-\k) \right. \\
&&\left. \times {\Gamma}_{\ell_1\ell_2\ell_3\ell_4} (\k,\k',\omega=0) a_{j}^{\ell_2}(\k')  a_{j}^{\ell_3}(-\k') \right] \nonumber
\end{eqnarray}
where the momenta $\k$ and $\k'$ are confined to the various
Fermi surface sheets with $\k \in C_i$  and $\k' \in C_j$, and $a_i^\ell(\k)$ are
orbital-band matrix elements.    The
orbital pairing vertices $\Gamma_{\ell_1\ell_2\ell_3\ell_4}$ describe the
particle-particle scattering of electrons in orbitals $\ell_1,\ell_4$ into $\ell_2,\ell_3$
 and in the fluctuation exchange
approximation \cite{n_bickers_89, Kubo} are given by
\begin{eqnarray}\label{eq:fullGamma}
&&{\Gamma}_{\ell_1\ell_2\ell_3\ell_4} (\k,\k',\omega) = \left[\frac{3}{2} \bar U^s
\chi_1^{\rm RPA}  (\k-\k',\omega) \bar U^s + \nonumber \right.\,~~~~~~\,\\
&&\,~~~~~\left.
 \frac{1}{2} \bar  U^s
 - \frac{1}{2}\bar U^c  \chi_0^{\rm RPA}  (\k-\k',\omega)
\bar U^c + \frac{1}{2} \bar U^c \right]_{\ell_1\ell_2\ell_3\ell_4},
\end{eqnarray}
where each of  $\bar U^s$, $\bar U^c$, $\chi_1$, etc
represent matrices in orbital space which depend on the interaction parameters.
Here  $\chi_1^{\rm RPA}$  describes the spin-fluctuation
contribution and  $\chi_0^{\rm RPA}$  the orbital
(charge)-fluctuation contribution, determined by Dyson-type
equations as
\begin{align}
\label{eqn:RPA} \chi_{1\,\ell_1\ell_2\ell_3\ell_4}^{\rm RPA} (\q) &= \left\{ \chi^0 (q) \left[1 -\bar U^s \chi^0 (q) \right]^{-1} \right\}_{\ell_1\ell_2\ell_3\ell_4},\\
 \chi_{0\,\ell_1\ell_2\ell_3\ell_4}^{\rm RPA} (\q) &= \left\{ \chi^0 (q) \left[1 +\bar U^c \chi^0 (q) \right]^{-1} \right\}_{\ell_1\ell_2\ell_3\ell_4}.
\end{align}
where repeated indices are summed over.  Here $\chi_{st}^{pq}$ is a generalized multiorbital susceptibility (see \onlinecite{Kemper_sensitivity}).


{\it Results of microscopic theory.}
The simplest goal of the microscopic approach is to calculate the critical
temperature $T_c$ via the instability equation and determine the symmetry of the pairing eigenfunction and the
pairing eigenvalue (which determines $T_c$) there.
If one writes the order parameter
$\Delta(\k)$ as $\Delta g(\k)$, with $g(\k)$ a dimensionless function
describing the momentum dependence on the Fermi
surface, then $g(\k)$ is given as the stationary solution of
the dimensionless pairing strength functional \cite{Graser09}
\begin{equation}
\lambda [g(\k)] = - \frac{\sum_{ij} \oint_{C_i} \frac{d k_\parallel}{v_F(\k)} \oint_{C_j}
\frac{d k_\parallel'}{v_F(\k')} g(\k) {\Gamma}_{ij} (\k,\k')
g(\k')}{ (2\pi)^2 \sum_i \oint_{C_i} \frac{d k_\parallel}{v_F(\k)} [g(\k)]^2 }
\label{eq:pairingstrength}
\end{equation}
with the largest eigenvalue $\lambda$,  a dimensionless measure of the pairing strength.
 Here  $\k$
and $\k'$ are restricted to the various Fermi surfaces $\k \in C_i$
and $\k' \in C_j$ and $v_{F}(\k) = |\nabla_\k E(\k)|$ is the
Fermi velocity on a given Fermi sheet $i$.   Note the band indices on $v_F(\k)$, etc. have been suppressed because specification of $\k$ also specifies the band.

 The most sophisticated calculations of this type are performed for the 122 systems, which are the most 3D of the FeSC due to the stronger overlap of the pnictogen/chalcogen between the layer containing Fe and the reservoir layers\cite{Ba122_Kemper09} , and a few other cases where 3D Fermi surface pockets exist, like LiFeAS.   In these cases 2D calculations are not sufficient, and one needs to account for 3D scattering processes.  Two  examples are shown in Fig. \ref{fig:3Dsf}, indicating the rather complex gap structures $g(\k)$ that can be obtained.

\subsection{LiFeAs: challenge for theory}
\label{subsec:LiFeAs}
In earlier classes of unconventional superconductors, consensus on a particular pairing state was generally reached only after different classes of experiments, including a) bulk thermodynamic and transport
 measurements that probe the averaged quasiparticle spectrum, b) phase sensitive experiments, typically requiring surfaces or interfaces, which probe the order parameter sign changes directly, and c) ARPES,
 which in principle measures the momentum-resolved gap structure $|\Delta_\k|$.  In the Fe-based systems, designing and fabricating devices for phase sensitive measurements has proven problematic.  Furthermore, while  ARPES experiments should in principle be the most direct measurements of  gap structure,  early ARPES measurements indicated  insignificant $\k$-space anisotropy in the 122 systems, despite the fact that thermodynamic probes all reported the presence of low-energy quasiparticles consistent with large anisotropy and even nodes in several cases, particularly Ba(Fe$_{1-x}$Co$_x$)$_2$As$_2$.   Poor surface quality may have limited angular resolution in these cases.

\begin{figure}[tbp]
\includegraphics[angle=0,width=0.9\linewidth]{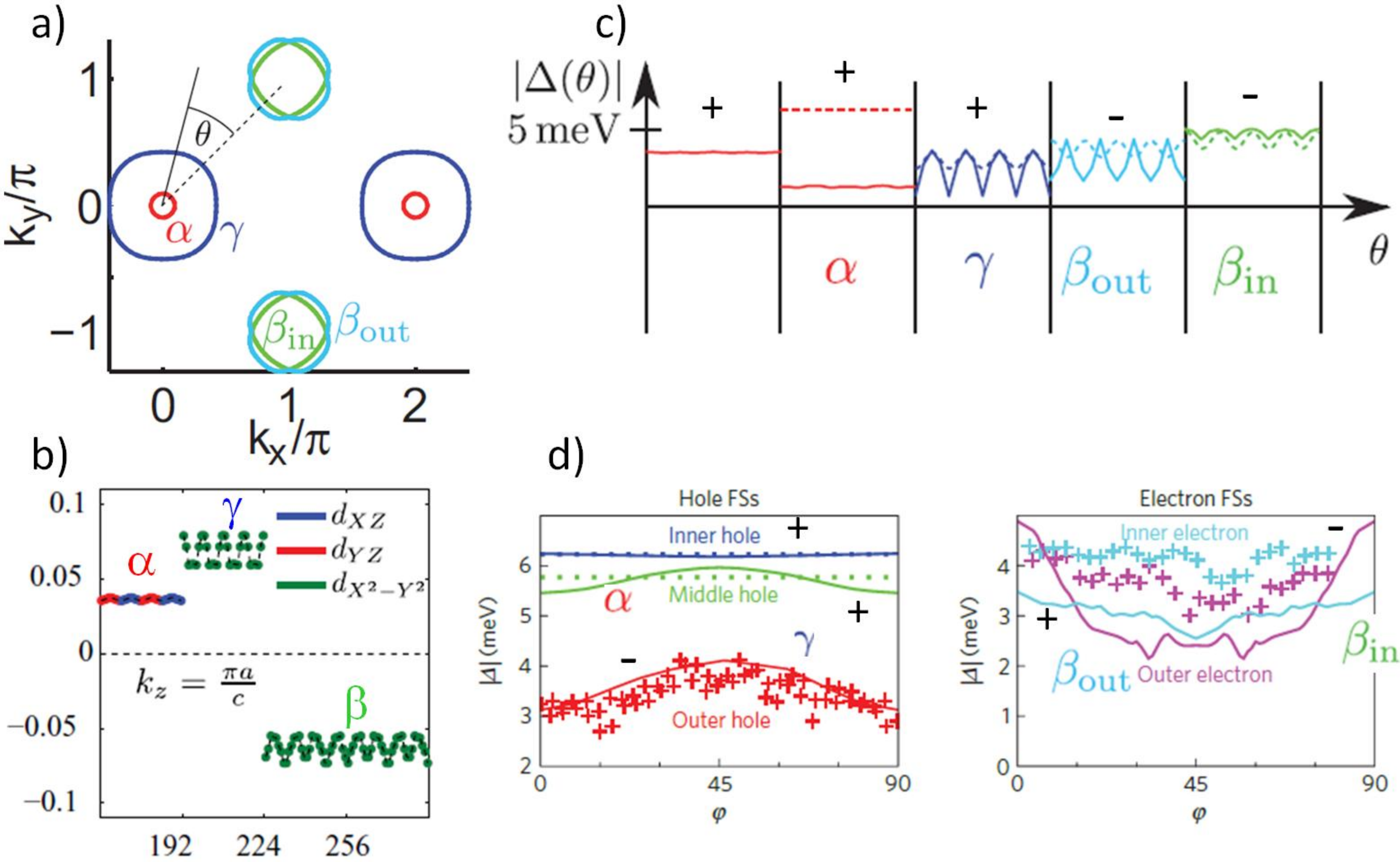}
\caption{(Color online.) (a)  LiFeAs Fermi surface at $k_z=\pi$ for ARPES-derived band structure used by Refs. \onlinecite{Wang13},\onlinecite{Ahnetal},\onlinecite{KontaniLiFeAs}.  (b) Functional renormalization group calculation of
$k_z=\pi$ SC gaps
from Ref. \onlinecite{ThomaleLiFeAs}, where DFT Fermi surface was used.    (c)  Gaps at $k_z=\pi$ predicted by RPA spin fluctuation calculation of Ref. \onlinecite{Wang13}.  Solid lines are theory; dashed lines are approximate sketches of experimental determination of same gaps\onlinecite{Borisenko_Symmetry12}.     (d) Comparison of gap predictions of Ref. \onlinecite{Yin2014} (LDA+DMFT-based  calculation of the pairing interaction) with same experimental data.
    }
\label{fig:LiFeAs}
\end{figure}

 Thus a good deal of attention was focussed on the first high-quality crystals of LiFeAs, which were shown to have excellent nonpolar surfaces without significant reconstruction\cite{Lankau10}.  Both detailed STM-quasiparticle interference\cite{DavisLiFeAs12} and ARPES measurements\cite{Borisenko_Symmetry12,Umezawa12}  of the momentum dependence of the superconducting gaps on the various pockets were reported, and found to be qualitatively consistent with one another where comparisions were possible.  Aspects of the measured  gap function in  LiFeAs, according to these measurements, were a) the gap was nodeless; b)  gap sizes varied from a minimum on the large $\Gamma$-centered hole pocket ($\gamma$)  of about 3 meV to a maximum of about 6meV on the inner hole pockets ($\alpha$); c) significant angular variation was reported on both the two electron pockets ($\beta$) and the outer hole pocket.  Note that the various ARPES groups disagreed somewhat on the details of the Fermi surface itself: according to Ref. \onlinecite{Borisenko_Symmetry12},  around the $\Gamma$ point of the zone there was only a single hole pocket $\gamma$; the $\alpha_1$ and $\alpha_2$ hole bands were below the Fermi level, whereas Ref. \onlinecite{Chi2014} found that $\alpha_2$ indeed crossed the Fermi level near $\Gamma$.    Some of these data are reproduced in Fig. \ref{fig:LiFeAs}.

This rough concordance of several experimental groups and methods on the detailed structure of the gap function
in one of the Fe-based materials led several theoretical groups\cite{Wang13,Ahnetal,Yin2014,KontaniLiFeAs}
to consider the challenge seriously.   In what follows, I give a somewhat detailed summary of these efforts, not because the details of the gap function in LiFeAs are fundamental to the physics of Fe-based superconductors, but because the agreements and discrepancies illustrate the level at which
 the field of materials-specific calculation of superconducting properties has arrived for these systems.
 The general reader can skip this discussion and glance at Fig. \ref{fig:LiFeAs}, where a subset of these
 results are presented, indicating that rather good agreement among the recent generation of calculations is obtained for the larger Fermi surface pockets in the system, and discrepancies remain for those tiny, possibly incipient  pockets where the interactions are harder to calculate.

 The earliest results of Thomale et al\cite{ThomaleLiFeAs} had been based on a 2D calculation using a tight-binding model derived from a density functional theory calculation.  However, the ARPES measurements had shown that the true Fermi surface in this material differed considerably from these DFT calculations, due to the small to negligible
size of the inner $\alpha_1, \alpha_2$  hole pockets observed by ARPES
compared to the relatively large sizes found in DFT.
Local-density approximation (LDA)+dynamical mean-field
theory (DMFT) calculations had some success showing that stronger electronic correlations
in the 111 family  lead to a shrinkage of the inner hole pockets
but maintenance of the electron pocket size and shape\cite{Yin2012,Ferber2012,Lee_etal_Kotliar2012}, but still did not really reproduce the ARPES Fermi surface accurately.

An improved approach adopted by several theoretical groups was to use a tight-binding model for the low-energy bands derived from  ARPES\cite{Wang13,Ahnetal,KontaniLiFeAs}.  These  calculations  represented  various versions of what may be called weak-coupling spin fluctuation theory, and all considered pairing only at the Fermi level.    The remarkable thing was that, despite their differences in implementing this theory, all arrived at momentum-dependent  gap functions  for the electron and outer hole pockets semiquantitatively in agreement with experiment, with the
tuning of at most one or two parameters within a reasonable range (Ref. \onlinecite{Ahnetal} was more phenomenological in approach).  They disagreed, however, on the size of the gaps on the rather small inner hole $xz/yz$ Fermi surface pockets reported by ARPES,   those which undergo Lifshitz transitions first upon electron doping.  Experiment reported these gaps to be the largest, but e.g. Ref. \onlinecite{Wang13} found them to be intermediate to small among the various gaps in the system.     In some works, better agreement with the gap sizes on the smaller hole pockets was found, based on claims of improved formulation of the problem in orbital space\cite{Yin2014,KotliarTremblay15}  or inclusion of Aslamasov-Larkin vertex corrections to the RPA theory\cite{KontaniLiFeAs}.  In the only fully 3D spin fluctuation pairing calculation, Ref. \onlinecite{Wang13},  the small gaps found on these tiny pockets were attributed to the neglect of states away from the Fermi level.  This point
  is discussed further in Sec. \ref{subsec:incipient}.

  A further point of contention among these theories was the distribution of gap signs over the various Fermi pockets.   While Ref. \onlinecite{Wang13} found a traditional $s_\pm$ state, with all hole pockets one sign and all electron pockets the other,   Yin et al.\cite{Yin2014}  reported that their leading pair state had the signs of the electron
  and outer hole pockets equal (``orbital antiphase $s$-state), while Saito et al.\cite{KontaniLiFeAs} found both the electron pockets and the outer hole pockets with sign opposite the ``traditional" $s_\pm$.  Recently, Nourafkan et al.\cite{KotliarTremblay15} reported a LDA+DMFT calculation, improved
relative to Yin et al.\cite{Yin2014},   which also found a leading $s_\pm$ state.

  Needless to say, experiments are not yet in a position to confirm any of these last predictions.   They arise in the theories because of relatively subtle differences in the treatment of the various interband interactions, which are beyond the scope of this review, and may also to some extent depend on errors in the band structure adopted, since there is still some controversy about the small hole pockets.    Nevertheless, one may be reasonably optimistic that, as experiments and crystal growth techniques improve, and some kind of consensus is reached about how to improve the simplest spin-fluctuation theories,  the role of these small pockets (and incipient bands) will be elucidated, and the pairing state in the  LiFeAs system will be understood.  If this turns out to be the case, it will be a remarkable triumph for
  theory, as this system is quite complicated, and (due to lack of nesting) relatively far from the ``standard" situation in the FeSC where strong $(\pi,0)$ magnetic fluctuations dominate the excitation spectrum.  Of course, it is possible that some  ingredients, e.g. phonons, have been left out of these approaches,  and may be essential to achieve a quantitative description.

\subsection{ Alternative pairing theories}
\label{subsec:alternative}
\subsubsection{Orbital fluctuations}
\label{subsubsec:orbital_fluctuations}
There are many theoretical approaches in addition to the weak-coupling spin fluctuation exchange approach described in Sec. \ref{subsec:sf_pairing},
of which I treat only two here, because they have received the most attention, and (apparently) involve the inclusion of different physics.
If one looks at Eq. \ref{eq:fullGamma}, one sees that instabilities are in principle possible both in the spin $\chi_1$ and charge $\chi_0$ channel.  Most of
the community has noted the proximity of long range magnetism to superconductivity, as well as the   strong magnetic fluctuations observed in NMR and inelastic neutron scattering,
 and concluded that spin fluctuations dominate the pairing in most FeSC.    From a theoretical standpoint, if one takes Eq. \ref{eq:H} with local interactions $U,J,U',J'$ as representing the essential physics,  one can easily show for realistic models that for sensible values of these parameters $\chi_1$ is singular and dominates over $\chi_0$.

 The  importance of orbital degrees of freedom  was noted  by several authors in the context of orbital ordering in the Fe $d$ states at the orthorhombic transition, usually in the context of localized models\cite{Zaanen_orborder,Sachdev_orborder,WeiKu_orborder,Barzykin_orborder}.
In itinerant models, orbital order usually requires concomitant spin stripe order or sufficiently strong fluctuations.   The competition between theoretical models of
``spin fluctuations" and ``orbital fluctuations"  is sometimes discussed as if it were an either-or situation, but in fact all models used contain both types of fluctuations.  The  RPA treatment of Sec. \ref{sec:standardmodel} yields the ``fluctuation exchange" expression (\ref{eq:H}) for the effective interaction, which manifestly contains the enhanced contribution of the charge/orbital fluctuations ($\sim\chi_0$), although it is sometimes referred to as ``spin fluctuation pairing theory".   The point is that for physical values of the {\it bare} interactions $U,J,U',J'$, the spin fluctuation terms ($\sim\chi_1$) are much larger since one is close to a magnetic instability.
It is  easy to check\cite{Kemper_sensitivity} that if one artificially makes $U'>U$ in (\ref{eq:H}), the singularity in the charge, or orbital channel, can be favored as well.
Ab initio calculations \cite{Imada} show, however, that $\bar U\gg \bar U'$, as one would expect on fundamental quantum mechanical grounds.    Pairing by  orbital fluctuations, as proposed by Kontani and co-workers \cite{Kontani_neutron,Kontani_Hubb_Holst,Kontani_bond_angle,Kontani_nodal,KontaniLiFeAs},  therefore focussed initially on finding  ways of enhancing the effective interorbital Coulomb interaction.      They showed that in the event such an enhancement occurred,  the ground state of such a system would be of the $s_{++}$ type, since the efffective interaction due to orbital fluctuations is attractive.

 Initially,  Kontani and Onari \cite{Kontani_Hubb_Holst} focussed on the role of certain in-plane Fe phonons that  can in principle enhance the interorbital scattering processes such that they dominate the spin fluctuation part of the interaction (the effect of a different phonon was discussed by Yanagi \cite{Yanagi}). The electron-phonon coupling to these phonons was included in an RPA-type calculation\cite{Kontani_Hubb_Holst}, suggesting that phonon-mediated enhancement of
 orbital interactions to drive $s_{++}$ pairing was feasible.  However a full first principles calculation of the same vertex by Nomura et al.\cite{Arita14} found that
 this effect was much too weak to overcome the dominant spin fluctuation interaction.

More recently, the enhancement of the orbital fluctuation interaction has been sought in extensions of the electronic model to include additional
vertex corrections beyond the RPA in the calculation of the Cooper vertex\cite{Kontani_vertex,KontaniLiFeAs}, in particular Aslamasov-Larkin type diagrams.
More accurate results are claimed, but the physical meaning of these additional processes is not transparent.
Thus, to  which extent this proposed  orbital-fluctuation
mechanism  is at work  in real materials remains open.  As discussed in Sec. \ref{subsec:spm_vs_spp}, there is some evidence in FeSC for sign-changing order parameter behavior, but it is still weak and not all tests can be performed in all materials.  For many of these tests, which have ``natural" explanations in the context of
spin fluctuation pairing, there are alternative explanations.  In many cases,  claims of weak impurity pairbreaking\cite{Onari} (see however Sec.\ref{sec:disorder}), the damped inelastic  neutron scattering resonance observed in most FeSC \cite{Kontani_neutron}, and a simple explanation \cite{Kontani_bond_angle} of the Lee plot (maximum of $T_c$ within family at tetrahedral Fe-As angle \cite{Leeplot}) keep interest in the orbital fluctuation pairing idea alive, or at least muddy the waters.

It is of course possible that, due to varying interactions strengths, orbital fluctuations dominate in some of the FeSC and spin fluctuation in the others.   Particularly in systems like bulk FeSe (Sec. \ref{subsubsec:bulkFeSe}), it is not clear that spin fluctuations are strong enough to dominate pairing.    Recent papers from the Kontani group indeed appear to emphasize that the properties of many FeSC, particularly the formation of nodal structures in superconducting gaps, depend on the interplay of both types of fluctuations\cite{Kontani_orb-sf1,Kontani_orb-sf2}.

\subsubsection{``Strong coupling" theories}
\label{subsubsec:strongcoupling}

Interactions in the FeSC are not in the weak coupling limit; even the systems showing perhaps the weakest electronic correlations, like LaFePO, manifest
renormalization of the low-energy dispersions of factors of 1.5-2 relative to DFT, while other reported mass renormalizations are somewhat higher,
and there are significant band shifts as well.    Nevertheless FeSC
are clearly also not Mott insulators,  so a lively debate has existed almost since the beginning of the field about which starting point is the most likely to capture
the essential physical aspects of these systems.   While most of this discussion has played out in the arena of the normal state, together with the related debate regarding the degree of itinerancy appropriate for  models of the magnetism, the debate has also spilled over into the neighboring stage where the origin of superconductivity is discussed.    I do not focus on issues of
correlation strength, magnetism, nematicity, or Mott selectivity, although I will touch on them elsewhere in the text, and in the conclusions.  These topics are all reviewed in up-to-date accompanying articles by Bascones et al.\cite{BasconesReview2015}, v. Roekeghem et al.\cite{BiermannCRAS}, and B\"ohmer and Meingast\cite{BohmerMeingastCRAS}.

The so-called ``strong-coupling'' approach to pairing, based on the $J_1-J_2$ generalized Heisenberg model for the magnetism in these systems \cite{SiAbrahams, CFang,Sachdev_orborder}  begins from the premise that that magnetism is nearly localized.   Since the FeSC are metals, one might ask whether or not
a localized description is  ruled out a priori,  but in fact magnetic excitations in metals are reasonably   well described by DFT, which for the FeSC gives a characteristic energy range for the magnetic interactions of order 100 meV or larger decaying in space as a power law, such   that  the moments are indeed quite localized.
However actual band structure calculations \cite{Mazin2008, Yaresko, Antropov} also show that the magnetic excitations in the FeSC cannot be mapped onto those of a generalized ($J_1-J_2-J_3...)$ Heisenberg Hamiltonian of any range. They can be mapped  onto a Heisenberg Hamiltonian with biquadratic exchange\cite{Antropov}, or possibly to a more
complicated Hamiltonian (such as ring exchange), but not onto a pure
Heisenberg model.     The biquadratic exchange eliminates the need for the unphysical in-plane $J_{1a,b}$ anisotropy of the nearest neighbor
exchange which had been used to fit magnon excitations in FeSC parent compounds\cite{DaiReview2012}.

The strong-coupling approach\cite{Seoetal} assumes $H=H_0 + H_{int}$, where $H_0$ is a multiband tight-binding model, and $H_{int}=\sum_{ij} J_{ij}{\bf S}_i\cdot {\bf S}_j$,
where typically only  $J_1$,  $J_2$ $J_3$ (1st-3rd neighbors) are kept, although recently the effect of the biquadratic term $\sum_{ij}K_{ij}({\bf S}_i\cdot {\bf S}_j)^2$
has also been considered\cite{YuNevidomskyy}.   Such models are of course reminiscent of the $t-J-J'$ model used in the cuprates, but since (for the common pnictides and chalcogenide-based systems with near 6 electrons/Fe) there is no proximity to a Mott insulator,  there is no corresponding strong coupling limit of a Hamiltonian with local Coulomb interactions.  Also unlike the
$t-J$ or $t-J-J'$ models, there is no justification to project the  kinetic energy onto a reduced Hilbert space, and this is generally not done (when it is done, of course it makes little difference).    There are related models, equally {\it ad hoc},  where $U$ and $J_{1,2}$ terms are treated as being independent before the mean field decoupling\cite{Daghofer1, Daghofer}.

 The interaction is then decoupled in the pairing channel in mean field theory,  such that the nearest neighbor exchange $J_1$ induces competing $\cos k_x + \cos k_y$ and $\cos k_x-\cos k_y$ ($s$- and $d_{x^2-y^2}$-wave) pairing harmonics, while the next nearest neighbor exchange leads to $\cos k_x \cos k_y$ and $\sin k_x \sin k_y$ ($s$- and $d_{xy}$-wave).   General phase diagrams for 2- and 5-orbital band models were worked out in Ref. \onlinecite{Goswami}, who found that with $J_2 \gtrsim J_1$, $\cos k_x \cos k_y$
  was the leading instability, leading to a nodeless $s_\pm$ state, but that $d$-wave was a close competitor.  The results of this approach intially showed
 an intriguing set of circumstantial agreement with the predictions of itinerant weak-coupling models, despite the lack of any fundamental connection between the two approaches.  Indeed, the pairing symmetry in any weak-coupling spin-fluctuation model is primarily determined  by optimizing the structure of the spin fluctuations to the given Fermi surface.  The Fermi surfaces are the same or similar in the two approaches, and   Wang \textit{et al}\cite{WangDHLee09}
 have shown that the  low energy spin and charge excitations in the fRG treatment of the 5-orbital model~(\ref{eq:H}) overlap very well with those of the $t-J_1-J_2$ model as well.   Since both are rather similar in the two approaches, not surprisingly, the main results seem to agree.

Despite these successes, I believe these approaches are less reliable, for the following reasons:

\begin{itemize}
\item they unphysically  separate the itinerant electrons and the local moments, as if the latter were coming from a separate atomic species. Of course  the moments are formed by exactly the same Fe $d$-electrons that form the band structure, which also mediate the magnetic interaction mapped onto the $J_1-J_2$ Heisenberg Hamiltonian.

\item  actual band structure calculations \cite{Mazin2008, Yaresko, Antropov} can never be mapped onto a Heisenberg Hamiltonian of any range.  In some cases  mapping  onto a Heisenberg Hamiltonian with biquadratic exchange is possible, but in others even such mappings lead to large deviations from DFT (see, e.g. Ref. \onlinecite{glasbrenner15}).

\item  by including low-order Brillouin zone harmonics,  ``strong-coupling" models specify the shape of the spin-fluctuation
induced interaction; therefore, the resulting solution to the gap equation specifies the structure of the gap nodes in momentum space, so that the amplitude,
anisotropy and possible nodes on actual FSs depends only on the  proximity of these FSs to the nodal lines in the Brillouin zone. This result is quite different from  the weak coupling calculations and is likely unphysical.    Another consequence of the low-order
harmonics is that it is impossible for such theory to capture the ``antiphase" $s$-wave states that are close competitors in some
systems such as LiFeAs, which in spin fluctuation theory arise due to small-$q$ repulsive interactions.   As discussed in Sec. \ref{subsec:LiFeAs}, these  involve sign changes among gaps on Fermi surfaces that are in close
proximity in ${\bf k}$ space,  and it seems unlikely that the low-order harmonics will ever vary fast enough to drive this effect.  In
  a similar vein, antiphase $d$-wave states (opposite sign $d$-wave order parameters on two $\Gamma$-centered hole pockets, for example)  which should be the ground state in a strongly overdoped system, are hard to
   understand in the strong coupling picture.
\end{itemize}

The partial agreement between fRG, RPA, FLEX and the $t-J_1-J_2$ (+ more recent approaches including biquadratic couplings) results is nevertheless an interesting open question that deserves further study, but is certainly related to
the fact that the spin excitations in all these models are peaked at roughly similar momenta.    Essentially this question maps onto the larger question of understanding when band structure calculations, including those incorportating correlations in various schemes, capture correctly the observed spin excitations in experiment, and deciding whether these excitations can ever be properly described by an effective localized model\cite{BasconesReview2015}.

\section{Pairing at the extremes}
\subsection{s vs. d}
\label{sec:s_vs_d}

\subsubsection{Proximity in spin fluctuation theory}
\label{subsec:sd_prox}
As discussed above, the ``standard model" arising from elementary considerations on pairing by repulsive interactions suggests an $s_\pm$ pair state in systems
with hole and electron pockets separated by wavevector $(\pi,0)$.    Nevertheless, there are competing pairing interactions and other effects that lead to anisotropy in the $s_\pm$
function, including orbital weights on the Fermi surface, the frustrating effect of $(\pi,\pi)$ pair scattering processes between electron pockets, and intraband Coulomb interactions\cite{Maieranisotropy,Skepnek09}.  Obviously the electronic structure, and how it changes with doping, will be the primary factor influencing the relative weights of these various
effects.   Kuroki et al., for example, famously noticed that the pnictogen height controlled a $d_{xy}$ band near the Fermi level,  whose existence bolstered the $(\pi,0)$ interactions and
therefore made the superconducting gap more isotropic\cite{Kuroki_pnictogen_ht_2009}.    However in cases of extreme overdoping, one type of pocket must disappear, electron or hole, leaving one with a situation where the usual $(\pi,0)$ pair scattering is strongly suppressed or absent. In the strongly electron overdoped case, $(\pi,\pi)$ scattering will be most prominent,
favoring $d$-wave pairing\cite{FWangdwave,Maierdwave} (Fig. \ref{fig:symmetry}(e)).  In the overdoped hole-pocket case,  the leading pairing instability is not intuitively clear from such arguments.

In fact, the $d$-wave pair channel was recognized early on as being a strong competitor for the $s_\pm$ state\cite{Kuroki08,Graser09}.  Graser et al.\cite{Graser09}, using a multiband RPA calculation of the spin fluctuation pairing vertex, even found that the $d$-wave pairing interaction dominated the $s$-wave for  smaller  values of the local Coulomb repulsion $U$ in the case of a band suitable for 1111 systems near 6 electrons/unit cell.    Maiti et al.\cite{Maiti}  then mapped the model of Graser et al.  onto a conceptually simpler one with low-order Fermi surface angular harmonics, and showed how the amplitudes of these harmonics evolved with doping for the 1111 band structure.  They concluded that the form of the  interaction in the highly overdoped case was consistent with a strong tendency towards $d$-wave pairing in both the  hole- and electron-doped extremes.

Interest in the competition between $s$ and $d$ states grew with measurements on the low-$T_c$ highly hole-doped superconductor KFe$_2$As$_2$.   Some experiments, in particular thermal transport,
were interpreted in favor of $d$-wave superconductivity\cite{taillefer}, and early functional renormalization group treatments of this system also predicted $d$-wave\cite{thomale_kfe2as2}. Such a picture was also consistent with other experiments clearly indicating the existing of gap nodes\cite{Hashimoto10}.
  However, such  bulk experiments lack the resolution to determine easily the position of these nodes, and shortly thereafter ARPES measurements, in principle a more
  direct measurement of the gap,  reported an $s$ wave gap with four symmetrically placed pairs of nodes on the intermediate hole band\cite{okazaki}.  While there are no obvious
  problems with the ARPES identification, reconciling the gap determined in this way with thermodynamics has proven difficult\cite{hardy_spht}, and this system remains controversial.  From the theoretical standpoint, a state with nodes of this type has been obtained phenomenologically in Ref. \onlinecite{ChubukovKFe2As2}, and was recently proposed to arise ``naturally" from a non-standard representation of $s_\pm$ states as ``orbital triplet" structures\cite{ColemanSchmalian}.

Perhaps even more interesting is the question of how, if the stoichiometric system KFe$_2$As$_2$ is $d$-wave, it makes a transition when doped with
Ba to the optimally doped Ba$_{1−x}$K$_x$Fe$_2$As$_2$ with $x\sim 0.4$, which is generally agreed to be fully gapped $s_\pm$.    This question is discussed in Sec. \ref{sec:Tbreaking} .

\begin{figure}[tbp]
\includegraphics[angle=0,width=0.6\linewidth]{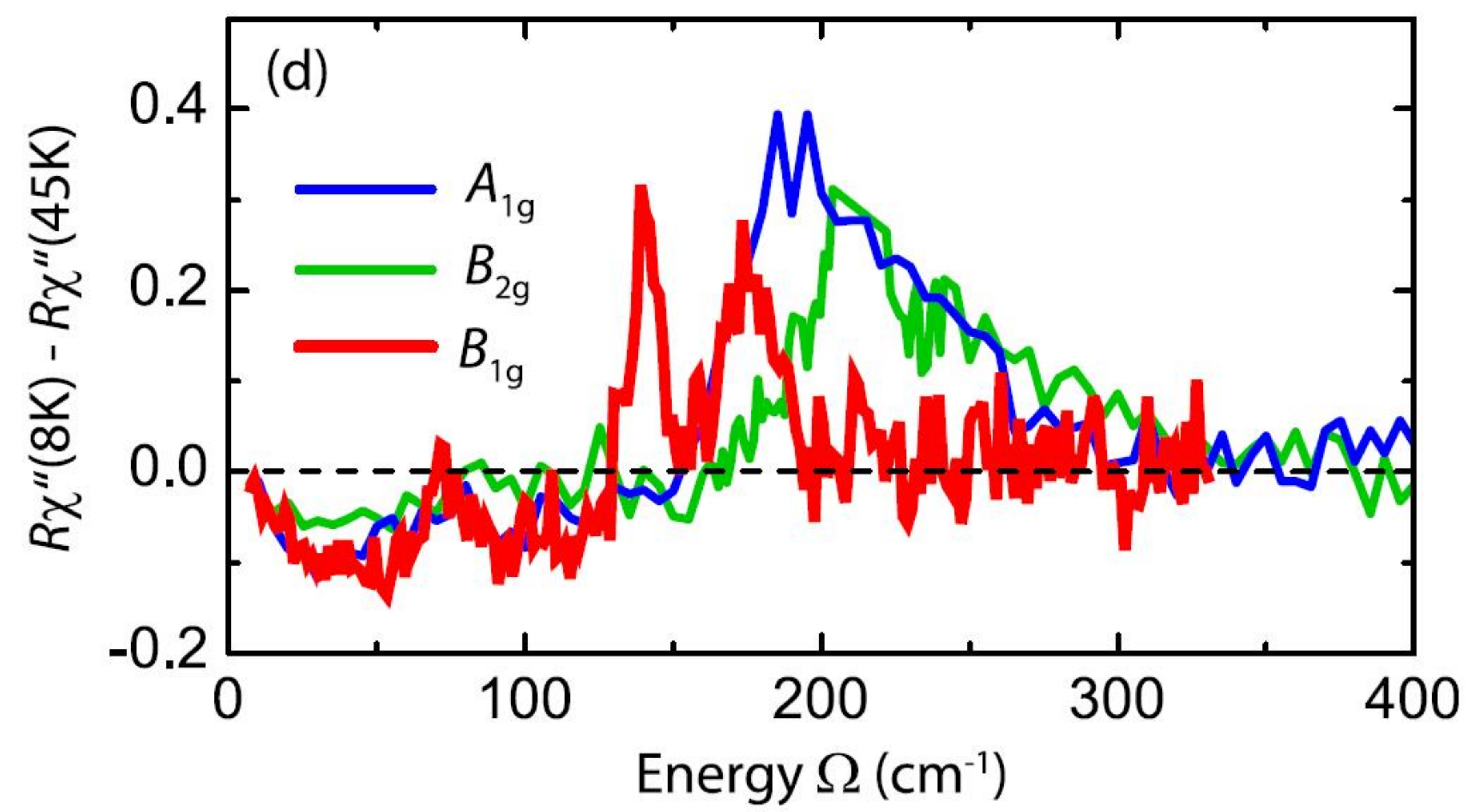}
\caption{(Color online.) Difference between superconducting and
normal-state Raman spectra on Ba$_{0.6}$K$_{0.4}$Fe$_2$As$_2$ in three different Raman polarization channels.  From Ref. \onlinecite{Hackl12}.
    }
\label{fig:Raman}
\end{figure}
\subsubsection{ Nematic order and superconductivity}
\label{subsec:nematic}
In some of the situations where  $s$- and $d$-wave interactions compete in the FeSC, electronic nematic order or fluctuations are also present.  For the most part I have ignored  the controversial and fascinating issues of nematicity in this article, primarily because excellent reasonably up-to-date reviews exist\cite{Fernandes_nematic_Review,BohmerMeingastCRAS,GallaisPaulCRAS}.  The issues covered there  are mostly related to   the origin of nematicity; here I focus on a side question, namely the interplay between nematic and superconducting degrees of freedom.  Ultimately it is probable that the same electronic states contribute to both superconductivity and nematic interactions, but one may separate the two effects  phenomenologically and  ask how symmetry determines  their interplay near $T_c$ when $s$- and $d$-wave interactions compete.  This was the approach adopted by Fernandes and Millis\cite{FernandesMillis_SC_nematic_coupling13}, who studied a Ginzburg-Landau free energy $F=F_{nem}(\phi^2) + F_{SC}(\Delta_s,\Delta_d,\theta)+F_{SC-nem}$, with the well-known $s-d$ part with competing  $s$- and $d$-wave orders
\begin{equation}
F_{SC}= {t_s\over 2}\Delta_s^2 +{t_d\over 2} \Delta_d^2 +{\beta_s\over 4} \Delta_s^4 +{\beta_d\over 4}\Delta_d^4 +{1\over 2} \Delta_s^2\Delta_d^2(\beta_{sd}+\alpha\cos 2\theta),
\end{equation}
as well as a nematic part $F_{nem}$, and a trilinear coupling between the two, $F_{sc-nem} \propto \phi\Delta_s\Delta_d \cos \theta$. Here $\phi$ is the nematic order parameter, and $\theta$ is the relative phase of the $s$- and $d$- components.

Taken by itself, the $s-d$ SC free energy is well known to have minima for pure $s$, pure $d$, and ${\cal T}$-breaking $s+id$ mixtures (see Sec. \ref{sec:Tbreaking}), depending on coefficients, so one obtains the usual $s-d$ phase diagram in the case of weak nematic order (Sec. \ref{sec:Tbreaking}).
In the case where nematic order forms at high $T$,  as in many of the Fe pnictides, the SC transition takes place in the presence of $C_4$ symmetry breaking and a robust expectation value of the nematic field $\phi_0$.  In this case the ground state is a real mixture $s+d$, i.e. $\Delta_\k=\Delta_s f_s(\k) + \Delta_d f_d(\k)$,
with $f_{s,d}$ real harmonics of $s,d$ symmetry,  and $\Delta_{s,d}$ real coefficients.   Clearly $C_4$ symmetry is broken simply because it is broken already before the SC condenses out of the normal state.    A qualitatively different situation is found in a fluctuating nematic phase, where  $F_{nem}= {1\over 2} \chi_{nem}^{-1} \phi^2$
accounts for the independent dynamics of the nematic field.  This now provides a nontrivial renormalization of the coupling coefficients $\beta_{sd}$ and $\alpha$,
leading to two new possibilities,  a  $s+d$ state forming in the competition region between pure $s$ and $d$ phases,  a {\it spontaneous} $C_4$ symmetry-breaking
in the sense that it condenses at high $T$ from a tetragonal phase.   When the nematic fluctuations become sufficiently strong, this phase disappears and is replaced by a 1st-order transition between $s$- and $d$-phases.    Thus the observation of properties (specific heat, elastic coefficients) of a nematic system like
\BFA ~or \NFA ~as a function of doping as $d$-correlations grow in strength can provide information on the trilinear coupling between nematic and SC fields.  Similar results were obtained in microscopic weak coupling theory by Livanas et al.\cite{Varelogiannis}.

Kang et al.\cite{Kang_etal_nematic_gap_nodes}  considered how $s+d$ states could be created and controlled starting from a tetragonal system
by applying uniaxial strain.  Starting with a model where strain coupled via an orbital ordering $\Delta_{oo}(n_{i,yz}-n_{i,yz})$, they showed how strain  mixed  a $d$-wave order parameter into a starting $s$-wave ground state.  In principle, if the unstrained $s$-state was fully gapped,  it is difficult to say if one could apply enough strain to create nodes by mixing in a $d$ amplitude comparable to $s$.  If one starts with a nodal $s$-state, however, addition of a small amount of $d$ via strain  suffices to move the nodes around the Fermi surface, either separating or coming together in pairs.   Nodes can also be created or annihilated
  in pairs as well.   They argued that one could in principle observe this process via the anisotropic current response of the system.

\subsubsection{Detection: Bardasis-Schrieffer mode}
\label{subsec:BS}

A recent experimental development of great interest has been the possible observation of a Bardasis-Schrieffer (BS) particle-particle exciton in several FeSC.  This collective mode of the
$s$-wave condensate in the presence of a $d$-wave subdominant pair interaction was predicted by Bardasis and Schrieffer in 1961\cite{BardasisSchrieffer61}.
The  frequency of the collective mode at $q\rightarrow 0$ frequency depends on the difference between the inverses of the two pairing interaction components, and is found
 below the pair-breaking edge of the condensed $s$-wave system.    This  mode was never observed in conventional superconductors, but may be present in FeSC due to the aforementioned near-degeneracy of the $s$ and $d$ channels.    Devereaux and Scalapino\cite{DevereauxScalapino09} proposed that Raman scattering would be a clear way to observe the mode,
  and provided a simple calculation of its frequency in the case of  an $s_\pm$ state.  Recently, BS mode-like features were identified  in Ba$_{1-x}$K$_x$Fe$_2$As$_2$\cite{Hackl12,Hackl14} and NaFe$_{1-x}$Co$_x$As,\cite{Blumberg14} in Raman scattering (Fig. \ref{fig:Raman}). The exact identification of these features with a BS mode is hindered by the fact that these systems posses multiple gaps and the exact nature of possible collective modes and their evolution across a typical doping phase diagram is not clearly known.   Collective modes in the particle-hole channel can also arise and need to be separated from the particle-particle excitons\cite{ChubukovRaman}. Nevertheless, these discoveries raise the prospect of systematic studies of the interaction strengths and collective modes in different channels in FeSC for the first time.

\subsection{Missing hole pockets}
\subsubsection{Background: bulk FeSe}
\label{subsubsec:bulkFeSe}
As I prepare to discuss three electron doped systems without hole pockets, all based on  layers of FeSe, it is perhaps
useful to review briefly what is known  about  superconductivity  of the bulk compound.
FeSe is   structurally rather simple, displaying a tetragonal to orthorhombic structural phase transition at $T_s\sim 90$ K, entering
a low-$T$ phase exhibiting strong electronic anisotropy, but without  SDW order. It makes a transition to a superconducting state only below a $T_c$ of about 9K.  Early scanning tunneling microscopy (STM) studies of thick FeSe films on SiC  or bilayer graphene substrates  found highly elongated vortices and
impurity states, and also measured a nodal superconducting gap\cite{SongFeSe}. Until recently, similar experiments on crystals were hindered
by sample quality, but these problems were overcome by T. Wolf and collaborators, who fabricated
very clean samples,  appropriate to study the details of the low-energy properties\cite{bohmer13}, using cold vapor deposition.  This discovery led to a small renaissance of studies on this
hitherto neglected material, and it has now become one of the most intensively investigated FeSC.

Central to the  mystery of FeSe is the existence of strong nematic
tendencies in the absence of obvious magnetism, leading to suggestions that nematicity in this system is driven by orbital order, unlike other FeSC.
 Recent $^{77}$Se NMR measurements have reported a  splitting of the NMR line shape beginning at $T_s$, with an order parameter-like $T$ dependence below $T_S$.\cite{baek14,bohmer15}    Unlike the well-documented case of \BFA,  at high $T$, the spin-lattice relaxation rate is found to be unaffected by the structural transition, and the  upturn at low $T$ signaling the onset of a strongly spin-fluctuating state appears only  close to the superconducting $T_c$\cite{ImaiCava2009,baek14,bohmer15}. The naive interpretation of these experiments is that nematic order is driven by orbital fluctuations in FeSe, but despite the apparent weakness of  local spin fluctuations near $T_s$, the spin-nematic
picture may still apply.   Several proposals, including  strongly frustrated long-range magnetic
order due to competing magnetic ground states\cite{glasbrenner15}, quantum paramagnetic fluctuations of large Fe spins\cite{kivelson15}, competition with charge-current density waves\cite{chubukovFeSe15}, or  quadrupolar spin order\cite{qsi15} have been discussed at this writing.  Recently, the first inelastic neutron scattering measurements on these samples appeared, showing robust low-energy $(\pi,0)$ fluctuations very similar to the Fe pnictides  persisiting up to $T_s$, thus deepening the mystery of why the system does not magnetize, and why NMR is apparently not sensitive to the spin fluctuations at higher $T$.
I do not discuss these
fascinating questions further here, as they are covered in more detail in the accompanying review by B\"ohmer and Meingast\cite{BohmerMeingastCRAS}.

 Other remarkable properties of bulk  FeSe include the significant enhancement of the superconducting critical temperature $T_c$  under pressure,\cite{medvedev09} , as well as the extremely small size of the Fermi surface pockets measured at low $T$\cite{Lubashevsky12,Okazaki14,Kasahara14},
 leading to suggestions of BEC-BCS condensation in these systems where the band extrema $E_F$ are of order the gap $\Delta$.
These large and $T$-dependent renormalizations of the Fermi surface needed as an input to pairing calculations have slowed theoretical understanding of the  superconducting state and its properties.  Although there are still some disagreements among ARPES measurements on the new crystals
\cite{nakayama14,maletz14,shimojima14,watson14,ZhangP15,SuzukiY15,ZhangY15}, general aspects of  recent studies include a Fermi surface (FS) above $T_S$ consisting of two small hole pockets of primarily $d_{xz}/d_{yz}$ character around the $\Gamma-Z$ line. The hole bands are split there  above $T_s$ by a spin-orbit coupling (SO) of order $\lambda_{\text{Fe}}\simeq 20\,\text{meV}$, but below $T_S$ this evolves  into a single  hole surface\cite{watson14}.  ARPES also finds an electron pocket at the $M$ point of mainly $d_{xz}/d_{yz}$ character.  Here the   $d_{xz}$/$d_{
yz}$ degeneracy  is lifted by $\sim 50\,\text{meV}$ at low $T$,  taken by several authors to indicate strong orbital order in FeSe.   As mentioned above, these results  are very different from those obtained from DFT calculations.\cite{maletz14,Eschrig2009}
For example, ARPES finds that the electronic bands in FeSe are  renormalized compared to DFT calculations by a factor of $\sim$3 for the $d_{xz}/d_{yz}$ bands and $\sim$9 for the $d_{xy}$ band, and hole and electron pockets are shifted such that pockets are much smaller than calculated by DFT\cite{maletz14,watson14}.
Quantum oscillations (QO) at low $T$ are consistent with the ARPES data in observing small, mostly 2D pockets, even though the amount of dispersion along $k_z$ is still disputed.\cite{terashima14,watson14}  Recently it was suggested  that the splittings at the high symmetry points were not consistent with a simple momentum independent ferro orbital ordering\cite{ZhangY15}, but were in fact of opposite signs at the $\Gamma$ and $M$ points, leading to suggestions of bond-centered (sometimes referred to as $d$-wave since the $d_{xz}/d_{yz}$ splitting has opposite signs at $M$ and $\bar M$).

Recently, Mukherjee et al.\cite{Shantanu15}  proposed circumventing the theoretical issues associated with the band structure by considering the pairing problem using
a tight binding model with site energies and hoppings renormalized with respect to DFT so as to reproduce the ARPES and QO data.   They showed that
adding phenomenological orbital ordering terms to the Hamiltonian allowed one to do so at both low and high $T$, and used the model to calculate NMR properties in excellent agreement with experiment.  They then took the tight-binding model and showed that it led via the conventional RPA spin-fluctuation pairing
approach to an anisotropic $s_\pm$  superconducting gap on the (strongly nematic) low $T$ Fermi surface with nodes on the $d_{xz}/d_{yz}$ electron bands, giving a penetration depth and $V$-shaped STM spectrum quite similar to experiment.   The nodes found were quite weak (shallow gap extrema),  consistent with
a recent experiment on FeSe twin boundaries where the nodes were shown to be easily lifted\cite{shibauchi_twins}.  Recently, this approach was also shown to account well for inelastic neutron scattering data on FeSe\cite{Kreisel_FeSe_neutron}.

\subsubsection{Alkali-intercalated FeSe}

 \begin{figure}[tbp]
\includegraphics[angle=0,width=\linewidth]{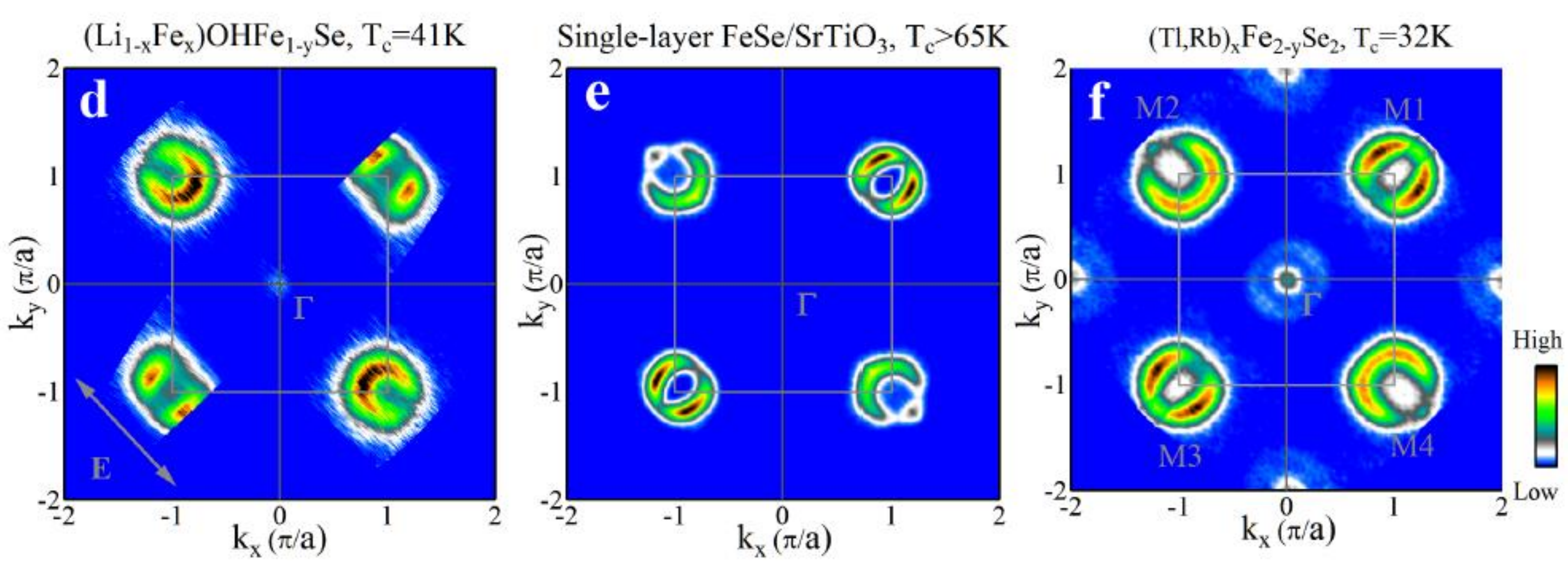}
\caption{(Color online.) Fermi surfaces of a) (Li$_{0.8}$Fe$_{0.2}$)OH FeSe  ; b) FeSe monolayer on STO. ; c)  (Tl,Rb)Fe$_2$Se$_2$ measured by ARPES.  Note the apparent weight of the $\Gamma$-centered hole band in (c) is due to a broad band whose centroid disperses below the Fermi surface.
 From Ref. \onlinecite{XJZhou_Common_Origin15}. }
\label{fig:3Fermisurfaces}
\end{figure}

  The standard model for $s_\pm$ pairing in Fe-based systems discussed above assumes   the existence of a hole pocket and   an electron pocket  separated in the Brillouin zone by large $\q$, as indeed found in most of the FeSC  systems, including bulk FeSe.     Within this picture, repulsive interband  Coulomb interactions yield a  peak at this wavevector in the magnetic  susceptibility, which in turn plays the essential role in  a spin-fluctuation
 exchange  pairing interaction.   The system can lower its energy however only  if the superconducting order parameter changes sign between the two Fermi surface sheets\cite{mazin08,kuroki08}.   The early expectation that all FeSC could be described by this single paradigm was gradually eroded  by the discovery of several Fe chalcogenide systems
 which are missing the characteristic $\Gamma$-centered Fermi surface holelike pockets altogether.  It was immediately suggested that a new mechanism of pairing must be at work, or at least a new symmetry or structure of the order parameter must result.

  The discussion of a new paradigm for pairing in the FeSC began in  2011 with the discovery of relatively high-temperature superconductivity ($T_c\gtrsim $40K) in the alkali-intercalated FeSe materials\cite{KFe2Se2}, which nominally correspond to the chemical formula AFe$_2$Se$_2$, with A=K,Rb,Cs.     As of this writing,  the superconducting samples of all  materials are available only in mixed-phase form and have relatively low superconducting volume fractions; thus the exact composition of the superconducting materials are unknown.    They nevertheless excited considerable interest due to their proximity to unusual high-moment block  antiferromagnetic phases, and because    ARPES\cite{KFe2Se2ARPES} measurements on KFe$_2$Se$_2$ reported that there were no $\Gamma$-centered hole pockets at the  Fermi level (although a small electron pocket  is found near the $Z$ point near the top of the Brillouin zone); at $\Gamma$ the hole band maximum is $\sim$ 50 meV below the Fermi surface.  An example of one of the ARPES-determined Fermi surfaces of these materials is shown in Fig. \ref{fig:3Fermisurfaces}c).  Note that there is some spectral weight at the center of the zone, but that the centroid of the band is well beneath the Fermi level.

 Several groups pointed out that despite the missing hole pockets, repulsive interactions at the Fermi level remained among the electron Fermi surface pockets, and could lead to $d$-wave pairing with
  significant critical temperatures\cite{Fangdwave,Maierdwave}.    In addition,  the expected hybridization of the two electron bands in the proper 122 body centered tetragonal crystal symmetry leads to 2 hybridized electron Fermi surface  pockets at the M point in the 2-Fe Brillouin zone, with concomitant strong  interactions between them.
  Mazin\cite{Mazin_bonding_antibonding,HKMReview2011}  suggested that these interactions could lead to a new type of $s$-wave state (``bonding-antibonding $s$-wave")  with sign change between
 inner and outer hybridized electron pockets.
  Such a state was not found to be competitive with the $d$-wave state in such a situation due to the weak hybridization in calculations based on the DFT bandstructure for  KFe$_2$Se$_2$\cite{KreiselKFe2Se2}, but the bonding-antibonding $s$-wave remains an interesting candidate in part because this system is apparently intrinsically inhomogeneous and its exact electronic structure is unknown.

 As mentioned briefly above,  inelastic neutron scattering measurements\cite{InosovKFe2Se2} agree rather well with the wave vector $\sim (\pi,\pi/2)$ for
 the inelastic neutron resonance predicted in Ref. \onlinecite{Maierdwave}, which correponds e.g. to the nesting of the
 internal sides of the electron pockets at $(\pi,0)$ and $(0,\pi)$ in the 1-Fe zone.  On the other hand, the $d$-wave state
  found here appears to disagree with the absence of nodes on the small $Z$-centered pocket reported by ARPES\cite{DLFengKFe2Se2}.
  Pandey et al.\cite{Khodas_swave_neutron} argued that the bonding-antibonding $s$ wave state would in fact support a resonance at roughly the wave vector observed in experiment.  While this is natural in 2D, since the bonding-antibonding state is essentially a folded $d$-wave state stabilized by hybridization, it is less obvious in 3D, and again requires significant hybridization that is not present, at least in DFT.

  The uncertainties associated with this system, including the materials issues, have left the question open.  Recently, yet another solution was proposed by Nica et al\cite{QSi_KFe2Se2} as a way of reconciling neutron and ARPES experiments, by essentially incorporating both $s$ and $d$-wave symmetry into different orbital channels, so that different symmetry channels effectively dominated different Fermi surface sheets. The resulting mixed state was shown to be stable within a $t-J_1-J_2$ mean field calculation over some range of parameters.

  \subsubsection{FeSe monolayers}
  \label{subsubsec:monolayer}
\begin{figure}[!h]
\renewcommand{\baselinestretch}{.8}
\includegraphics[width=\columnwidth]{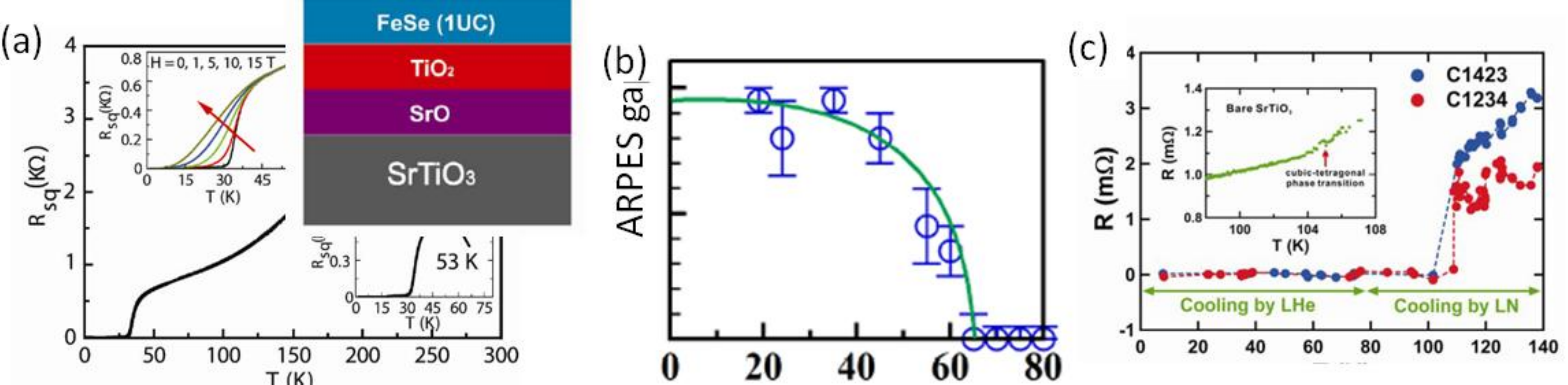}
\caption{ (Color online.)  (a) Resistivity of monolayer FeSe films on STO\cite{Xue_monolayer}; (b) Spectral gap measured by
ARPES on such films. From Ref.\onlinecite{Zhou_monolayer}; (c) Resistivity of newer films of FeSe/STO\cite{FeSeSTO100K}.}
\label{fig:FeSe_mono}
\end{figure}

  The same states discussed in the alkali intercalates of FeSe  have arisen  in discussions of FeSe monolayers grown epitaxially on strontium titanate, STO\cite{Xue_monolayer}.   Of course the spectacular aspect of these films is their high $T_c$, as much as 70K in ARPES measurements\cite{Zhou_monolayer},
  and 110K in {\it in situ} transport measurements\cite{FeSeSTO100K}.    Fig. \ref{fig:FeSe_mono} (a) shows the epitaxial
   structure of the film used originally to obtain a transport $T_c$ four times higher than bulk FeSe (8K), (b)  the subsequent
   ARPES gap closing temperature of 65K measured on similar samples, and (c)  the more recent $T_c>$ 100K transport
   measurement showing an abrupt loss of resistivity.   The very abruptness of the drop at $T_c$ in the last case is
    an important hint, since a true 2D system would be expected to show the thermal broadening at a Kosterlitz-Thouless transition.
    In their annealed state, the STO films are conducting, and it is suspected that   the high temperature 3D current distribution becomes abruptly 2D as the SC transition is passed.    This is one of many indications that the STO substrate is playing a crucial role.
  Others include the fact that FeSe on other substrates (graphene, for example) produce superconductivity very similar to the bulk\cite{SongFeSe}, and the remarkable observation that, while a one layer FeSe/STO system is a high-$T_c$ superconductor, a 2-layer FeSe/STO system is not a superconductor at all\cite{Xue_monolayer,Liu_bilayer} .

    The ARPES measurements have carefully studied how the
  high-$T_c$ superconducting state evolves out of the as-grown sample, and how the requisite Fermi surface evolves with it.
  At the beginning of the annealing process, the central  hole band imaged by ARPES is at the Fermi level,  but by the end
 it has moved 80 meV below, and the electron pockets have enlarged.  It is believed that this effective doping arises mostly from O vacancies in the STO\cite{Gong2014}, since Se vacancies in
 the film itself appear unlikely\cite{Berlijn}.  The Fermi surface has only (apparently unhybridized\cite{Shen_FeSe-STO}) electron pockets at the M points (Fig. \ref{fig:3Fermisurfaces}(c)).     Note that the Fermi surface obtained by standard DFT calculations is  found to be incorrect, even if the system is electron doped ``by hand" using rigid band shift or virtual crystal methods, and even accounting for the strained lattice constant imposed by the STO.  So a pressing challenge is how to understand the origin of the  electronic structure in this case, presumably by incorporating O defects in a proper way including disorder effects and/or by properly including correlations.

 An important clue to the physics of these systems, and the influence of the substrate, was recently presented by new ARPES measurements\cite{Shen_FeSe-STO}, which showed the existence of ``replica bands"  (nearly exact ``shadow" copies
  of bands both at $\Gamma$ and M shifted downward in energy by $\sim$ 100 meV).  These experiments indicate the presence of a large electron-phonon interaction, probably originating from the substrate\cite{Shen_FeSe-STO}.  It has recently been argued in addition that the electron-phonon interaction must be quite strongly peaked near momentum transfer $q$=0 to explain this observation\cite{Johnston_FeSe-STO}.

Both STM and ARPES agree that the monolayer system has a full superconducting gap, of order 20 meV.    It is intriguing
that STM reports two coherence-like peaks in the spectrum, although the hole band is presumed not to participate in
 superconductivity, and the two electron pockets at M in ARPES show
no evidence of hybridization\cite{Shen_FeSe-STO}.      It has been suggested that the two scales might indicate some substantial
anisotropy (but no nodes) on the electron pockets, but, at least within BCS theory, the  gap minimum does not lead to a singularity
in the DOS.  For this and other reasons,
if one considers only the band structure at the  Fermi surface,  high-$T_c$   and the form of the SC gap are puzzling.  Despite the apparent role of phonons in the electronic structure, electron-phonon interactions in the FeSe are unlikely to be strong enough to explain a $T_c$ of 70K or above\cite{Li_etal_2014} (a calculation that disagrees, and finds a much higher $T_c^{ph}$ than others under some
 rather generous assumptions is  Ref. \onlinecite{Cohen}).

 Thus a ``plain" $s$-wave from attractive interactions alone seems improbable, even if  soft STO phonons play a role\cite{DHLee2012,Shen_FeSe-STO}.   The forward scattering character of the relevant phonon processes\cite{Johnston_FeSe-STO} then implies that phonons cannot  contribute to the interband interaction.   Taken by itself, however, the interband  spin fluctuation interaction  should lead  to nodeless $d$-wave (since $\chi({\bf q},\omega)$ will be roughly  peaked at the momentum connecting the electron pockets),  similar to the arguments given for alkali-intercalates\cite{Fangdwave,Maierdwave}.   On the other hand, there is some evidence that the system does not have a sign-changing gap. In  STM measurements by Fan et al. \cite{Fanetal_impurities_Fe-STO15} $T_c$ and the gap were reported to be   suppressed only by magnetic impurities, as one might indeed expect from a ``plain" $s$-wave SC.  These arguments, if correct, would also rule out states of the ``bonding-antibonding $s$-wave" type\cite{HKMReview2011}.

 A final possibility is the ``incipient $s_\pm$" state, which naiviely seems disfavored by energetic arguments.  Recently Bang\cite{Bang} and Chen et al.\cite{Xiao_incipient2015} revisited these arguments and found that that in the presence of preexisting Fermi surface superconductivity (e.g. a phonon attraction in the electron pockets of the monolayer) this state was quite strongly favored.
 D.H. Lee and co-workers have elaborated on a scenario\cite{DHLee2012,Shen_FeSe-STO,DHLee2015} for high-$T_c$ in
  FeSe monolayers in which STO  phonons assist spin fluctuations, but the origin of the spin fluctuations near $(\pi,0)$  was never
  clear due to the incipient hole pockets.  With the current picture of Chen et al.\cite{Xiao_incipient2015} this now becomes a clear candidate, along with the $d$-wave.     The bonding-antibonding $s$-wave state is also a possibility, but ARPES sees no hybridization of the two electron pockets that would stabilize this state\cite{Shen_FeSe-STO}.   Some details of these arguments are reviewed in Sec. \ref{subsec:incipient}.

  Of course,  one still needs to understand the objections of Ref. \onlinecite{Fanetal_impurities_Fe-STO15}  regarding the effect of impurities.   All the above candidates involving spin fluctuations at either $(\pi,0)$ or $(\pi,\pi)$ exhibit a sign change.    It may be, however,  that the expected pairbreaking effects in a sign-changing incipient $s$-wave states are weaker because impurity scattering is elastic\cite{Xiao_incipient2015}.

  \subsubsection{Other FeSe intercalates}
Speculation on the origin of the higher $T_c$ in the alkali-intercalated FeSe  centered on
the intriguing possibility that enhancing the FeSe layer spacing improves the two-dimensionality of the band
structure and therefore might enhance Fermi surface nesting.
To explore this effect, organic molecular complexes,  including alkali atoms, were successfully
intercalated between the FeSe layers~\cite{AmmoniaPoorNature, AmmoniaRichJACS,
LithiumIronHydroxideSelenides, IntercalateLayerDistance,
AlkalineEarthIntercalates, ScheidtPureAmmonia,
PyridineIntercalate}.   The extremely air-sensitive powders synthesized  had transition temperatures up to 46 K. Until
recently, the most intensively studied materials intercalated ammonia,
e.g. Li$_{0.56}$(NH$_2$)$_{0.53}$(NH$_3$)$_{1.19}$Fe$_2$Se$_2$ with $T_c =
39~\mathrm{K}$~\cite{AmmoniaRichJACS} with
Li$_{0.6}$(NH$_2$)$_{0.2}$(NH$_3$)$_{0.8}$Fe$_2$Se$_2$ with $T_c =
44~\mathrm{K}$~\cite{AmmoniaPoorNature}.
 Noji \textit{et al.}~\cite{IntercalateLayerDistance} correlated
data on a wide variety of FeSe intercalates and noted a strong
correlation of $T_c$  with inter- FeSe layer spacing,  with a nearly
linear increase between 5 to $9~\mathrm{\AA}$,  which then saturated   between 9 to $12~\mathrm{\AA}$.
This tendency was attributed to a combination of doping and changes in nesting with increasing two-dimensionality
by Guterding et al.\cite{Guterding_intercalate}.

\vskip .2cm

While the ammoniated FeSe intercalates are fascinating, their air sensitivity prevented many important experimental
 probes and limited their utility.    Recently the discovery of a new class of air-stable FeSe intercalates, lithium iron selenide hydroxides, was
 reported in Ref. \onlinecite{XHChen2014}.   Remarkably    Li$_{1-x}$Fe$_x$(OH)Fe$_{1-y}$Se has also been shown to have a Fermi surface without $\Gamma$-centered hole pockets.\cite{Zhao2015} (Fig. \ref{fig:3Fermisurfaces} (a)
 ).     It was recently pointed out that another similarity of these systems with the FeSe/STO monolayers is a double coherence peak in STM\cite{HHWen_LFOHFS}, with extremely large inferred gap-$T_c$ ratios of order 8.  In addition, these authors reported detailed QPI measurements consistent with two hybridized electron pockets, which they associated with the two gaps.  The further observation of an in-gap impurity resonance at a native (Fe-centered) defect site suggests that the gap is sign changing.  This is
because the defect is either an Fe vacancy or possibly a chemical substituent on this site, presumably not magnetic.  The SC order
parameter therefore appears to  have  a bonding-antibonding $s$-wave structure.    Further research on this system is certainly needed, and may indeed provide insight into the other systems shown in Fig. \ref{fig:3Fermisurfaces}.

\subsection{Incipient band pairing}
\label{subsec:incipient}

The observation of high-$T_c$ in systems with missing hole pockets has been a challenge to the established $s_\pm$ paradigm, which naively requires (at least) two Fermi surface pockets separated by large $\q$.   In their study of KFe$_2$Se$_2$,
   Wang et al.\cite{Fangdwave} found however that pairing in an $s_\pm$
  channel, with gap sign change between a gap on the regular electron band at the Fermi surface and the incipient hole band $\sim$80 meV below the Fermi level was surprisingly competitive with the expected\cite{Maiti}
   $d$-wave state.  This "incipient $s_\pm$" state was treated  as a candidate  for pairing in these systems in Ref. \onlinecite{HKMReview2011}, along with the (quasi-nodeless) $d$-wave state and the ``bonding-antibonding" $s$-wave state that changed sign between two hybridized
  electron pockets taken in the 2-Fe zone.  It did not receive a great deal of attention, however, presumably because of the general prejudice  that  incipient bands (those that nearly cross the Fermi level)  do not play a significant role in superconductivity.

Interest in pairing in incipient bands in Fe-based superconductors was revived by the experiments on
 FeSe/STO monloayers showing an absence of $\Gamma$-centered hole pockets\cite{XJZhou2013}, and later by the observation by He et al \cite{HongDingLiFeAs2015} of the persistence of the superconducting gap on one of the hole bands of LiFeAs as it sank below the Fermi level with electron doping by Co.    The gap on this band was only
   weakly suppressed in this process, at least up to band extremum values of $E_g\sim 10$ meV (Fig. \ref{fig:incipient2}), and probably significantly further.    Moreover, there was no indication of any abrupt change in either $T_c$ or the gap on the incipient band through the Lifshitz transition, and the gap on the incipient band was the largest gap on the Fermi surface.

 \begin{figure}[!h]
\renewcommand{\baselinestretch}{.8}
\includegraphics[width=\columnwidth]{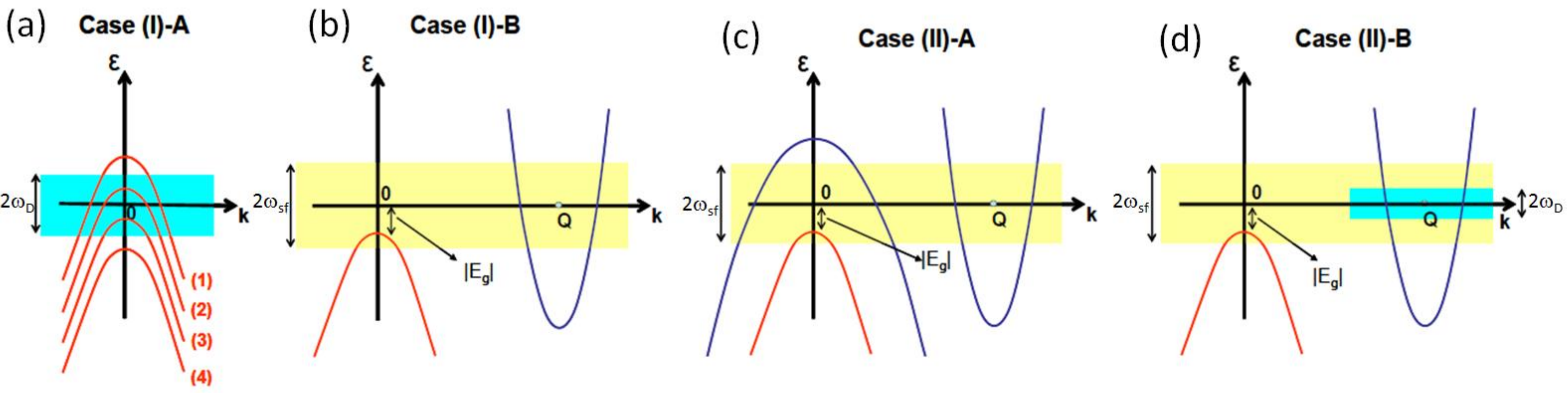}
\caption{ ((Color online.) a) Case(I)-A: Four examples of a single  hole band corresponding to (1) regular band (2) shallow band (3) incipient
band (4) vegetable band depending on position of band maximum with respect to Fermi level and range of attractive
energies between  $\pm\omega_D$ (blue
region). (b) Case (I)-B: Spin fluctuation driven pairing on incipient band (cut-off $\omega_{sf}$-yellow region) SC. (c) Case (II)-A:
SC is driven by spin fluctuations in the regular (blue) bands, and  SC in
the incipient band is induced by the same interaction. (d) Case(II)-B:
SC  driven by phonons in the regular (blue) band, and induced via spin fluctuations  in the incipient band.}
\label{fig:incipient}
\end{figure}

 Bang had earlier pointed out that $T_c$ could remain large in a situation where superconductivity was stabilized by an intraband attractive interaction\cite{Bang}.  In an attempt to explain the results in LiFeAs,  Hu et al. \cite{FCZhang2015} calculated the effect of the incipient band using  a 3-orbital model with  next-nearest neighbor  BCS pairing ansatz, and found that the qualitative aspects of the experiment could be understood, but only if the pairing interaction exceeded a critical value.  Chen et al.\cite{Xiao_incipient2015} criticized this approach
 as a nongeneric formlation of the usual multiband problem, and argued that weak coupling -- without a threshold interaction -- was sufficient to explain the experiment.   Within Eliashberg theory,
  Leong and Phillips\cite{Phillips2015} studied a 5-band model for LiFeAs  where the incipient band was coupled to the others only by Coulomb interactions, and other Coulomb interactions were neglected; using this approach, they showed that the energy dependence of the Coulomb interactions gave rise to a large gap on the incipient band.  However, general conclusions were hard to draw due to the complexity of the model.

Here I follow the simple approach of Ref. \onlinecite{Xiao_incipient2015} and consider (Fig. \ref{fig:incipient}) several different situations with a minimal number of bands, showing how they differ qualitatively.    The first, depicted in Fig. \ref{fig:incipient}(a), is the well-known case of a single incipient band with band extremum $E_g$ within the BCS cutoff energy $\omega_D$, with gap equation
\bea
1&=&-\frac{mV_{\text{ph}}}{2\pi}\left[\int^{E_g}_{-\omega_D}\frac{d\epsilon}{2E}\tanh\frac{E}{2T}\right],
\eea
 where $V_{ph}$ is a putative attractive interaction due to phonons.
 The  single band case also  has a critical value of  $E_g$ where SC disappears before the lower cutoff $-\omega_D$ is reached, given by
\beq\label{eq:Ec}
E_g^{\text{crit}}= -\omega_D e^{\frac{2}{v_{ph}}},
\eeq
 with $v_{\text{ph}} = mV_{\text{ph}}/2\pi$.    In fact, since the spacing between pairs in the 1-band case becomes
larger than the Cooper pair radius at some point approaching the Lifschitz transition in case (I)A, 
 one enters the BCS-BEC crossover regime in the low density limit.  If this occurs, $T_c$ should vanish at the Lifschitz point if
 the problem is treated properly\cite{GorkovMelik-Barkhudarov,Mazin_vanderMarel}.

 Cases (I)B and (II)A are shown in Fig. \ref{fig:incipient}(b,c) respectively,   and can be discussed with a simple multiband model of the sort discussed in section \ref{sec:multiband}:
\bea\label{eq:gaps}
\D_{{ e}} & =&-V_{sf1}\D_{{ h}_{1}}L_{{ h}_{1}}-V_{sf2}\D_{{ h}_{2}}L_{{ h}_{2}},\\
\D_{{ h}_{1}} & =&-2V_{sf1}\D_{{ e}}L_{{ e}},\\
\D_{{ h}_{2}} & =&-2V_{sf2}\D_{{ e}}L_{{ e}}.
\eea
where
\bea\label{eq:def}
L_{h_1} & =&\int_{-\omega_{sf}}^{\omega_{sf}}{\rm d}\varepsilon N_{h_1}\frac{{\rm tanh}\frac{E_{h_1}}{2T}}{2E_{h_1}},\nonumber\\
L_{e} & =&\int_{-\omega_{sf}}^{\omega_{sf}}{\rm d}\varepsilon N_{e}\frac{{\rm tanh}\frac{E_e}{2T}}{2E_e},\nonumber\\
L_{h_2} & =&\int_{-\omega_{sf}}^{E_{{\rm g}}}{\rm d}\varepsilon \frac{m}{2\pi}\frac{{\rm tanh}\frac{E_{h_2}}{2T}}{2E_{h_2}},
\eea
$m$ is the mass of the incipient band $h_2$.  Here $V_{sf\alpha}$ with $\alpha=1,2$  is the (spin-fluctuation) interband interaction between the electron band $e$ and the two hole bands $h_1$ and $h_2$ of Fig. \ref{fig:incipient}(c).   Note that the definition of $L_\nu$ here includes the density of states $N_\nu$.  The case depicted in
Fig. \ref{fig:incipient}(b) can be obtained from these equations by dropping  the hole band $h_1$.   Case (II)B shown in     Fig. \ref{fig:incipient2}(d) is slightly more complicated since in principle it involves two energy scales for pairing, which may be thought of for convenience as a spin fluctuation cutoff and phonon cutoff.  Ref. \onlinecite{Xiao_incipient2015} obtained analytical and numerical solutions for these equations and showed that
\begin{itemize}
\item in the single band incipient case $T_c$ is strongly suppressed as soon as $E_g$ falls below the Fermi level.  There is a critical value of $E_g$ below which no superconductivity is possible.
    \item  these thresholds do not exist in other cases.  However, in case (I)B, where superconductivity is driven by an interband  repulsion  connected to the incipient band, $T_c$ is still strongly suppressed as the band falls below the Fermi level.  This is the intuition that seems to have guided the conclusions of Ref. \onlinecite{HongDingLiFeAs2015}.
        \item when superconductivity existed before the incipient band was considered, either due to a finite $q$ spin-fluctuation
        interaction ((II)A), or due to an intraband phonon-like attraction ((II)B),  the introduction of the incipient band substantially assisted $T_c$, and induced a large gap on the incipient band.
        \item In the latter case, there is no abrupt change in $T_c$ or gaps as the Lifshitz transition is crossed, and the induced gap on the incipient band can easily be of the same magnitude or larger than gaps on regular bands at the Fermi surface, depending on details of interactions.
\end{itemize}
In Figure \ref{fig:incipient2}, the essential data from the experiment Ref. \onlinecite{HongDingLiFeAs2015} are plotted alongside the $T_c$ suppression for different $E_g$ as the Lifshitz transition is crossed found from the equations of Ref. \onlinecite{Xiao_incipient2015} for cases (I)A, (I)B, and (II)A.
 \begin{figure}[!h]
\renewcommand{\baselinestretch}{.8}
\includegraphics[width=0.9\columnwidth]{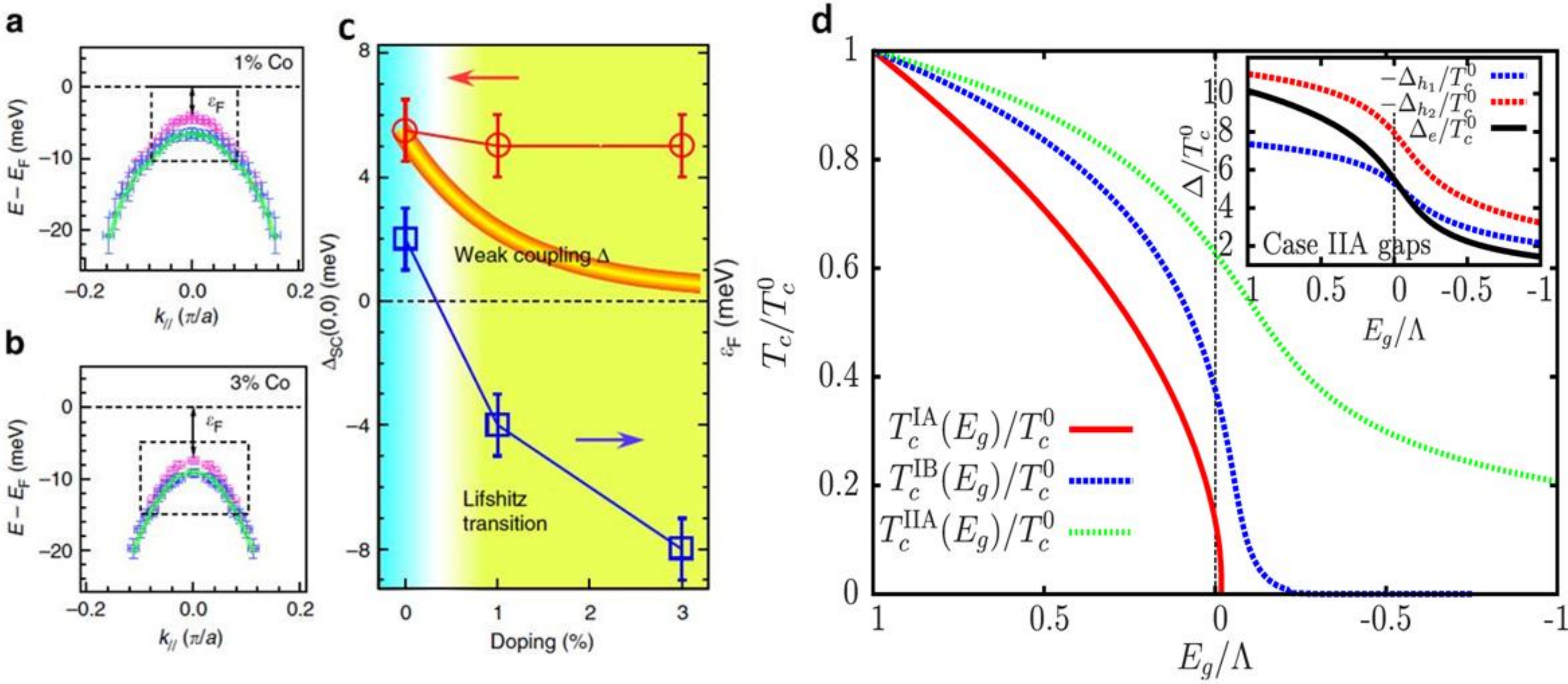}
\caption{ (Color online.) (a)-(c) adapted from Ref. \onlinecite{HongDingLiFeAs2015}.  (a,b) Band measured by ARPES near $\Gamma$ point in LiFe$_{1-x}$As at high (30K) and low (8K) for 1\% (a) and 3\% Co.  (c) Plot of gap measured by pullback of leading edge of ARPES
EDC intensity vs. Co doping, together with energy of band maximum below the Fermi level $E_g$ vs. doping.  Line labelled ``weak coupling" is apparently sketched under the assumption that incipient gap should behave as in case (I) in Fig. \ref{fig:incipient}.  (d) adapted from Ref. \onlinecite{Xiao_incipient2015}.  Plot of $T_c/T_c^0$ for interactions  $v_{ph}=0.5$ (red), $v_{sf} = 0.3$ (blue), $v_{sf1}$= 0.2; $v_{sf2}$=0.3 (green).  Inset:  gap on both Fermi surface pockets ($h_1$ and $e$), together with gap on incipient band ($h_2$), vs. $E_g/\Lambda$.
 }
\label{fig:incipient2}
\end{figure}

 Case (II)B is quite similar to (II)A, but requires the specification of
additional parameters.   Chen et al.\cite{Xiao_incipient2015} proposed that an $s_\pm$ state of this type might represent a possible candidate for the ground state of FeSe monolayers on STO, with a large gap on both the electron pockets at the  Fermi surface band and the incipient hole band well below it.  The spin fluctuations were shown capable of significantly (factor 10, depending on
ratios of phonon and spin fluctuation interactions and cutoffs)  enhancing a weak phonon $T_c$.  Due to the sign change,  the $s_\pm$  state naively has a difficulty with the results of Ref. \onlinecite{Fanetal_impurities_Fe-STO15}.  However, Chen et al. anticipated that since impurities scatter elastically,  pairbreaking effects in an incipient band might not be so effective.

\section{Multiphase gaps and time reversal symmetry breaking}
\label{sec:Tbreaking}
The close proximity of $s$ and $d$ pairing channels in spin fluctuation theory led Lee et al.\cite{Zhang_sid} to propose that a mixed symmetry
ground state, a so-called $s+id$ state, could be stablilized, reviving a proposal that had arisen in the cuprate context some years before, but had
never been realized.  Note that such a state is a mixture of two distinct irreducible representations of the symmetry group of the system, and as such should lead generically to two $T_c$'s except at a single point of doping or whatever parameter tunes the transition.    The Ginzburg-Landau free energy studied by Lee et al., identical  to the the earlier works, was shown to have a minimum with 90$^\circ$ phase difference
between $s$ and $d$, depending on the sign of the $\beta_4$ quartic mixing coefficient.   And of course the same interesting phenomena were identified: locally
$C_4$ symmetry-breaking spontaneous current patterns near impurities and edges,  and the prediction of a resonance mode corresponding to the oscillation
of the relative phase of $s$ and $d$, predicted to appear in the $B_{1g}$ polarization of a Raman scattering experiment.   The spontaneous currents are
 difficult to measure, because they are not globally chiral and there is no direct coupling to observables measurable by STM.  On the other hand,
 the Raman mode should be observable, and is in fact  exactly analogous  to the mode predicted by Balatsky et al.\cite{Balatsky_mode} in the cuprate context.
 Neither of these calculations went beyond the simplest GL approximation to work in the low-$T$ phase where $s+id$ states are actually stable, however.
 The microscopic origin of the phase stability  and the Raman-active mode itself were therefore unclear.  Using a realistic five-orbital model appropriate for 1111 systems, and local Hubbard-Hund interactions treated
 with the functional renormalization group, Platt et al.\cite{Hanke_sid}   showed that an $s+id$ state could
 be obtained as the ground state for intermediate electron doping range close to where the hole pockets disappear.     There appears to be no
 analogous calculation for the strongly overdoped hole case, where, as discussed above, there is evidence from Raman scattering of a strong
 $d$-wave interaction implied by the observation of a subgap Bardasis-Schrieffer type mode.

 These studies considered as the ``$s$" component of $s+id$ the standard $s_\pm$ state found in standard spin fluctuation calculations for systems with
 hole and electron pockets.  However, from the symmetry point of view any $s$-type representation is possible, and
 Khodas and Chubukov studied the combination with the bonding-antibonding $s_\pm$  state arising from the hybridization of two electron pockets in the context of KFe$_2$Se$_2$\cite{KhodasChubukov}.  They found the bonding-antibonding $s_\pm$ stabilized with sufficently large hybridization, which they considered as a parameter.   In the weak hybridization limit, they also found that the $d$-wave was stable; and at the border between these two phases,  an $s+id$ state was favored.     The identification of the symmetry of this and related systems is still controversial.  On the one hand, inelastic neutron scattering measurements\cite{InosovKFe2Se2} appear to give a resonance peak very close to the incommensurate wave vector predicted by the spin fluctuation theory calculations giving $d$-wave\cite{Maierdwave}; on the other, the nodes on the tiny $Z$-centered electron pocket expected for the $d$-wave case were not observed by ARPES\cite{DLFengKFe2Se2}.

\begin{figure}[tbp]
\includegraphics[angle=0,width=\linewidth]{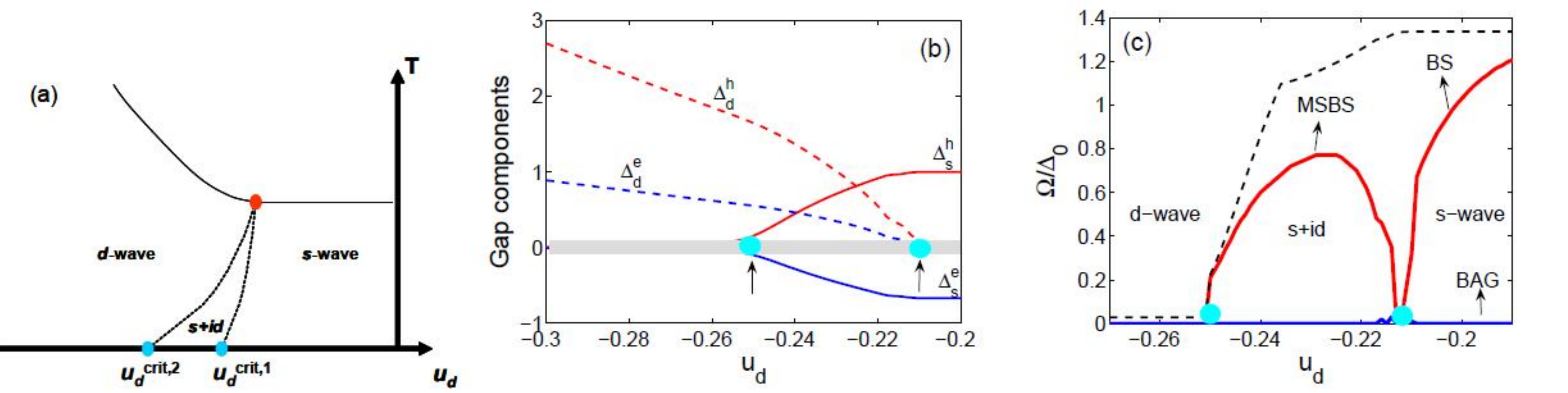}
\caption{  (Color online.) (a) Phase diagram of a  3-pocket model appropriate for hole-overdoped FeSC, from Ref. \onlinecite{MH_coll_mode}.  (b) Evolution of the gap components with fixed $u_s$ and varying  $u_d$ in the 3-pocket model at $T=0$.  Red is for the hole pocket and blue is for the electron pocket(s). The solid line is the $s-$wave part and dashed is the $d-$wave part. The grey line at zero is the size of error in the numerical calculation due to the the choice of grid and resolution parameters. (c): The  (undamped) collective modes across the phase diagram in  different channels.   ``MSBS"= mixed symmetry Bardasis Schrieffer mode; ``BAG" = Bogoliubov-Anderson-Goldstone mode (neutral analog system).  The dashed black line is $2\Delta_{\text{min}}$ in the system. The dots indicate the $s+id$ phase boundaries at $T=0$. In all the figures $z = 1/2$ and $u_s\nu_2D=0.2$.
  }
\label{fig:Tbreaking}
\end{figure}

 Finally, again following the cuprate example, the notion that an $s+id$ state can be induced as a result of gradient terms in the free energy that mix the two representations even if only one
 is stable in the homogeneous case, has been proposed.  High quality FeSe crystals, orthorhombic at low $T$, exhibit twin boundaries that can be imaged in STM.  Recently Watashige et al.\cite{ShibauchiTwin} measured the STM spectrum in the superconducting state as a function of distance to the twin boundary.  They found a $V$-shaped gap in the bulk, indicative of nodes, but a $U$-shaped full gap that opened smoothly as the twin boundary was approached.  This full gap was largest when measurements were
  made between two parallel twin boundaries; the simultaneous proximity to both evidently magnified the effect.  On the other hand, this interference of the two twin
  boundaries took place over a length scale corresponding to nearly 7 times the low-temperature bulk coherence length, as measured by standard means.   Under normal
  circumstances,  perturbations to the electronic structure should influence superconductivity over a few coherence lengths at most.  Watashige et al. appealed to an idea
  put forward in the cuprate context by Sigrist et al.\cite{SigristTwin}, that a new length scale can appear close to the critical point where an $s+id$ state is
  created from the pure $s$ or $d$ state, corresponding to the coherence length of the subdominant order parameter component which appears at the 2nd order transition,
  and argued that their results were indirect evidence for ${\cal T}$-breaking.

  It is interesting to pose the question how best to identify the transition from an $s$-state to an $s+id$ state, both
   of which are fully gapped, as might be realized, e.g. in
  K$_{1-x}$Ba$_x$Fe$_2$As$_2$.    The quasiparticle spectrum will
  not change in any distinctive way at the transition, the impurity signatures are weak\cite{Zhang_sid}, and there are no edge states or other chiral phenomena predicted.
  Perhaps the best method is that proposed in Ref. \onlinecite{MH_coll_mode}, to look in a Raman scattering experiment for the softening of the Bardasis-Schrieffer mode at the phase boundary. In the $s$-ground state, the BS mode is a pure $d$-exciton, and should show up only in the $B_{1g}$ Raman polarization; however, because it is a mixed symmetry mode in the $s+id$ phase, it should begin to show up in both $A_{1g}$ and $B_{2g}$ polarizations as the $s+id$ phase is entered, and subsequently harden\cite{MH_coll_mode} (See Fig. \ref{fig:Tbreaking}).

\begin{figure}[tbp]
\includegraphics[angle=0,width=\linewidth]{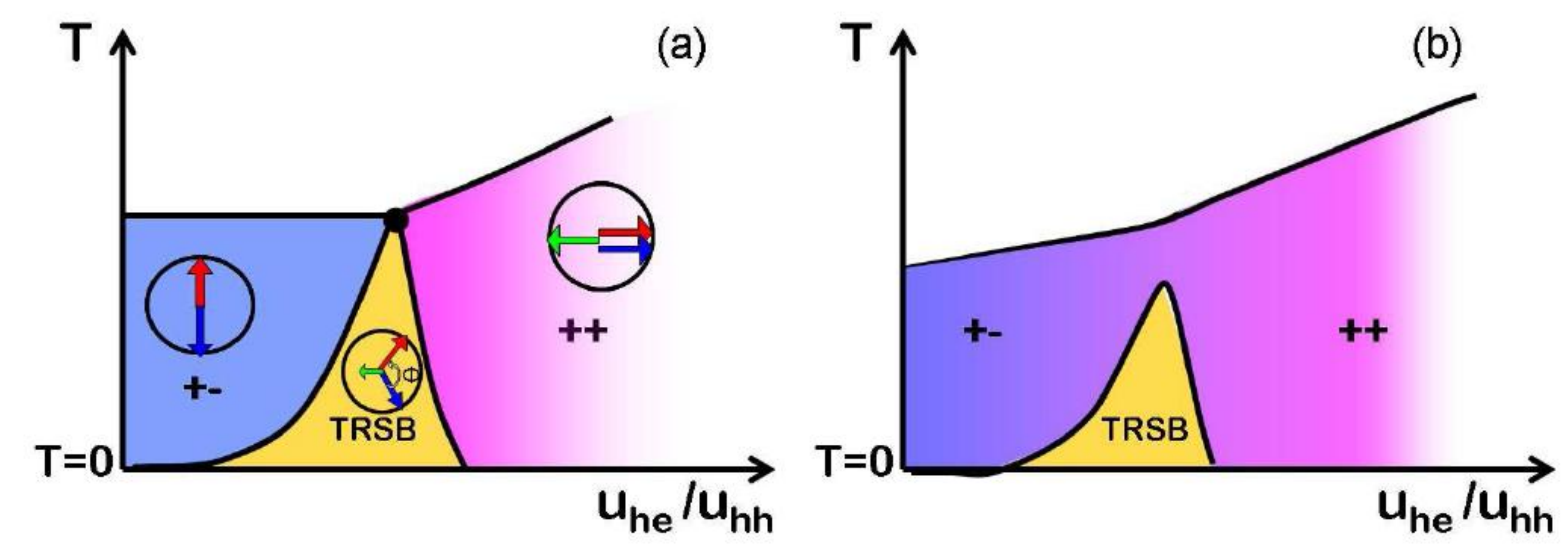}
\caption{(Color online.) Qualitative phase diagram for Ba$_{1−x}$K$_x$Fe$_2$As$_2$,
 from Ref. \onlinecite{Maiti_sis}.  (a) phase diagram as function of ratio $U_{he}/U_{hh}$ of interband to intraband interactions, for case where two hole pockets
 $h_1$ and $h_2$ are degenerate.  ``TSRB" indicates time-reversal symmetry broken $s+is$ phase.  Arrows indicate phases of $\Delta_1$ (red), $\Delta_2$ (blue), and $\Delta_3$ (green). (b)  Phase diagram with nondegenerate hole pockets.     }
\label{fig:Tbreaking_sis}
\end{figure}

 In parallel with the discussion regarding $s+id$, it was realized that the multiband aspects of the Fe-based system allowed for a new type of ${\cal T}$-breaking state.
 Neglecting all angular dependence of the effective interband repulsions,  Stanev and Tesanovic\cite{Stanev} considered a simple model with three bands, and showed that it
 had, in addition to ``conventional" $s$-wave generalizations of the two-band $s_\pm$, a possible ground state which minimized the frustration associated with the additional
 pocket by acquiring an internal nontrivial phases between the different gaps.  A possible ``$s+is$" phase of this type is shown in Fig. \ref{fig:symmetry} (c),
 where the gaps on the three bands are  $\Delta_1\sim e^{i\phi/2}$,  $\Delta_2\sim e^{-i\phi/2}$ are complex, but $\Delta_3$ is real.  Here $\phi$ is a phase angle that depends on
 interactions, and evolves between 0 and $\pi$  as a function of these interactions, or doping.    Such a state was proposed\cite{Maiti_sis} as a natural ``intermediary" between a standard $s_\pm$ state for the optimally doped Ba$_{1-x}$K$_x$Fe$_s$As$_2$,
 where both central hole pockets have the same sign, and  the state found from from spin fluctuation phenomenology (low-order harmonic parametrization of
 the interband interactions) for KFe$_2$As$_2$, where the two gaps on the hole pockets have opposite sign\cite{ChubukovKFe2As2}.    In Fig. \ref{fig:Tbreaking_sis},
 their conclusions are sketched in the form of a generic phase diagram for a simple multiband  model.   This state was also one of several considered by Ahn et al.\cite{Ahnetal} in the context of LiFeAs (Sec. \ref{subsec:LiFeAs}).

   Since the expected local
 current signatures of a spatial perturbation in an $s+is$ state are even weaker than in $s+id$ (the net current vanishes in both states since they are not chiral, but it
 vanishes even locally in the $s+is$ case unless orthorhombic symmetry is broken), distinguishing such states will be difficult.  The best possibility at the present time
 appears to be measuring characteristic collective modes.   Since the system is by assumption at least 3-band, there are several possible Leggett-type modes
 that soften at the phase boundaries to the pure (real) $s$-wave states\cite{Maiti_sis,Benfatto}.    It has also been proposed that $s+is$ states would
 have magnetic signatures at domain walls that could be detected by scanning SQUID probes\cite{Babaev}.

 In a realistic system with several hole bands at the Fermi surface like LiFeAs, the ground state is a subject of considerable controversy even among groups that
 share basic assumptions about its electronic structure and the fundamental origin of the pairing.   This disagreement is a good illustration of the sensitivity of
 theory in a multiband system to small changes in electronic structure.

\section{ $\eta$-pairing}
\label{sec:eta}

Although much theoretical work on the pairing state of FeSC has been done using simplified models based on a 1-Fe unit cell containing 5 $d$ orbitals, it has not always been clear why this procedure works quite well.  In fact,
there are several claims in the literature that the reconstruction of the near-Fermi level states by the As(P,Se...) potential which breaks the
 translational symmetry of the Fe lattice has important consequences for
the pairing of electrons, despite the fact that the As states are nominally 2eV below.   These works often imply that the early pairing calculations missed an essential symmetry  aspect of the problem because
they did not explicitly account for the glide plane symmetry of the Fe-As layer (space group P4/nmm)\cite{Lee2008} .  In particular, it was suggested by Hu et al.\cite{Hu2013,Hao2014,Hu2012} that the ``standard" spin fluctuation approach misses a
spin singlet odd-parity pairing  state, which may
 represent the ground state of FeSC in general.

  A pure version of a state of this kind, as discussed in Sec. \ref{sec:symmetry},  would indeed be quite different from any states observed in unconventional superconductors to this
point, and would be of great interest if actually realized.     In Hu's work,  such states are shown to occur naturally if one properly respects
 the glide symmetry,  allowing  nonzero center of mass crystal momentum Cooper pairs (called $\eta$-pairing in the cuprate
 context by C.N. Yang\cite{Yang1989}, see also Refs. \onlinecite{Scalettar1991},\onlinecite{Bickers1992}).   Casula and Sorella\cite{Sorella2013} pointed out that the two-orbital models
used in Hu's work were insufficient to recover the full generality of pairing states formally possible in the $P4/nmm$
 symmetry of a single FeAs plane, since these models neglected pairing of  orbitals of unlike z-parity under reflection through the Fe plane.   Including such
 states in a variational Monte Carlo calculation led to a linear combination of  ``planar" $d$- and $s-$ states, i.e.
 those corresponding to the irreducible representations of the tetragonal group conventionally used; such admixtures
 provide, in their picture, a natural explanation for so-called ``nematic" $C_4$ symmetry-breaking in the superconducting
 state, as seen in STM on FeSe\cite{Song2012} (see however Sec. \ref{subsec:nematic}).  Finally, there are even recent  theoretical suggestions\cite{Lin2014} suggesting that the odd-parity  singlet state can
 explain the so-called ``ARPES paradox"\cite{HKMReview2011}, the apparent disagreement between  ARPES and a variety of thermodynamic
  and transport probes concerning the anisotropy of the superconducting gap in 122 systems,   and may in addition break time reversal symmetry\cite{Yuan2014,Lin2014}.

  The implications of the glide plane symmetry of a single Fe-pnictogen/chalcogen plane for various normal state properties as well as for pairing was discussed in several  works\cite{Lee2008,Eschrig2009,Andersen11,Sorella2013,FeSCsymmetry,Cvetkovic2013,Tomic14}: The Fe lattice is symmetric under the glide plane  operation $P_z=T_r\sigma_z$, i.e., a single translation along the $x$- or $y$-direction $T_r$ by 1 Fe-Fe distance, together with a reflection $\sigma_z$ along $z$. As a result, the  intra-sub-lattice hopping between $d$ orbitals  ``even" under $P_z$ ($xy,\,x^2-y^2,\,3z^2-r^2$) and ``odd" orbitals  ($xz,\,yz$) changes sign betwen the A- and B- Fe sublattices. If one labels states by  the ``physical" 1-Fe crystal momentum $\bk$, it forces a mixing\cite{Lee2008} between momenta $\bk$ and $\bk+\bQ$ with $\bQ=(\pi,\pi)$ of the type $\sum_{\bk,\sigma} \left[t^{xz,xy}(\bk)c^\dagger_{xz,\sigma, \bk+\bQ}c^{\phantom\dagger}_{xy,\sigma, \bk}+{\rm h.c.}\right]$, and similar terms, between even and odd (with respect to $z$-parity) orbitals. As a result, one is forced to consider off-diagonal propagators with both even and odd orbital states, evaluated at momenta $\bk$ and $\bk+\bQ$. At first glance, this is an important difference in the representation of pair states compared to the usual BCS ansatz, because one now has  nonzero total momentum $\eta$-pairs $\langle c_{\ell_1,\uparrow,\bk} c_{\ell_2,\downarrow,-\bk+\bQ}\rangle$ for $\ell_1$ even, $\ell_2$ odd or vice versa, in addition to the standard zero center of mass momentum pairs $\langle c_{\ell_1,\uparrow,\bk} c_{\ell_2,\downarrow,-\bk}\rangle$ for $\ell_1,\ell_2$ either both even or both odd.~\cite{Lin2014}.
However, this mixing is absent if one uses the eigenvalues of $P_z$, i.e., the pseudo-crystal momentum $\tbk$ to classify the states \cite{Lee2008,Eschrig2009,Andersen11,Lin2014}, which simply amounts to shifting the momentum of either the even or the odd orbitals by $\bQ$.

In Ref. \onlinecite{Wang_glideplane2015}, Wang et al. studied the problem of $\eta$-pairing in models of a FeAs layer  with both a 5 and 10 $d$-orbital basis,  assuming  spin-fluctuation driven superconductivity with  pairing interaction (\ref{eq:Gam_ij}).
They  chose the shift $\bQ$ in the even orbitals, so that states labelled by  pseudo-crystal momentum $\tbk$ are related to the states labelled by physical crystal momentum $\bk$ via
\begin{align}
	\label{eq:pcm} \tilde{c}_{\ell,\sigma,\tbk} =
	\begin{cases}
		c_{\ell,\sigma,\bk}, & \text{if $\ell$ odd,} \\
		c_{\ell,\sigma,\bk+\bQ}, & \text{if $\ell$ even.}
	\end{cases}
\end{align}
The tilde (${\tilde \,} $) basis is now identical to that used to perform pairing calculations in the 5-orbital model, where
the exact\cite{Eschrig2009} unfolding symmetry was assumed.  The Hamiltonian  automatically  accounts for
the additional terms arising from the mixing between $\bk$ and $\bk+\bQ$ in the physical 1-Fe crystal momentum $\bk$ space.
\vskip .2cm
One now simply uses the basis transformation to relate the pairing condensate eigenfunction $g(\k)$  in the two representations:
\begin{align}
	\label{eq:Delta_N_orb} g^N_{\ell_1\ell_2}(\bk)= \langle c_{\ell_1\uparrow,\bk}&c_{\ell_2\downarrow,-\bk}-c_{\ell_1\downarrow,\bk}c_{\ell_2\uparrow,-\bk}\rangle =\\
	&
	\begin{cases}
		\tilde{a}^{\ell_1}_{\nu,\bk}\tilde{a}^{\ell_2}_{\nu,-\bk}\,\tilde{g}_\nu(\bk), & \text{$\ell_1,\ell_2$ odd,}\\
		\tilde{a}^{\ell_1}_{\nu,\bk-{\bf Q}}\tilde{a}^{\ell_2}_{\nu,-\bk+{\bf Q}}\,\tilde{g}_\nu(\bk-{\bf Q}), & \text{$\ell_1,\ell_2$ even,}\\
		0, & \text{otherwise,}\nonumber
	\end{cases}
\end{align}
and $\eta$-pairing terms shifted by  center of  mass momentum $\bQ$,
\begin{align}
	\label{eq:Delta_eta_orb}  g^\eta_{\ell_1\ell_2}(\bk)= \langle c_{\ell_1\uparrow,\bk}&c_{\ell_2\downarrow,-\bk+{\bf Q}}-c_{\ell_1\downarrow,\bk}c_{\ell_2\uparrow,-\bk+{\bf Q}}\rangle  =\\
	&
	\begin{cases}
		\tilde{a}^{\ell_1}_{\nu,\bk}\tilde{a}^{\ell_2}_{\nu,-\bk}\,\tilde{g}_\nu(\bk), & \text{$\ell_1$ odd, $\ell_2$ even,}\\
		\tilde{a}^{\ell_1}_{\nu,\bk-{\bf Q}}\tilde{a}^{\ell_2}_{\nu,-\bk+{\bf Q}}\,\tilde{g}_\nu(\bk-{\bf Q}), & \text{$\ell_1$ even, $\ell_2$ odd,}\\
		0, & \text{otherwise.}\nonumber
	\end{cases}
\end{align}

Using ~(\ref{eq:Delta_N_orb}) and (\ref{eq:Delta_eta_orb})  Wang et al.\cite{Wang_glideplane2015}
 showed that the gap function $\tilde{\Delta}_\nu(\bk)\propto g_{\nu}(\bk)$ obtained  in the 5-orbital model in the pseudo-crystal momentum representation contains both normal and $\eta$-pairing terms in the physical crystal momentum space, i.e.,
\begin{align}
	\tilde{\Delta}_\nu(\bk)&=\Delta^N_{\rm odd}(\bk)+\Delta^N_{\rm even}(\bk+\bQ)\nonumber\\
	&+\Delta^\eta_\text{odd-even}(\bk)+\Delta^\eta_\text{even-odd}(\bk+\bQ)\,.
\end{align}
Thus the $\eta$-pairs are seen to be  indeed present in FeSC, but they form a natural part of the usual gap function that has been discussed thus far in the
literature.    Closer examination of the $\eta$ terms show that they are imaginary and odd parity singlet states {\it only when viewed  in the orbital basis}; they are real and have the usual even parity when viewed in band space.
\vskip .4cm
To see explicitly that there are no direct observable consequences beyond the models discussed already in Sec. \ref{sec:standardmodel}, one can construct the
one-particle spectral function in the physical momentum space, as observed by ARPES\cite{Lin2014},
\begin{align}
	\label{eq:Akw} A(\bk,\omega) = \sum_\nu &\left[ \sum_{\ell\,\text{odd}} |\tilde{a}^\ell_{\nu,\bk}|^2 \tilde{A}_\nu(\bk,\omega)\right.\nonumber\\
	&+\left.\sum_{\ell\,\text{even}}|\tilde{a}^\ell_{\nu,\bk-\bQ}|^2 \tilde{A}_\nu(\bk-\bQ,\omega)\right],
\end{align}
where $\tilde A$ is the spectral function corresponding to the pseudomomentum basis, i.e. containing poles at the energies $E_\nu({\tbk})=\sqrt{\epsilon^2_\nu(\tbk)+\tilde{\Delta}^2_\nu(\tbk)}$.  Therefore, the superconducting gap that enters $A(\bk,\omega)$, as measured in ARPES experiments is given by the gap function $\tilde{\Delta}_\nu(\tbk)$ calculated in the 5-orbital 1-Fe zone in pseudo-crystal momentum space. $\tilde{\Delta}_\nu(\tbk)$ thus accounts completely for  the symmetry breaking potential caused by the pnictogen/chalcogen atom\cite{Wang_glideplane2015}.

\vskip .2cm

Thus Wang et al. concluded that  $\eta$-pairing is indeed an important ingredient in the superconducting condensate, when properly defined. These terms occur as a natural part of the gap in the standard 5-orbital theories, which coincide identically (for such single layer models) with 10-orbital calculations\cite{Wang_glideplane2015}.  They contribute, however, with the standard  even parity spin singlet symmetry {\it in band space},  and time reversal symmetry is not broken.
\vskip .2cm


\section{Disorder}
\label{sec:disorder}
In principle, disorder is a phase-sensitive probe of unconventional superconductivity, because nonmagnetic impurities are pairbreaking only if they scatter electrons between portions of the Fermi surface with gaps of different sign.    Bound states of electrons near strong nonmagnetic impurities (Andreev bound states) can similarly be formed only in the sign-changing case.   This subject has a long history in the context of $d$-wave superconductivity in the cuprates which has been reviewed several times\cite{JLTPreview,Balatskyreview,Alloulreview}.   Many of the ideas worked out in this
context have found interesting extensions in the FeSC field, because of the added aspect of multiband physics.
\subsection{Intra- vs. interband scattering}

In  2-band superconductors,
nonmagnetic impurities can either scatter quasiparticles between
bands or within the same band.  Intraband processes tend to average
the gaps and can thus lead to some initial $T_c$ suppression, but eventually $T_c$
will saturate, according to Anderson's theorem\cite{Markowitz}.   However interband scattering
is  pairbreaking in a system with two gaps of different signs\cite{Muzikar, Golubov, kulic}, just as it is
in single-band systems where the order parameter has nodes, e.g. the $d$-wave case.
In the presence of any nonzero interband scattering amplitude, $T_c$ will eventually be suppressed to zero at some critical concentration, just
as in the more familiar theory of scattering by magnetic impurities in a 1-band $s$-wave superconductor\cite{AG}. The effect of interband scattering in the FeSC was discussed early on by
several groups \cite{Parker2008, Chubukovdisorder, SengaKontani, Bang_disorder}.

The simplest toy model of a chemical impurity in 2-band system  characterizes the scatterer crudely by an effective interband potential $u$ and an intraband potential $v$, such that results for various quantities in the superconducting state will depend crucially on the size and relative strengths of these two quantities. Most calculations are performed using the $T$-matrix approximation to the disorder self-energy for pointlike scatterers $\hat{\Sigma}^{imp}(\omega)$,
\begin{equation}
 \hat{\Sigma}^{imp}(\omega) =n_{imp}  \hat{\mathbf{U}} + \hat{\mathbf{U}} \hat{{G}}(\omega) \hat{\Sigma}^{imp}(\omega ), \label{eq:tmatrix}
\end{equation}
where $\hat{\mathbf{U}} = \mathbf{U} \otimes \hat\tau_3$, $n_{imp}$ is the impurity concentration, and the $\tau_i$ are Pauli matrices in particle-hole space. Here ${\mathbf{U}}$ is a matrix in band space $(\mathbf{U})_{\alpha \beta} = (v-u) \delta_{\alpha \beta} + u$, and $\alpha,\beta=1,2$. With the expression for the $\k$-integrated Green's function $\hat G (\omega) = \sum_\k {\hat G}(\k, \omega)$, this completes the specification of the equations which determine the  Green's functions

\begin{equation}
\hat{G}(\k, \omega)^{-1} =  \hat{G^0}(\k, \omega)^{-1} - \hat\Sigma^{imp}(\k, \omega),
\label{Greensfctn}
\end{equation}
where $\hat{G}^0$ is the Green's function for the pure system. The self-consistent $T$-matrix approximation includes
those scattering processes corresponding to multiple scattering from a single impurity, and may be considered correct for
dilute impurities (except possibly for nodal systems in 2D, where it has some singularities at low energies\cite{JLTPreview}).

\subsection{$T_c$ suppression}
\label{subsec:Tcsuppression}
Several groups \cite{SengaKontani,Onari09,Lietal12} have pointed out what is claimed to be  a ``slow"
decrease of $T_c$ in systematic chemical substitution measurements \cite{Li10,Nakajima10,Tropeano10,Lietal12}, which is
then claimed to be related  to the natural robustness  of an $s_{++}$ superconductor to disorder.  The meaning of ``fast" and "slow" $T_c$
suppression can be deceptive, however, especially in a situation when only large momentum scattering is pairbreaking, as
expected for an $s_\pm$ state.   Note that chemical substitution can cause pairbreaking, but also affect  the electronic structure
of the system by doping with carriers or by chemical pressure.  It is therefore useful to calculate a measure of disorder that is simultaneously observable along with $T_c$,
such as the residual resistivity, against which to compare $T_c$ suppression rates.  This problem was addressed particularly by
Wang et al.\cite{YWang_Tcsuppression13}, who argued that almost any rate of $T_c$ suppression was possible in an $s_\pm$ system, depending
on the ratio of intra- to interband scattering.  This conclusion is implicit in the discussion by  Golubov and Mazin\cite{Golubov} who discussed the general case of pairbreaking in a multiband superconductor.

To see this, I note that near $T_c$, the gap equation can be  linearized,
\begin{equation}
\Delta_{\mu}{(\mathbf{k})} = -2 T \!\sum_{{\bf k}^{\prime},\nu,\omega_{n} > 0}^{\omega_{n} = \omega_{c}}\!
V^{\mu \nu}_{{\bf k} {\bf k}^{\prime}} \frac{\tilde{\Delta}_{\nu}({\bf
k}^{\prime})}{\tilde{\omega}^{2}_{\nu}+\xi^{2}_{\nu}}, \label{eq:gpetc}
\end{equation}
where $\xi_{\nu}$ is the  dispersion of band $\nu$ near the Fermi level, and the renormalized gaps and
frequencies are $\tilde{\Delta}_{\nu}({\bf k}^{\prime})={\Delta}_{\nu}({\bf k}^{\prime})+\Sigma_\nu^{(imp,1)}$
and ${\tilde{\omega}_{\nu}}=\omega_n+{i\Sigma_{\nu}^{(imp,0)}}$.

To investigate both the effects of sign-changing gaps, and that of possible gap anisotropy,  let us consider a model  similar to that obtained from spin
fluctuation theories, where  the gap on   pocket $a$ is isotropic, $\Delta_a$, and the gap on the
 pocket $b$ may be anisotropic, $\Delta_b=\Delta_b^0 +\Delta_b^1(\theta)$, where $\theta$ is the
 angle of $\k$  around the $b$ pocket, with $\int d\theta \Delta_b^1(\theta)=0$. The pairing interaction is then
assumed to be  $V^{\mu\nu}_{\vk\vk'}=V_{\mu\nu}\phi_\mu(\vk)\phi_\nu(\vk')$, with $\phi_\mu=1
+r\delta_{\mu,b}\cos 2\theta$.  Note that the choice of $\cos 2\theta$ harmonic implicitly means that $b$ is identified
with the electron pockets at X and Y, but this is easily generalizable.  The parameter $r>0$ controls the anisotropy on $b$,
and yields gap nodes if $r>1$.  This ansatz then  gives coupled  equations for {$(\Delta_a,\Delta_b^0,\Delta_b^1)^\text{T}\equiv
\underline{\Delta}$} which may be written
as
\begin{equation}
 \underline{\Delta}={\ln}\Bigl(1.13\frac{\omega_{c}}{T_{c}}\Bigr) \underline{\mathcal{M}}\,\underline{\Delta}
  {\equiv \mathcal{L}_0\underline{\mathcal{M}}\,\underline{\Delta}},
\label{eq:tcc1}
\end{equation}
where the matrix
$\underline{\mathcal{M}}=({1}+\underline{V}\,\underline{\mathcal{R}}^{-1}\underline{\mathcal{X}}\,\underline{\mathcal{R}})^{-1}\,\underline{V}$
{and the constant $\mathcal{L}_0=\ln\bigl(1.13\frac{\omega_{c}}{T_{c}}\bigr)$} were introduced.
$\underline{V}$ is the interaction matrix in the above basis, and $\underline{\mathcal{R}}$ is the unitary
matrix which diagonalizes
\begin{equation}
 \underline{\Phi}=\frac{\pi n_\text{imp}}{\mathcal D_N} \left[\begin{array}{ccc}
                      N_{b}u^{2} & {-}N_{b}u^{2} & 0 \\
                      {-}N_{a}u^{2} & N_{a}u^{2} & 0 \\
                      0 & 0 & {N_{b}v^{2}+N_au^2} \\
                     \end{array}\right].
\label{eq:Lambda}
\end{equation}
Here $N_\mu$ is the density of states of  band $\mu$ at the Fermi energy, $\mathcal D_N$ is the determinant of the $T$-matrix  in the normal state, and $\underline{\mathcal{X}}$ is a  diagonal matrix,
\begin{equation}
{\mathcal{X}}_{ii}=
{ \mathcal{L}_0-\left[\Psi\Bigl(\frac{1}{2}+\frac{\omega_c}{2\pi T_c}+\frac{\lambda_{i}}{2\pi
T_c}\Bigr)-\Psi\Bigl(\frac{1}{2}+\frac{\lambda_{i}}{2\pi T_c}\Bigr) \right]} \label{eq:Xd},
\end{equation}
where $\Psi$ is the digamma function and  $\lambda_{i}$ are the eigenvalues of matrix
$\underline{\Phi}$. The critical temperature is now determined by finding the largest ($\lambda_\text{max}(T_c)$) of the matrix
$\underline{\mathcal{M}}$, $
T_{c}=1.13\;\omega_{c}\exp\left[-1/\lambda_\text{max}(T_c)\right]$.

\vskip .2cm
\begin{figure}[tbp]
\includegraphics[angle=0,width=0.5\linewidth]{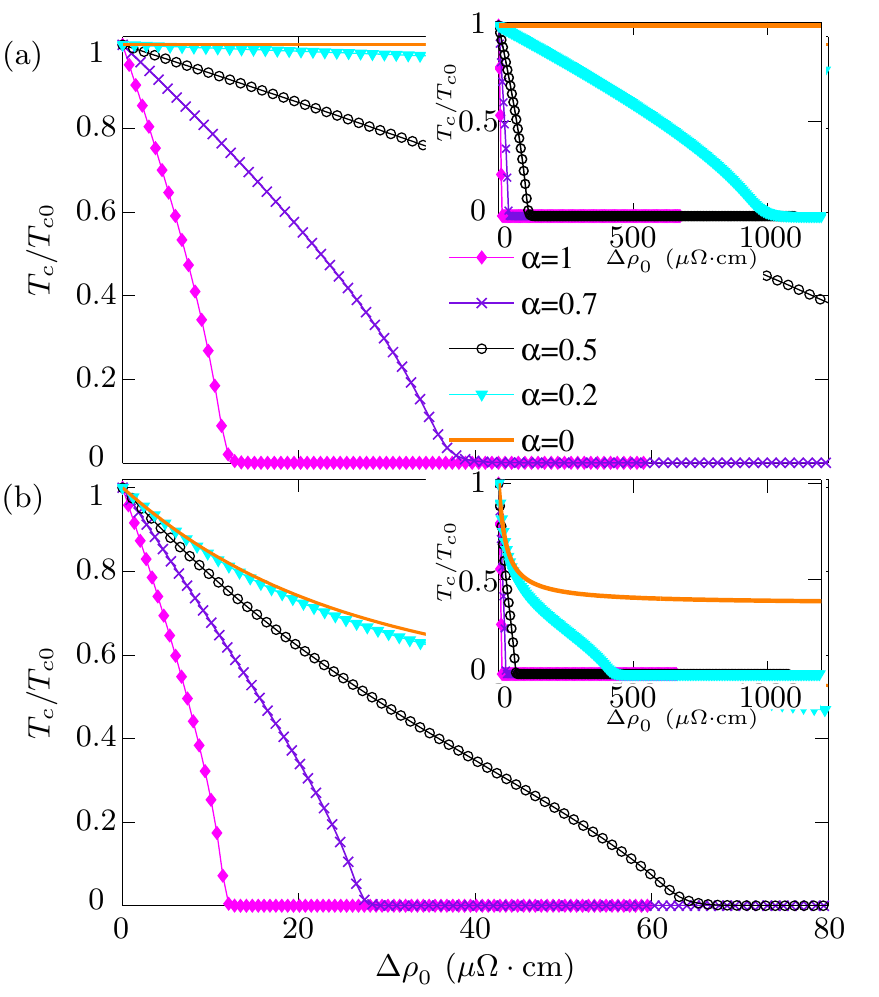}
\includegraphics[angle=0,width=0.4\linewidth]{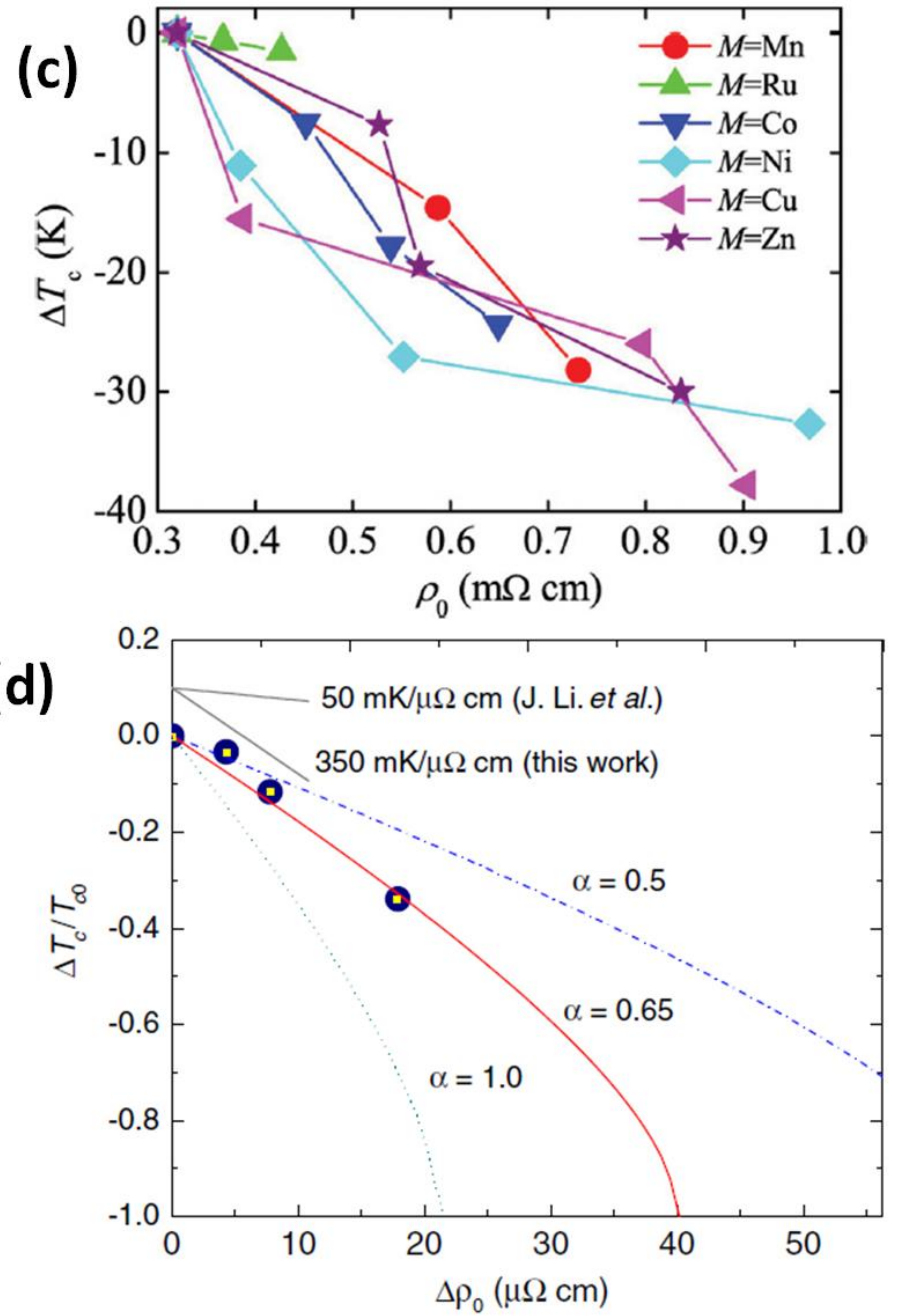}
\caption(Color online.) {a) Normalized critical temperature $T_c/T_{c0}$ vs. disorder-induced resistivity
change $\Delta\rho_0$ for isotropic $s\pm$~wave paring for various values of the inter/intraband scattering
ratio $\alpha\equiv u/v$. Insert:  same quantity plotted over a larger $\Delta\rho_0$ scale.  b) Same as a)
but for anisotropic (nodal) gap with anisotropy parameter $r=1.3$. From Ref. \onlinecite{YWang_Tcsuppression13}.  (c)
$T_c$ suppression vs. residual resistivity in $m\Omega$-cm for various transition metal impurities\cite{Lietal12}.  (d) $T_c$ suppression vs. residual resistivity
in Ru-doped Ba122 irradiated with 2.5meV electrons\cite{prozorov_e-irr}.    }
\label{fig:Tc-rho}
\end{figure}

These expressions now suffice to determine $T_c$ at any given impurity concentration $n_i$, with intra- and interband scattering strengths $v$ and $u$ given.  However,
since $u$ and $v$ are not known in general, it is not particularly useful when comparing to experiment to present calculations of $T_c$ vs. $n_i$, nor vs. some scattering rate component, simple to calculate but difficult to measure directly.  Claims in the literature that $T_c$ suppression is particularly slow, implying no
gap sign change, or particularly fast, implying one, abound.  They are mostly not relevant, however;  in order to draw conclusions in multiband systems, it is crucial to present results for $T_c$ vs. the residual resistivity $\Delta\rho_0$ due to disorder.
The total  DC conductivity  is the sum
of the Drude  conductivities of the two bands, $\sigma=\sigma_a+\sigma_b,$ with $\sigma_\mu=2e^2N_{\mu}
\langle v_{\mu,x}^2\rangle \tau_\mu$, where $v_{\mu,x}$ is the  Fermi velocity component in
the $x$-direction and $\tau_\mu$ the corresponding one-particle relaxation time,  $\tau_\mu^{-1}=-2\mathop{\mathrm
{Im}}\Sigma_{\mu}^{(imp,0)}$. The transport time and single-particle lifetime are identical if the
scatterers are $\delta$-function like, such that corrections
to the current vertex vanish.

In Fig. \ref{fig:Tc-rho} (a),(b), $T_c$ vs. residual resistivity is plotted for various gap structures and ratios of intra- to interband scattering.
The extreme cases include i) purely isotropic $s_\pm$ gaps with intraband scattering, where Anderson's theorem implies that there can be no pairbreaking until localization begins to occur; ii) $\alpha\equiv u/v=1$, where the rapid $T_c$ suppression follows the universal curve predicted by Abrikosov and Gor'kov for
magnetic impurities in an isotropic $s$-wave system.    It is obvious that essentially any other rate of $T_c$ suppression can be obtained for some particular value of the ratio $\alpha=u/v$, such that the observed rate will depend on the range (and orbital character)  of the impurity potential itself.   The role of
anisotropy is also interesting, if not surprising.  If one of the gaps is anisotropic, this speeds the initial rate of $T_c$ suppression even in the case of dominant intraband scattering, as the system averages out the gap by scattering until it is roughly isotropic.  Subsequent pairbreaking is due to the interband component of the scattering, which is sensitive to the gap sign change between the two pockets.  This sequence will be discussed further in Sec. \ref{sec:gapevolution} below.

In this context, the data on $T_c$ suppression by chemical substitution, e.g. those of Li et al\cite{Lietal12} (Fig. \ref{fig:Tc-rho} (c))  have to be treated with caution.  While the critical resistivity for which $T_c$
vanishes is
indeed large compared to the small $\Delta\rho_0$ scale one might find for an $s_\pm$ state and large interband scattering,  the impurities are certainly doping the system,
at least in some cases.  It is also somewhat bizarre to find in a putative $s_{++}$ state that Mn, certainly a magnetic impurity, suppresses $T_c$ at the
same rate as Co, certainly not; such behavior is rather characteristic of an $s_\pm$ state, where the magnetic character of the impurity is generically irrelevant.
What is needed are measurements  of $T_c$ as a function of some quantity tied directly  to pairbreaking.  This test is in fact provided by electron irradiation\cite{rullierdisorder,prozorov_e-irr,Shibauchi14}, since these works have shown that the main effect of low-energy electron irradiation is the creation of pointlike Frenkel-type
pairs of Fe vacancies and interstitials, which act as nonmagnetic scatterers that do not dope the system.    This is, incidentally, not the case with proton or heavy ion irradiation\cite{prozorov_e-irr}.

In Figure \ref{fig:Tc-rho}(d),  I show the $T_c$ data from the Ames group on Ru-doped BaFe$_2$As$_2$ under electron irradiation, plotted vs. the residual resistivity $\Delta\rho_0$.
The data clearly show a much more rapid suppression of $T_c$ with a much smaller critical resistivity for the destruction of the superconducting state, roughly 40 $\mu\Omega $-cm,
in fact very close to the value predicted but not measured by Ref. \onlinecite{Lietal12} for the ``symmetric" $s_\pm$ state under optimal pairbreaking conditions.  While an $s_\pm$ state
can show a much slower rate of $T_c$ suppression due to large intraband scattering, there is no way an $s_{++}$ state can show such a rapid rate of suppression unless the
impurities are purely magnetic.  Ref. \onlinecite{prozorov_e-irr} was able to rule out this possibility by measurements of the low-$T$ magnetic penetration depth, which showed
no indication of the upturns that would be caused by localized magnetic states.

Here I have mostly discussed the role of nonmagnetic impurities, but
I close this the subsection with a remark or two about some novel phenomena associated with magnetic
ones.  A rather general summary of the properties of magnetic impurities in multiband superconductors has recently been
given by Korshunov et al.\cite{Korshunov_magnetic_imp}, who found somewhat surprisingly that there are several cases
where one's intuition based on standard AG theory \cite{AG} breaks down, for example situations in which $T_c$ saturates in the limit of large disorder, instead of falling to zero.    These effects may be responsible for the weak suppression of $T_c$ by Mn along
with nonmagnetic transition metal impurities in \BFA\cite{Li_etal_2014} (Fig. \ref{fig:Tc-rho}(c)), but do not seem useful in understanding why all transition metal impurities (nominally magnetic and nonmagnetic)  suppress $T_c$ at essentially the same rate.    While this result, taken in isolation, suggests that the system must have a sign changing gap,  in none but the simplest toy models could this result for different impurities with different potentials give such similar $T_c$ suppressions.

Another interesting aspect of Mn in the \BFA ~system was pointed out by Ref. \onlinecite{McQueeneyGoldmanMn}, namely that Mn has the effect of actually enhancing the Ne\'el temperature in  parts of the phase diagram.   By combining neutron and $\mu$SR data, Inosov et al \cite{InosovMn} observed that this impurity-induced high-temperature magnetic state was actually an inhomogeneous ($\pi,0$)
stripe state, and proposed that each Mn nucleated a small stripe region around it.   A plausible model version of this effect based on the
effect of antiferromagnetic correlations on the impurity state was developed by  Gastiasoro and Andersen\cite{Gastiasoro_magnetic}, generalizing an ``order by disorder" scenario developed for the cuprates\cite{AndersenHirschfeldKampf}.

Whereas Mn suppressed $T_c$ rather slowly in \BFA,  Hammerath et al.\cite{Hammerath} reported the remarkable result that in
 optimally F-doped La-1111, a concentration of Mn of only 0.2\% sufficed to drive $T_c$ from about 30K to zero.  Simultaneously,
 spin fluctuations measured by NMR were observed to increase with Mn concentration (indicating, among other things, that the Mn
  impurities were not simply introducing a magnetic pairbreaking potential) until a state with quasi-long range magnetic
 order was reached at a concentration of around 1\%.  The authors attributed this extreme sensitivity of $T_c$ to disorder
 to the proximity of this sample to the AF critical point.    While the theoretical problem of the behavior of the local
 spin fluctuation spectrum around an impurity embedded in a system near a quantum critical point has been studied briefly
 in the cuprate context\cite{MorrMillis}, where there is little hard evidence of quantum critical behavior near the
  low temperature AF transition, it has received relatively little attention in the FeSC field where critical points are much better established.

\subsection{Impurity bound states}
\label{subsec:boundstates}

The observation of a localized resonance near a nonmagnetic impurity in STM is one of the clearest indications of a sign-changing order parameter.   Establishing that a given impurity is not magnetic is not  always trivial, particularly since in the presence of electronic correlations, a bare nonmagnetic potential can induce a local moment\cite{Alloulreview}.  However in the case of chemical substituents or irradiation,  creation of such moments should lead to other observable signals, such as Schottky anomalies in specific heat or upturns in the London penetration depth, so induced moments of this type can usually be excluded.
On the other hand,  if no bound state is seen,  one may  {\it not}  conclude that the superconducting gap does not change sign.
This is because in the multiband system the existence of an impurity bound state requires a fine tuning of the impurity potential, as pointed out in Ref. \onlinecite{HKMReview2011}, and discussed extensively in Ref. \onlinecite{BeairdVekhter}.      This applies not only to weak potentials, which never give rise to bound states even in single band systems, but also to strong potentials in general.      The bound state can also be obscured if the gap function of the system in question has
deep minima or nodes, in which case the impurity pole may occur at an energy less than the coherence peaks but still be so broadened by its coupling to extended states at energies greater than the gap minimum that it is not observable in a practical sense.
In the simplest case, that of a 2-band isotropic $s_\pm$ superconductor,  it is clear that the intraband scattering does not contribute to bound state formation.
This point will be relevant in the next section.

There are a few relatively clear observations of subgap nonmagnetic  impurity bound states in the FeSC by STM, the only tool allowing direct measurement.  Grothe et al.\cite{Grothe2012} observed clear in-gap bound states at native defect sites on the surface of LiFeAs,
some with remarkable shapes which broke tetragonal symmetry.        H. Yang et al.\cite{HHWen_impurity_bound_state} found such a state for Cu in another 111 material,  Na(Fe$_{0.97-x}$Co$_{0.03}$)As,  and
Song et al. reported a bound state around a native defect site and Fe adatom in FeSe\cite{Song2012}.  These impurity resonances do constitute strong evidence in favor of $s_\pm$ pairing, but careful bulk studies are still needed to rule out magnetic character, although this seems unlikely.

These observations reported above all have the qualitative form more or less expected for standard theories of multiband impurity resonances
in these systems\cite{OgataImp,GastiasoroImp13,ZhangImp13}: the state occurs at nonzero energy $\pm \Omega_0$ somewhere in the gap, reflecting the
fine-tuning of the impurity potential relative to the given band structure, and tends to have very different weights at positive and negative energies, such that it is often visible only at one or the other.   Often such bound state energies are found to be slightly different depending on the local disorder environment, an effect expected on the basis of the splitting of the states due to interference processes.    On the other hand, a rather unusual impurity state that does not seem to behave in this way was reported recently by Yin et al.\cite{SHPan_ZBA}, around what was believed to be an Fe interstitial in Fe(Se,Te) ($T_c$=12K).    This state was located at exactly zero energy, to within experimental resolution,  and was unsplit and unmodified by a magnetic field up to 8T, by disorder environment, and by doping over some range.   This smells like some kind of protection of the state by some kind of energy barrier, and the authors of Ref. \onlinecite{SHPan_ZBA} speculated that it might be a Majorana fermion, without providing further details.

\subsection{Gap evolution vs. disorder}
\label{sec:gapevolution}
\begin{figure}[tbp]
\includegraphics[angle=0,width=\linewidth]{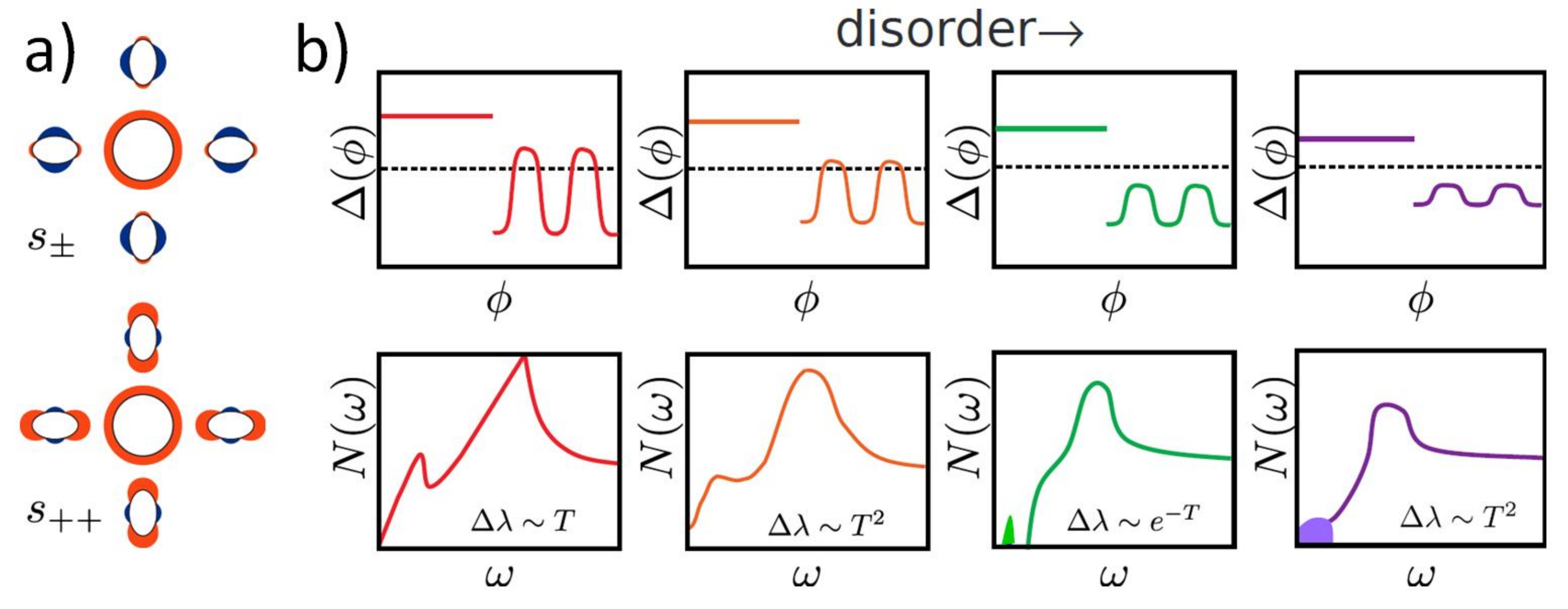}
\caption {(Color online.) a), Schematic of possible $s_\pm$ and $s_{++}$ states. Large circles and small ellipsoids (black lines) are hole and electron Fermi surfaces, respectively. Red and blue represents the superconducting order parameter with different signs; b),  Schematic of $s_\pm$ order parameter vs. azimuthal angle $\phi$ (top row) and density of states vs. energy $\omega$ (bottom row) with increasing irradiation dosage. Dotted lines (top row) represent zero gap. From Ref.\onlinecite{Shibauchi14}.}
\label{fig:gapevolution_theory}
\end{figure}
As disorder is added to an $s_\pm$ superconductor, various qualitative  evolutions of the gaps are possible, depending both on the intra- vs. interband character of the impurity scattering, but also on the intra- and interband character of the interactions.
Remarkably, in the case of an isotropic  $s_\pm$ system with two gaps $\Delta_a$ and $\Delta_b$, one can classify these evolutions in a rather simple way. Efremov et al.\cite{Efremov11} make  use of the  average
interaction strength $\langle \Lambda \rangle \equiv \left[ (\Lambda_{aa}+\Lambda_{ab})N_{a}/N + (\Lambda_{ba}+\Lambda_{bb})N_{b}/N\right]$ with $N=N_{a}+N_{b}$ the total density of states\cite{footnoteEPC}.
Depending on $\langle \Lambda \rangle$,
there turn out to be  two classes of $s_\pm$ superconductivity. The first, for $\langle \Lambda\rangle <0$, is the
usual case, for which $T_c$ is suppressed as disorder, including some amount of interband potential scattering,  is increased.  In this case $T_c$ decreases
 until it vanishes at a critical scattering rate.    There is however also a second class with $\langle \Lambda\rangle >0$ for which $T_c$ tends to a nonzero value as disorder increases.      At the same time, intraband scattering has no pairbreaking effect on the gaps(because they are assumed isotropic), and
the interband scattering mixes the + and - gaps.   In this case, the smaller of the two in magnitude  changes its sign, leading to an $s_{++}$ state.
This transition, which should be accompanied by a passage through a gapless regime, has never been observed to my knowledge.

\vskip .2cm
One may ask what should happen to a gap with accidental nodes upon adding disorder.    Here again two possible scenarios obtain: the first, in the case that interband scattering dominates, leads to the formation of an impurity band that overlaps the Fermi level, with concomitant increase of the Fermi level density of states with disorder.    This evolution is then similar to the $d$-wave one-band case, as in cuprates\cite{Balatskyreview}.   On the other hand, if intraband scattering dominates (as is usually the case for screened Coulomb impurities), the effect will be primarily to average the anisotropy in the gaps themselves.  In such a case the averaging can lead to the lifting of the nodes altogether\cite{Mishra_09}.
\vskip .2cm
\begin{figure}[tbp]
\includegraphics[angle=0,width=0.5\linewidth]{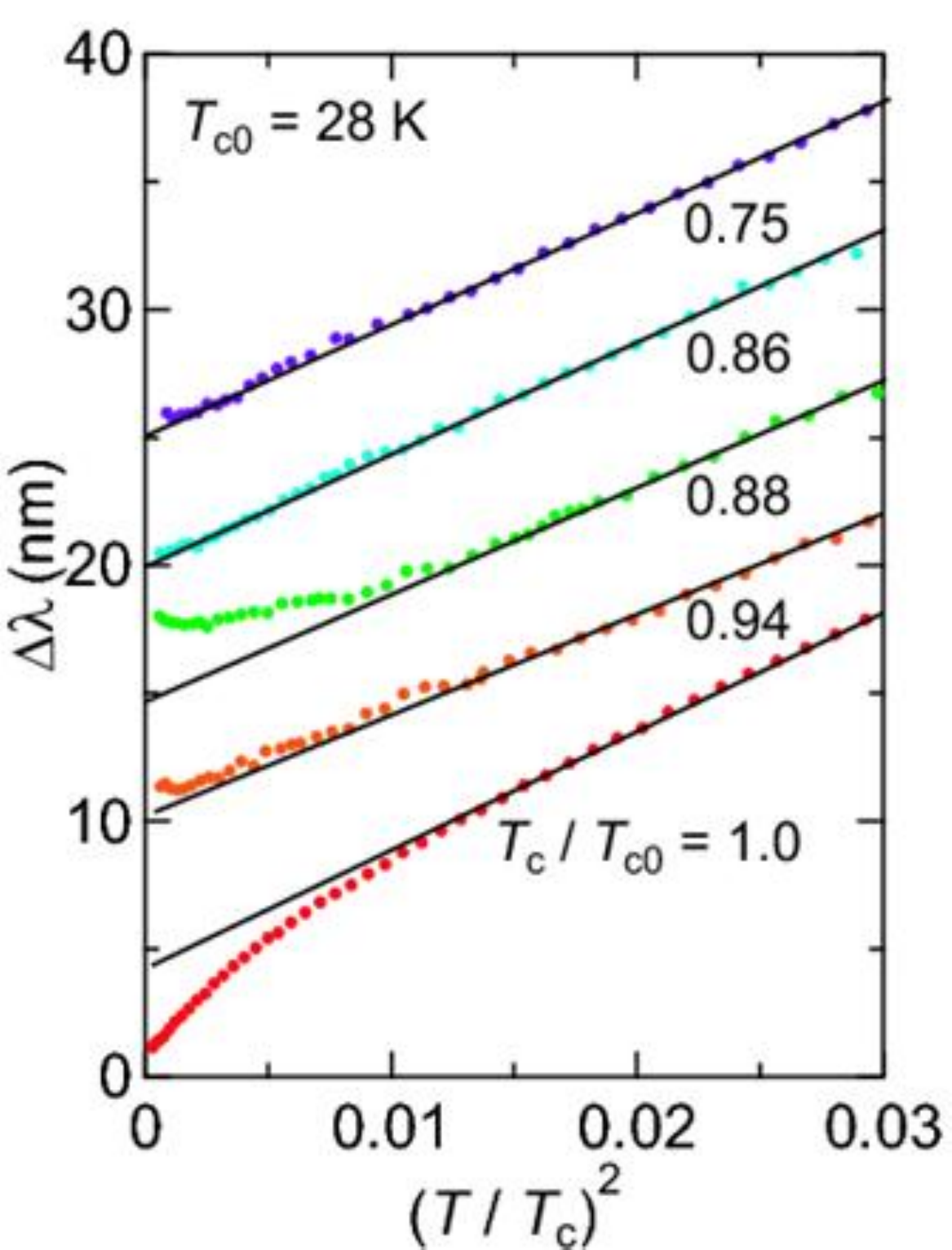}
\caption{(Color online.) Effect of electron irradiation on the low-temperature penetration depth in BaFe$_2$(As$_{1-x}$P$_x$)$_2$ single crystals:
 change in the magnetic penetration depth $\Delta\lambda$ plotted against $(T/T_{\rm c})^2$ for the  $T_{\rm c0}=28$\,K sample.   Lines are the $T^2$ dependence fits at high temperatures. From Ref. \onlinecite{Shibauchi14}.}
\label{fig:gapevolution_expt}
\end{figure}
\vskip .2cm
Note that node lifting can occur either in nodal $s_\pm$ or $s_{++}$ states (see Fig. \ref{fig:gapevolution_theory}a)), so one may ask how they are to be distinguished.  If the gap averaging continues as disorder is increased, eventually both gaps will be isotropic, and intraband scattering will no longer play a role.  In such a case,  the remaining smaller interband scattering potential will begin to dominate.  Then -- only in an $s_\pm$ state -- it may create bound states and an impurity band within the full gap created when the initial nodes in the clean system were lifted, as shown in Fig.  \ref{fig:gapevolution_theory}b).   As complicated as this scenario may seem, this effect was apparently observed   in penetration depth measurements on electron-irradiated P-doped BaFe$_2$As$_2$ by Mizukami et al.\cite{Shibauchi14}, shown in Fig. \ref{fig:gapevolution_expt}.  These authors concluded that this system must have a nodal  $s_\pm$-wave ground state due to the apparent observation of the impurity band appearance at the highest irradiation doses.  Note that there should be various signatures of this
effect in a nodal $s_\pm$ system\cite{YWang_Tcsuppression13}, e.g.  the analogous evolution for NMR spin lattice relaxation $1/(T_1T)$ should be $T^3\rightarrow
T\rightarrow \exp(-\Omega_G/T) \rightarrow  T$. Similarly, the residual linear $T$ term in the thermal conductivity
$\kappa(T\rightarrow 0)/T$ should initially be nonzero, disappear when the nodes are lifted, and then reappear with increasing disorder.  In the $s_{++}$
case, the last step in each sequence is missing,  because interband scattering cannot give rise to
low-energy bound states.
\vskip .2cm

\subsection{Quasiparticle interference}
\label{subsec:qpi}

   \begin{figure}[tbp]
\includegraphics[angle=0,width=0.8\linewidth]{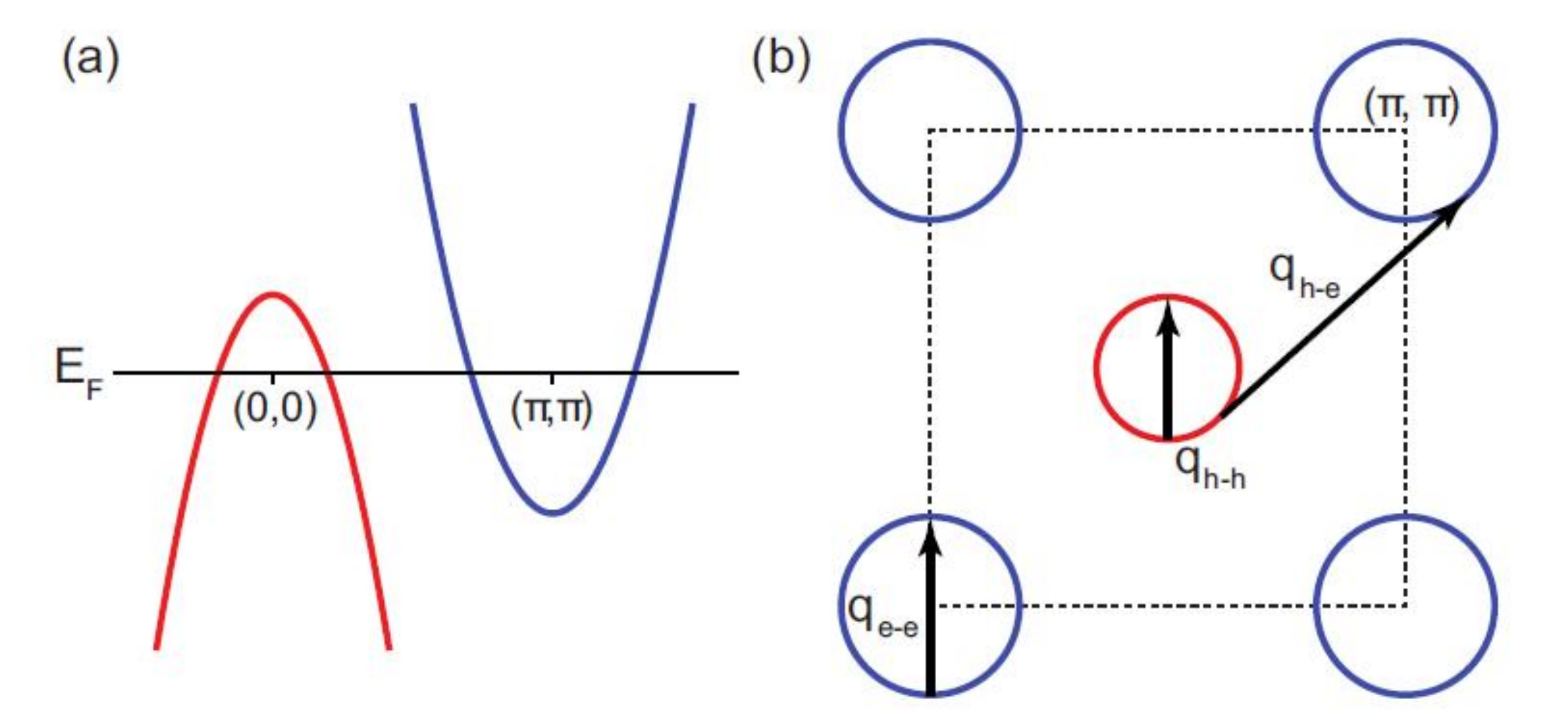}
\caption{(Color online.) Model Fermi surface and scattering processes for analysis of QPI data in LiFeAs.  From Chi et al. \cite{Chi2014}}
\label{fig:QPIgeometry}
\end{figure}
Quasiparticle interference (QPI), or Fourier transform STM spectroscopy\cite{Sprunger},
 is sensitive to  the wavelengths of Friedel oscillations caused by defects present in a metal or superconductor.  In the absence of disorder, the quasiparticle signal disappears; yet if disorder is not too strong,  information on  the electronic structure of the {\it pure} system can be obtained.   The wavelengths of the Friedel oscillations  appear in the form of peaks in the Fourier transform conductance maps at particular wavevectors $\vq(\omega)$ which disperse as a function of  STM bias $V=\omega/e$.    The interplay between theory and experiment in this
 niche field has a checkered history\cite{HEM15}.    The usual method has been to calculate a QPI pattern at a particular bias, given   a theory of the superconducting state, and then compare with the
  expermental maps.   However typically the sources of disorder and exact impurity potentials are unknown, so the weights of the various $\vq$ are imperfectly reproduced, and the process of comparison turns out to be somewhat subjective, to say the least.  In addition, the STM tip measures the LDOS at the tip position several \AA  \, above the surface,
   not necessarily where superconductivity occurs\cite{Kreisel15}. Fortunately,  the positions of the $\vq$ do not depend on these effects\cite{Capriotti03} and are related only to the band structure, including the superconducting gap function, so considerable information can be semi-empirically extracted without depending on detailed theoretical calculations.

\begin{figure}[tbp]
\includegraphics[angle=0,width=0.44\linewidth]{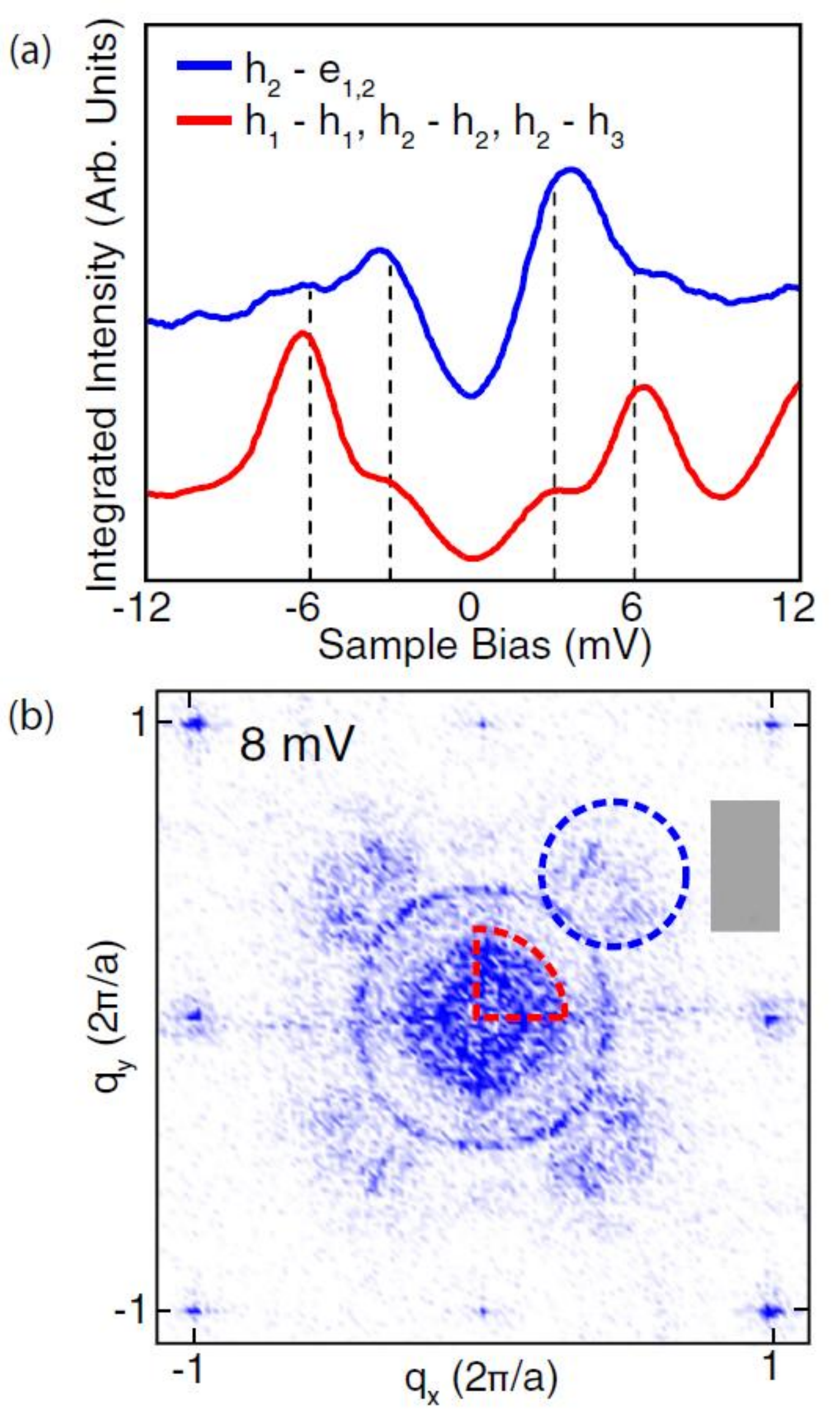}
\includegraphics[angle=0,width=0.45\linewidth]{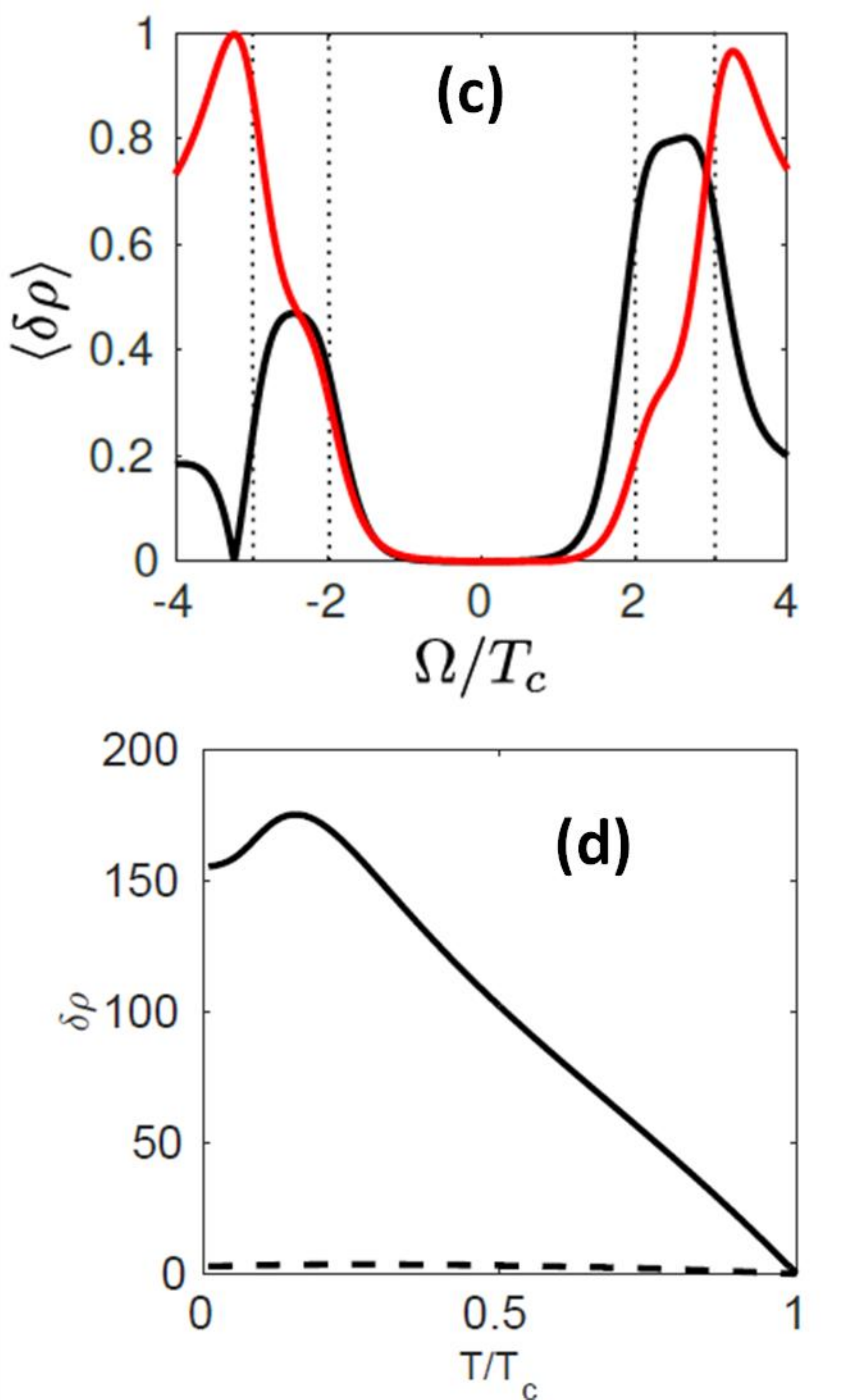}
\caption{(Color online.) (a) q-scan indicating area integrated to define weight in intraband (red) and interband(blue) peaks.  (b) Low temperature bias scan of weights defined as in (a). From Chi et al. \cite{Chi2014} (c) Calculation of intra- (red) and interband (black) weights from Ref. \onlinecite{HEM15}; (d) Proposal of Ref. \onlinecite{HEM15} to distinguish $s_\pm$ (solid) and $s_{++}$ (dashed)
by measuring temperature dependence of antisymmetrized interband LDOS $\langle \delta \rho_{inter}^-(\Omega;T) \rangle$ vs. $T$ for $\Omega$ between two gap scales $|\Delta_1|$ and $|\Delta_2|$. }
\label{fig:QPI}
\end{figure}

\vskip .2cm
  Can QPI information be used to give information on the sign changes of the gap over the Fermi surface?   Some information can be gleaned by noting that subsets of these $\vq$
connecting  gaps of equal or opposite sign on the Fermi surface can be enhanced or not,  according to the type of scatterer\cite{Nunner2006}.  In Ref. \onlinecite{PeregBarnea2008}
it was suggested that  a  disordered vortex lattice should act  like a scattering center of the Andreev type  for quasiparticles  (scattering from order parameter gradients); increasing the
field was therefore in this sense equivalent to introducing  disorder in a controlled way.   This analysis  was  first performed  on the cuprate NaCCOC  by Hanaguri et al.\cite{Hanaguri2009}, and more recently on the Fe-based superconductor
 Fe(Se,Te)\cite{Hanaguri2010}.   These authors identified three $\vq$ related to hole-electron scattering  and two associated with two different electron-electron $\vq$ . The last two
peaks were enhanced by field with respect to the first one,  which led them to
conclude that the hole  and electron pockets have opposite signs of the order
parameter, consistent with a $s_{\pm}$ state\cite{Wang2010,Akbari2010}.  Recently Chi et al.\cite{Chi2014}  did a QPI analysis of their data on LiFeAs in  zero external field.  They identified two $\vq$ peaks,   a large one  corresponding to interband and a small
one corresponding to intraband scattering (see Fig. \ref{fig:QPIgeometry}).  They observed that with decreasing bias, one peak was suppressed while the other enhanced, and argued  that this  was possible only with $s_\pm$ pairing (Fig. \ref{fig:QPI}(a),(b)).

To make their claims, both experiments used reasoning  based on heuristic arguments by Maltseva and Coleman\cite{Maltseva2009} stating that the scattering processes between
wave vectors $\k,\k'$   were weighted in the Fourier transform density of states by the BCS coherence factors, $(u_\k u_{\k'}\pm v_\k v_{\k'} )^2$, where $u_\k,v_\k$ are the
  usual Bogoliubov variational coefficients.  This elegant argument is unfortunately not strictly correct, as shown in Ref. \onlinecite{HEM15}, which gave examples where this notion can in fact be quite misleading.    For example,  by analogy with coherence factors in BCS, one would expect the interband and intraband QPI weights, measured as a function of temperature,
  to increase or decrease just below $T_c$, by analogy with NMR relaxation or ultrasound attenuation\cite{Tinkham}.    This is not the case\cite{HEM15}.  However, a) the
  argument that the different sign slopes of the conductance weights at the upper gap edge implied an order parameter sign change probably is correct, as shown in
  Fig. \ref{fig:QPI}(c).  Still the claim should be treated with caution, and a much more definitive criterion proposed by Ref. \onlinecite{HEM15}, namely the temperature dependence of the
  intraband antisymmetrized conductance weight (Fig. \ref{fig:QPI} (d)), has not yet been tested.

\section{Conclusions}

This review has been targeted at experimental and theoretical developments  related to the   iron-based superconducting state, mostly since 2011 when a previous review\cite{HKMReview2011} examined the state of the art.  Here  I've focussed on a set of phenomena which seem to me the most interesting and important currently.  In 2011, the  the familiar 1111, 111 and 122 systems possessing both hole and electron Fermi surface pockets dominated the discussion, and  the picture of  $s_\pm$ superconducting states formed by $(\pi,0)$ spin fluctuations was both intuitive and appealing.   The pressing questions at that time seemed to be deciding if some or all FeSC could be shown to have
sign-changing $s$-wave ground states, and understanding the origin of gap nodes in some systems.

At the present writing, the field  seems just as vigorous, in fact infused with enthusiasm to answer a new set of pressing questions.   These are mostly related to a new class of systems which seem to belong to a different paradigm than those discussed in 2011, namely
those lacking either electron or hole pockets.     Does the existence of such systems with high critical temperatures imply that a completely new physical mechanism for superconductivity is at work, or is the ``standard model" capable of subsuming these materials
into a different subclass, also driven by repulsive, primarily interband interactions?  Is there any evidence for an important role for $d$-wave interactions in stabilizing these systems, and how can one best detect this?      What is the role of phonons in ``bootstrapping" the spin fluctuation interaction, particular in the fascinating FeSe/STO monolayers?    Under what circumstances can phonons help or hurt the critical temperature in a strongly spin-fluctuating system?  All these questions are under intense debate.

There are other questions which affect the more familiar systems with both hole and electron pockets as well.  Can one find a system where a transition between two channels of superconducting order, e.g. $s$ and $d$ takes place as a function of doping or pressure?
What is the nature of this transition,  does it involve time-reversal symmetry breaking states and if so, how does one  detect them?  What new types of pairing are possible utilizing the orbital degrees of freedom of the Fe $d$-electrons, and are there systems with interactions that can stabilize such exotic states?   Do such states necessarily involve pairing of electrons at different energies, and if so is this
 necessarily fatal, given new understanding of the role of incipient bands in pairing?   Is nematicity a passive bystander in superconductivity, or a driver?  I have described some important contributions to these questions, but they are not settled.

 In the introduction I pointed to the fact that the lack of universality in many of the properties of the Fe-based superconductors, while disturbing to a certain mindset reared on cuprate physics, can be viewed as an advantage.  The hope is that, since electronic correlations are less pronounced in these systems than the cuprates, a quantitative theory of superconductivity is not out of reach.  In this context, the diversity of these systems can therefore be thought of as a challenge to build a theory  explaining this diversity using first principles input, ultimately helping in the search for new superconducting materials.  It should be clear from this article, for example the description of the situation regarding LiFeAs,  that theoretical methods are not quite refined enough to  assist in this effort.

 My own view is that theoretical progress is needed in two directions.  First, the tight-binding models taken from DFT are not adequate for the description of the most interesting systems, which tend to be more strongly correlated.    Methods are needed to accurately reproduce the
 correlated electronic structure over a few hundred meV near the Fermi surface accurately.  While LDA+DMFT has done well enough to
 reproduce some qualitative trends, it still produces inaccurate Fermi surfaces for LiFeAs and FeSe, to name two.  It is clear, however, that  electronic structure methodology is advancing rapidly\cite{BiermannCRAS}, and I have little doubt that this situation will be vastly improved within a few years.  The second, more difficult, question is whether the treatment of electronic excitations involved in pairing is adequate.   It is possible that the simple Fermi-surface based RPA treatment of spin fluctuations will be sufficient to predict semiquantitative  trends in pairing correctly once adequate tight-binding models for the more correlated systems  are available.  More likely is that the current machinery will need to be improved at least by accounting more carefully for the contribution of the near-Fermi level states to pairing, and possibly to include further electronic vertex corrections and, in some cases, phonons.     All these developments will require significant increase in computational power.

 Looking carefully at cases like LiFeAs where various theoretical approaches have been brought to bear, however,
 I would say that there is considerable reason for optimism.  It is remarkable that the detailed gap structure over much
 of the Fermi surface, and the relative sizes of gaps, can be obtained accurately without fine tuning.  Experimental progress since 2011, particularly the
 discovery of new systems with nonpolar surfaces accessible to high resolution STM and ARPES, has allowed theory and
 experiment to follow each other in a tight circle,  an ideal scientific situation which seems destined  to lead to a solution to many of these fascinating
  problems in the not-too-distant future.

\section{Acknowledgements}
I would like to thank A. Chubukov, I. Eremin, R. Fernandes,  A. Linscheid, T. A. Maier, S. Maiti,  I.I. Mazin,  V. Mishra, D.J. Scalapino, R. Valenti, and  Y. Wang for critical readings of the manuscript.    Research was partially supported
 by the Department of Energy under Grant No. DE-FG02-05ER46236, and by the National Science Foundation
 under Grant NSF-DMR-1005625.

\bibliographystyle{apsrev4-1}

\clearpage

\end{document}